\documentclass[aps,prd,onecolumn,superscriptaddress,nofootinbib,floatfix,10pt]{revtex4-2}

\usepackage{amsmath,amssymb,amsthm,mathtools}
\usepackage{graphicx}
\usepackage{longtable}
\usepackage{bm}
\usepackage{silence}
\usepackage[colorlinks=true,linkcolor=blue,citecolor=blue,urlcolor=blue]{hyperref}
\allowdisplaybreaks

\makeatletter
\renewcommand{\p@subsection}{}
\makeatother

\numberwithin{equation}{section}

\newcommand{\dd}{\mathrm{d}}

\newcommand{\lhat}{\hat l}
\newcommand{\Om}{\Omega}

\begin{document}

\title{Axial-metric/polar-gauge perturbations of the Maldacena--Milekhin--Popov wormhole}

\author{Abhishake Sadhukhan}
\email{abhishake.physics@presiuniv.ac.in}
\affiliation{Department of Physics, Presidency University, Kolkata - 700073, India}

\begin{abstract}
We study linear perturbations of the Maldacena-Milekhin-Popov (MMP) wormhole in the parity sector pairing axial metric and polar gauge perturbations. The lowest Landau level (LLL) of the fermions responds through its charge. Its exact Schwinger current gives the longitudinal electric field the MMP screening mass $m_S^2=\alpha q^2$, and the axial metric perturbation shifts the worldsheet gauge field of every LLL state at relative order $l(l+1)/q$. We assume that the Casimir stress tensor follows the tilting field lines. Force balance and charge conservation then fix a polarisation current, the counterpart of the polar Hall current, up to self-conserved currents. The higher Landau levels enter through the one-loop Euler-Heisenberg action, to first order in $\eta_{\rm H}=g_c^2/12\pi^2$. Virtual transitions to them could enter at the order of the $l(l+1)/q$ shift and are not estimated. At first order in the fermion backreaction $\alpha$ the system reduces to three coupled master equations in closed form, and mouth and throat match. Consistency does not select the fermion constitutive relation. Among local closures without derivatives, covariance fixes it up to one kernel. At $\alpha=0$ the throat has a screened, a light massive and a massless neutral channel, and the Schwinger response already breaks the isospectrality of the two parities. The higher levels raise the squared light mass by $\eta_{\rm H}l(l+1)$ in AdS$_2$ units and break it further. At the formal point $m_S^2=0$ the massive channels have real $O(\alpha)$ shifts on every closed-throat level, as does the light channel at the physical $m_S^2$ on every level below the screened channel. The neutral channel has an exact zero-energy solution, a persistent current along the field lines, which the $O(\alpha)$ mouth terms and the higher levels leave intact. Finite-element runs find no bound state below it, and its stability is open.
\end{abstract}

\maketitle
\raggedbottom
\tableofcontents
\section{Introduction}
\label{sec:intro}

A traversable wormhole requires a violation of the null energy condition at its throat \cite{Morris:1988cz}, and quantum fields can provide it \cite{Gao:2016bin,Maldacena:2018lmt,Fu:2018oaq}. The construction of Maldacena, Milekhin and Popov (MMP) \cite{Maldacena:2018gjk} takes the negative energy from massless charged fermions in an asymptotically flat solution of four-dimensional Einstein--Maxwell theory. Its two mouths are near-extremal magnetic Reissner-Nordstrom black holes of opposite charge, and a long AdS$_2\times S^2$ throat joins them. In the monopole field the fermions fill a lowest Landau level (LLL) of $q$ states, one per flux quantum \cite{Dirac:1931kp,Atiyah:1963zz,Wu:1976ge}, and these states move along the field lines as two-dimensional massless fermions \cite{Alford:1992yx}. The field lines close through the ambient region, and the Casimir energy of this two-dimensional conformal field theory (CFT) on the closed lines holds the throat open. The fermionic backreaction is measured by $\alpha=g_c^2/4\pi^2q$, and we expand the perturbations to first order in their amplitude $\epsilon$. No such throat can be built within effective field theory from higher-derivative corrections alone \cite{Kanai:2025zgq}. Charged traversable throats can form dynamically \cite{Koga:2025bqw}, and the transmission of low-frequency scalar and charged-fermion probes through a wormhole of this kind has been computed in \cite{Freivogel:2026ujn}.

Without a horizon, the long throat of the MMP wormhole is a nearly reflecting cavity and a possible source of gravitational-wave echoes \cite{Cardoso:2016rao,Cardoso:2017cqb,Bueno:2017hyj,Konoplya:2016hmd,Sharma:2026gvu}. Its echoes were computed for scalar and gauge test fields in \cite{Mondal:2025tht}, and the transmission analysis of throat spacetimes in \cite{Riley:2026avn} finds the stronger echoes at higher $l$ reported there consistent with its $l$-dependent barrier heights. Quasinormal modes of wormholes held open by a phantom scalar field are known in both parities. For the Ellis-Bronnikov wormhole the scalar and axial modes approach those of a Schwarzschild black hole of equal mass when the mass is large at fixed $l$, with a difference that falls inversely with the mass \cite{Blazquez-Salcedo:2018ipc}. The polar and axial modes of the same wormhole are isospectral when it is massless and split into distinct branches when it has mass \cite{Azad:2022qqn}. At the fiducial parameters of \cite{Mondal:2025tht} light needs $\pi\ell/c\simeq1.5\times10^{25}$\,s to cross the throat, where $\ell\simeq1.4\times10^{35}$\,cm is the length scale of the throat defined in Sec.~\ref{sec:sector}. The question is therefore one of structure and not of observation. Test fields leave out the coupled gravitational and electromagnetic perturbations and the response of the fermions. The companion paper \cite{Sadhukhan:2026czl} reduced the linearised Einstein--Maxwell--fermion system to master equations of Zerilli--Moncrief type in one of the two parity sectors. The master-equation program for spherically symmetric backgrounds goes back to Regge, Wheeler, Zerilli, Vishveshwara, Moncrief and Chandrasekhar \cite{Regge:1957td,Zerilli:1970se,Vishveshwara:1970cc,Zerilli:1974ai,Moncrief:1974am,Moncrief:1974gw,Moncrief:1974ng,Chandrasekhar:1979iz,Chandra,Martel:2005ir}. Master functions now also cover new branches, the Bertotti--Robinson limit of the extreme Reissner-Nordstrom black hole, and a Hamiltonian setting in which Darboux transformations are canonical \cite{Lenzi:2021wpc,Liu:2025reu,deCesare:2024csp,Lenzi:2026aay}. One-loop corrections to a charged background couple the master equations and spoil the isospectrality of the two parities \cite{delRio:2024kms}, and magnetically charged black holes are also studied beyond Einstein--Maxwell theory \cite{Stashko:2026kja}. The present paper treats the second parity sector at $l\ge2$.

The monopole field strength is odd under parity. The linearised field equations therefore couple a metric perturbation of one parity to a gauge perturbation of the other \cite{Pereniguez:2023wxf,DeFelice:2023rra}. The companion paper treated polar metric perturbations, which couple to axial gauge perturbations. There the linearised Bianchi identities hold in the throat only when the LLL stress tensor responds on the moving field lines through its conformal Ward identity and an azimuthal Hall current is induced. Here we treat axial metric perturbations, which carry the other gravitational degree of freedom and couple to polar gauge perturbations. The polar gauge perturbation carries the electric field along the magnetic field lines. Together the two papers cover both parity sectors of the $l\ge2$ perturbations, at first order in $\alpha$ and for the lowest Landau level.

The fermion response is different in this sector, and it is the main subject of the paper. In the companion sector the perturbation deforms the two-dimensional metric of each field line, through the metric and through the local field strength, and the conformal anomaly fixes the response. Here the two-dimensional metric is untouched. The axial metric perturbation has no $(\tau,\rho)$ block and does not perturb the sphere radius. A two-dimensional stress tensor has no linear response to a gauge field by charge conjugation. The fermions respond through their charge instead. The components $a_\tau,a_\rho\propto Y(\theta)$ pull back to a two-dimensional gauge field on the worldsheet of every field line. The exact linear response of a massless two-dimensional charged fermion is the Schwinger current \cite{Schwinger:1962tp}. In bosonised form \cite{Coleman:1975pw,Mandelstam:1975hb} it reads $j^a=\frac1\pi\tilde\epsilon^{\,ab}\partial_b\chi$ with $\Box_2\chi=E$. Here $\Box_2$ is the wave operator of the worldsheet metric, $\tilde\epsilon^{\,ab}$ the alternating symbol and $E$ the worldsheet electric field. Lifted over the flux-threaded sphere, this gives the electric master field the mass
\begin{equation}
m_S^2 \;=\; \frac{g_c^2 q}{4\pi^2} \;=\; \alpha\,q^2 .
\end{equation}
This is the Schwinger mass of the two-dimensional gauge field in Appendix~A of MMP, the screening term of their Appendix~G and the Schwinger mass of eq.~(4.82) of \cite{Freivogel:2026ujn} (Sec.~\ref{sec:linesum}). With $q=g_c^2/4\pi^2\alpha$ one has $m_S^2=g_c^4/16\pi^4\alpha$, and at fixed $g_c$ the Schwinger mass is large whenever $\alpha$ is small. The response does not switch off in the mouth, where it is far larger than the response of the stress tensor (Sec.~\ref{sec:whynot}).

In the throat two further effects appear. The transverse components $\delta F_{a\theta}\propto Y'$ are an electric field across the field lines for $a=\tau$ and a twist of them for $a=\rho$, and $h_{a\phi}$ displaces the lines azimuthally. In the background the LLL fermions move on the timelike invariant plane of $F$, the $(\tau,\rho)$ plane. At $O(\epsilon)$ this plane tilts toward $\partial_\phi$. We assume that the Casimir stress tensor $T^Q$ of the LLL fermions follows the tilted plane. Its covariant lift then acquires mixed components $\delta T^Q_{a\phi}$. The $\tau$, $\rho$ and $\theta$ components of the force balance $\nabla_\mu T_Q^{\mu\nu}=-F^{\nu\lambda}J_\lambda$ are then satisfied identically, but the $\phi$ component is not. The tilted Casimir stress tensor carries azimuthal momentum, and the only way the LLL fermions can supply the required torque is a current across the field lines. The force balance determines this current $J^\theta$, and charge conservation fixes the polarisation charge and current $J^a_{\rm pol}\propto l(l+1)$. This is the counterpart of the Hall current of \cite{Sadhukhan:2026czl} with the roles of $\theta$ and $\phi$ exchanged. Each guiding centre moves by $O(1/q)$, and summed over the $q$ field lines the current is $O(1)$ and independent of $q$.

With all three ingredients the linearised Bianchi identity and charge conservation hold identically at $O(\alpha)$, and the system reduces to three coupled master equations of Regge--Wheeler type. In both sectors these identities reject truncated responses and do not select the lift. Control experiments show how little they fix here. The Schwinger current is identically conserved and exerts no force at $O(\epsilon)$, and the Bianchi tests pass with or without it. The response-free closure $\delta T^Q_{a\phi}=\mathcal P\,h_{a\phi}$, with $\mathcal P$ the sphere pressure of the Casimir stress tensor, also passes every test. The closure with no fermion response at all passes in the throat at $O(\alpha)$ and fails only in the mouth. Among local closures without derivatives, covariance under axial diffeomorphisms reduces the closures to a family with one response kernel (Sec.~\ref{sec:lll}), and the assumption stated above fixes the kernel. The drift lift is then the local Lorentz transformation of the background stress tensor. The charge response, by contrast, we compute on the exact LLL states. Beyond the sum over independent field lines it has two corrections of relative order $\lhat/q$. The finite width of the states renormalises $m_S^2$, and their kinetic azimuthal momentum and the spin connection couple them to $h_{a\phi}$ as an extra worldsheet gauge field. The second is of order $1/q=4\pi^2\alpha/g_c^2$, the order of the $O(\alpha)$ terms, and we include it.

The remaining results are the following. In the mouth the Einstein--Maxwell part of the system reduces to two coupled master equations with an identically vanishing first-derivative matrix and a frequency-independent potential whose decoupled components are Chandrasekhar's axial Reissner-Nordstrom potentials $V^{(-)}_{1,2}$ \cite{Chandrasekhar:1979iz,Chandra}. Here they are derived on the magnetic background with a polar gauge field, and the duality between the sectors is an output. With the Schwinger response the mouth becomes a three-channel system. In the throat at $\alpha=0$ the mass matrix has one screened mode, one light massive mode and one exactly massless neutral mode for every $m_S^2>0$, with $E=0$. The squared mass of the light mode tends at large $m_S^2$ to $l^2+l+2$, which is not of the form $k(k+1)$ with integer $k$. The neutral mode carries the bosonised charge and current of the field lines with the angular profile $Y(\theta)$ and no net charge. It is the gravitationally dressed multipole of the massless relative sector of MMP's Appendix~A. The $O(\alpha)$ corrections to all three channels are computed in closed form. The mouth carries the drift lift and the polarisation currents at $O(\alpha)$ as well, and the two regions agree on every coefficient that both calculations fix, the drift terms of order $\alpha\rho^2$ included. At the fiducial parameters of \cite{Mondal:2025tht} the full potential matrices agree at the junction to $(\alpha\rho)^2$ in all three channels. At the physical $m_S^2$ the light massive channel has real first-order shifts on every level below the screened channel. At the formal point $m_S^2=0$ the Schwinger sector decouples, the massive channels are Poschl-Teller problems with the integer masses $l(l-1)$ and $(l+1)(l+2)$ of the companion sector, and their $O(\alpha)$ shifts are real on every level. Their differences from the shifts of the companion sector show that the two parities are not isospectral at $O(\alpha)$ in the closed throat. Part of the difference is fixed by the Einstein--Maxwell mouths, and the rest comes from the fermion response. At the physical $m_S^2$ the Schwinger response breaks it already at $\alpha^0$, in the throat and in the mouths. On the levels we computed all these $O(\alpha)$ shifts are negative for the clock of a static observer at the centre of the throat. On the lowest light levels at $g_c\sim1$ the finite value of $m_S^2$ lowers the levels by far more, and the coupling to $h_{a\phi}$ raises them by the fraction $g_c^2/4\pi^2$ of that amount. The null-energy contraction along radial null rays is unchanged at $O(\epsilon)$. The neutral channel needs a global treatment. At $\alpha^0$ it has an exact zero-energy solution in each mouth that continues as a constant through the throat. In the time domain it is a persistent current along the field lines with no charge density. The finite-element runs find no bound state at their resolution, and the channel sits at the threshold for one. The $O(\alpha)$ terms of the mouth and the coupling to $h_{a\phi}$ leave this solution intact. Whether the next order turns it into a bound state is not decided here. The one-loop Euler-Heisenberg action of the higher Landau levels, which makes the coupling run, brings a second small parameter $\eta_{\rm H}=g_c^2/12\pi^2$ that $1/q$ does not suppress (Sec.~\ref{sec:hll}). These levels act on the electric field along the field lines as a dielectric. In units of the AdS$_2$ radius they raise the squared light mass by $\eta_{\rm H}l(l+1)$ at large $m_S^2$, although on the clock of the static observer at the centre of the throat the light levels move down. They split the levels shared by the massive channels at $m_S^2=0$ and leave the zero-energy solution of the neutral channel exact. Their response to an electric field along the lines differs from their response to a change of $|F|$, and they break the isospectrality of the two sectors even without the lowest level.

Section~\ref{sec:sector} fixes the ansatz and the kinematics of the sector. Section~\ref{sec:algorithm} describes the reduction algorithm and the consistency conditions. Section~\ref{sec:mouthS} derives the Schwinger response of the LLL in bosonised form, which both regions use, and the coupling of the LLL states to $h_{a\phi}$. Section~\ref{sec:lll} constructs the LLL response in the throat and states the drift lift as an assumption. Section~\ref{sec:alpha} derives the throat master equations to $O(\alpha)$ and reports the control experiments. It computes the $O(\alpha)$ shifts at $m_S^2=0$ and the $O(\alpha)$ potential of the neutral channel, takes the $\alpha\to0$ limit, and computes the shifts of the light massive channel at the physical $m_S^2$. Section~\ref{sec:mouth} treats the mouth to $O(\alpha)$. Its leading part is the Einstein--Maxwell mouth with its Schwinger extension and the light neutral sector, and the section ends with the matching to the throat. Section~\ref{sec:iso} compares the spectra of the two parity sectors. Section~\ref{sec:stability} analyses the energy and linear stability. Section~\ref{sec:hll} adds the one-loop action of the higher Landau levels. Section~\ref{sec:summary} collects the results and assumptions. The complete radial equations and the $O(\alpha)$ coefficient matrices are collected in the appendices.

\section{Ansatz and kinematics of the sector}
\label{sec:sector}

\subsection{Background and conventions}

The theory and background are those of MMP \cite{Maldacena:2018gjk} and of the companion paper \cite{Sadhukhan:2026czl}, whose conventions we adopt. The bosonic action is
\begin{equation}
\mathcal I=\int\dd^4x\sqrt{-g}\Big[\frac{{}^{(4)}\!R}{16\pi G}-\frac{F_{\mu\nu}F^{\mu\nu}}{4g_c^2}\Big]+\mathcal I_{\rm fermion},\qquad \kappa\equiv8\pi G,
\end{equation}
with field equations
\begin{equation}
G_{\mu\nu}=\kappa\big(T^{\rm mag}_{\mu\nu}+T^{Q}_{\mu\nu}\big),\qquad \nabla_\mu F^{\mu\nu}=g_c^2\,J^\nu .
\label{eq:eom}
\end{equation}
Here $J^\nu\equiv-(1/\sqrt{-g})\,\delta \mathcal I_{\rm fermion}/\delta A_\nu$. It is minus the current $(1/\sqrt{-g})\,\delta \mathcal I_{\rm fermion}/\delta A_\nu$ that couples to $A_\nu$ in the action, and all currents below are given in the convention of \eqref{eq:eom}. The background is static and spherically symmetric,
\begin{equation}
\dd s^2=-A(r)\,\dd t^2+B(r)\,\dd r^2+R(r)^2\,\mathring g,\qquad \mathring g=\dd\theta^2+\sin^2\theta\,\dd\phi^2,\qquad
\bar A=\frac q2\cos\theta\,\dd\phi,\qquad \bar F_{\theta\phi}=-\frac q2\sin\theta ,
\label{eq:background}
\end{equation}
with $q>0$ an integer \cite{Dirac:1931kp} and $r_e^2=\kappa q^2/8g_c^2$ (MMP eq.~(2.3), numbered as in its arXiv version). With the orientation $\dd\theta\wedge\dd\phi$ one has $\frac1{2\pi}\int_{S^2}F=-q$. The density of magnetic field lines is $|F_{\theta\phi}|/2\pi$. It counts the LLL modes and integrates to $q$. The two regions are
\begin{align}
\text{mouth}\quad& A=\frac1B=1-\frac{2GM}{r}+\frac{r_e^2}{r^2}=\frac{\Delta}{r^2},\qquad R=r,\qquad \Delta\equiv r^2-2GMr+r_e^2,\\
\text{throat}\quad& A=r_e^2\big(f+\alpha\gamma(\rho)\big),\qquad B=\frac{r_e^2}{f+\alpha\gamma(\rho)},\qquad R^2=r_e^2\big(1+\alpha\varphi(\rho)\big),\qquad f\equiv1+\rho^2 .
\label{eq:throatbg}
\end{align}
Here $GM=r_e+\delta_M$ with $\delta_M<0$, there is no horizon, and $\alpha=g_c^2/4\pi^2q$. At $O(\alpha)$ the extremal mass of the mouth moves to $r_e(1-\alpha/18)$ (Sec.~\ref{sec:mouthalphasource}), and $\delta_M$ is measured from that value. With the higher Landau levels it is $r_e(1+\eta_{\rm H}-\alpha/18)$ at first order (Sec.~\ref{sec:hll-action}). The dimensionless throat coordinates are $\tau=t/\ell$ and $\rho=\ell(r-r_e)/r_e^2$ with $\ell=4r_e/\pi\alpha$. The mass defect is second order in the coupling, $\delta_M/r_e=-\pi^2\alpha^2/32$ (MMP eq.~(5.42)). It is too small to enter any coefficient at the order we work, and its sign removes the horizon. The corrected background functions of \cite{Sadhukhan:2026czl} satisfy the background equations of the present sector as well,
\begin{equation}
\gamma(\rho)=-\frac{4\rho^2}{3}-\rho\,(\rho^2+3)\arctan\rho+\log f,
\qquad
\varphi(\rho)=\rho\arctan\rho+\frac43 .
\label{eq:gamphi}
\end{equation}
At large $\rho$ one has $\varphi\simeq\frac\pi2\rho+\frac13$ and $\gamma\simeq-\frac\pi2\rho^3-\frac{\rho^2}3-\frac{3\pi}2\rho+2\log\rho+\frac83$. As in \cite{Sadhukhan:2026czl} the additive constant of $\gamma$ is fixed by $\gamma(0)=0$. For any constant $\gamma_0$ the substitution $\tau\to\tau(1+\alpha\gamma_0/2)$, $\rho\to\rho(1-\alpha\gamma_0/2)$ turns the metric with $\gamma$ into the metric with $\gamma+\gamma_0$ at this order. With $\gamma(0)=0$ one has $A(0)=r_e^2$, and $\Om$ is the frequency measured by a static clock at the centre of the throat, in units of $1/r_e$.  This choice is what we call the clock. A different constant $\gamma_0$ is a different clock and adds $\gamma_0\Om_0^2$ to every first-order shift $\varsigma$ of Secs.~\ref{sec:shifts} and \ref{sec:lightchannel}. In the throat all displayed equations are written in the coordinates $(\tau,\rho)$. A dot denotes $\partial_\tau$ and a prime denotes $\partial_\rho$ there, and $\partial_t$ and $\partial_r$ in the mouth.

\subsection{The parity pairing and the ansatz}
\label{sec:ansatz}

Under the parity map $(\theta,\phi)\mapsto(\pi-\theta,\phi+\pi)$ the monopole field strength is odd, $\bar F\mapsto-\bar F$. The linearised cross terms carry one factor of $\bar F$ and enforce the selection rule $\mathsf P[h]=-\mathsf P[\delta F]$ between the parity eigenvalues of the metric and gauge perturbations \cite{Sadhukhan:2026czl,Pereniguez:2023wxf,DeFelice:2023rra}. The sector treated here pairs an axial metric with a polar gauge field. We work in Regge--Wheeler gauge \cite{Regge:1957td} at azimuthal index $m=0$, which loses no generality since the background is spherically symmetric. With $Y(\theta)\equiv P_l(\cos\theta)$ and $\lhat\equiv l(l+1)$ the ansatz is
\begin{align}
h_{t\phi}&=h_0(t,r)\,\sin\theta\,Y'(\theta),\qquad
h_{r\phi}=h_1(t,r)\,\sin\theta\,Y'(\theta),\qquad \text{all other } h_{\mu\nu}=0,
\label{eq:hansatz}\\
a_t&=a_0(t,r)\,Y,\qquad a_r=a_1(t,r)\,Y,\qquad a_\theta=a_2(t,r)\,Y',\qquad a_\phi=0 .
\label{eq:aansatz}
\end{align}
Here $h_{\mu\nu}$ and $a_\mu$ are the coefficients of $\epsilon$ in $g_{\mu\nu}=\bar g_{\mu\nu}+\epsilon h_{\mu\nu}$ and $A_\mu=\bar A_\mu+\epsilon a_\mu$. The angular profile $\sin\theta\,Y'$ is the odd-parity vector harmonic at $m=0$. The general axial metric perturbation carries a third amplitude $h_2$ multiplying the odd tensor harmonic. Regge--Wheeler gauge uses the single axial gauge function to set $h_2=0$, which is a complete gauge fixing for $l\ge2$. The axial perturbation is traceless and $\sqrt{-g}$ is unchanged at $O(\epsilon)$. The polar gauge perturbation carries all three even components. The residual $U(1)$ freedom $a_\mu\to a_\mu+\partial_\mu(\Lambda(t,r)Y)$ shifts
\begin{equation}
(a_0,a_1,a_2)\;\longrightarrow\;(a_0+\dot\Lambda,\;a_1+\Lambda',\;a_2+\Lambda),
\end{equation}
and every equation can depend only on gauge-invariant combinations. In terms of the two invariants
\begin{equation}
\mathsf b_0\equiv a_0-\dot a_2,\qquad \mathsf b_1\equiv a_1-a_2',
\label{eq:beta}
\end{equation}
the perturbed field strength is
\begin{equation}
\delta F_{tr}=E\,Y,\qquad E\equiv\dot a_1-a_0'=\dot{\mathsf b}_1-\mathsf b_0',\qquad
\delta F_{t\theta}=-\mathsf b_0\,Y',\qquad
\delta F_{r\theta}=-\mathsf b_1\,Y',\qquad \delta F_{\theta\phi}=\delta F_{t\phi}=\delta F_{r\phi}=0 .
\label{eq:fstr}
\end{equation}
$E$ is the gauge-invariant radial electric field along the flux lines and will be one of the master fields. The transverse components $\delta F_{a\theta}=-\mathsf b_aY'$ are the electric field across the lines and their twist. They drive the drift response of Sec.~\ref{sec:lll}. The flux density $\delta F_{\theta\phi}$ vanishes, and the number of field lines through any patch of the sphere is unchanged. All six radial equations depend on $a_0,a_1,a_2$ only through $\mathsf b_0$, $\mathsf b_1$ and $E$. We therefore carry out the reduction in the gauge $a_2=0$, where $\mathsf b_0=a_0$ and $\mathsf b_1=a_1$. This choice uses up the residual $U(1)$ completely, since $\Lambda=0$ is the only transformation that preserves it. As in \cite{Sadhukhan:2026czl} we rescale the gauge amplitudes,
\begin{equation}
a_i=\frac q4\,w_i\quad(i=0,1),\qquad E\to\frac q4\,E,
\end{equation}
after which $q$ cancels from every Einstein--Maxwell equation. We pass to the frequency domain, $\{h_i,w_i,E,X\}(t,r)=\{h_i,w_i,E,X\}(r)\,e^{pt}$ with $p=-i\omega$, where $X$ is the bosonised LLL scalar introduced in Sec.~\ref{sec:linesum}, and $p=-i\Om$ with $\Om=\omega\ell$ in the throat. A growing mode has real positive $p$. The reduction divides by $p$ and assumes $p\neq0$. Where we set $p=0$ below, we set it in the coefficients of the reduced system. Where a worldsheet index $a\in\{\tau,\rho\}$ appears on a radial amplitude, $h_\tau\equiv h_0$, $h_\rho\equiv h_1$, $\mathsf b_\tau\equiv\mathsf b_0$ and $\mathsf b_\rho\equiv\mathsf b_1$.

We take $l\ge2$ throughout, since the odd tensor harmonic vanishes for $l=0,1$ and Regge--Wheeler gauge is complete only for $l\ge2$. The excluded multipoles have a different content from the companion sector. At $l=0$ one has $Y'=0$. The metric perturbation and $a_\theta$ vanish identically, and the sector reduces to the spherically symmetric electric perturbation $a_t(t,r)$, $a_r(t,r)$, an infinitesimal electric charge on the magnetic background. In the electrovacuum mouth this is the standard one-parameter family of solutions. With the Schwinger response of Sec.~\ref{sec:mouthS} the two Maxwell equations make the displacement flux $\mathcal F_D$ of \eqref{eq:flux} constant, and the anomaly equation \eqref{eq:Xeq} with $E=\sqrt{AB}\,(\mathcal F_D+m_S^2X)/R^2$ then gives the bosonised charge $X$ a mass proportional to $m_S^2$. This is MMP's statement that the two-dimensional gauge field is Higgsed and the Wilson line is fixed at the value minimising the Casimir energy (their Sec.~5.2 and Appendix~A). The only $l=0$ mode of this sector is this massive one. The spherically symmetric metric perturbations are polar, and neither paper treats them. For wormholes held open by phantom scalar fields this is the unstable sector \cite{Gonzalez:2008wd,Bronnikov:2012ch}. At $l=1$ the axial metric perturbation has $\sin\theta\,Y'\propto\sin^2\theta$, and $h_{t\phi}\propto\sin^2\theta$ is the slow-rotation mode. In vacuum it is non-radiative and fixed by the conserved angular momentum, as in the Schwarzschild and Reissner-Nordstrom cases \cite{Regge:1957td,Zerilli:1974ai}. The polar gauge perturbation at $l=1$ is a radiative electric-dipole mode. The tensor amplitude is absent at $l=1$, Regge--Wheeler gauge leaves a residual freedom, and the $l=1$ system requires a separate gauge fixing, which we do not carry out. In the $\alpha=0$ throat the mass $l(l-1)$ of one of the two Einstein--Maxwell channels of Sec.~\ref{sec:alpha0} vanishes at $l=1$, while the other keeps the mass $(l+1)(l+2)=6$.
\section{Radial equations and the reduction algorithm}
\label{sec:algorithm}

\subsection{Harmonic separation}

We expand every field to first order, $g_{\mu\nu}=\bar g_{\mu\nu}+\epsilon\,h_{\mu\nu}$ and $A_\mu=\bar A_\mu+\epsilon\,a_\mu$, and write $\delta$ in front of a quantity for its coefficient of $\epsilon$. The background solves the field equations \eqref{eq:eom} at $O(\epsilon^0)$. At $O(\epsilon)$ we move all terms of each field equation to one side and call the result its residual,
\begin{equation}
\mathcal E_{\mu\nu}\equiv\delta G_{\mu\nu}-\kappa\big(\delta T^{\rm mag}_{\mu\nu}+\delta T^{Q}_{\mu\nu}\big),
\qquad
\mathcal M^\nu\equiv\delta\big[\partial_\mu\big(\sqrt{-g}\,F^{\mu\nu}\big)\big]-g_c^2\,\delta\big(\sqrt{-g}\,J^\nu\big),
\label{eq:residuals}
\end{equation}
and the first-order field equations are
\begin{equation}
\mathcal E_{\mu\nu}=0,\qquad \mathcal M^\nu=0 .
\label{eq:linearised}
\end{equation}
A residual vanishes on a solution when the unknowns satisfy the equations. It vanishes identically when it is zero for any functions once relations derived from the other equations are inserted. An identically vanishing residual carries no new information, and the corresponding equation is redundant. Step~4 of the algorithm below tests exactly this. The Maxwell equation is written in the divergence form, which is equivalent to $\nabla_\mu F^{\mu\nu}=g_c^2J^\nu$ and makes charge conservation an identity. $\delta G_{\mu\nu}$ is the linearised Einstein tensor, and $\delta T^{\rm mag}_{\mu\nu}$ is linear in $h_{\mu\nu}$ and $\delta F_{\mu\nu}$. The fermion terms $\delta T^Q_{\mu\nu}$ and $\delta J^\nu$ are the responses constructed in Secs.~\ref{sec:mouthS} and \ref{sec:lll}. In the Einstein--Maxwell mouth both vanish. With the Schwinger response $\delta J^\nu$ is the current \eqref{eq:JS}, and the anomaly equation with residual $\mathcal X$ of Sec.~\ref{sec:Xmaster} joins the system. In the throat, and in the mouth at $O(\alpha)$, $\delta T^Q_{\mu\nu}$ is the drift lift and $\delta J^\nu$ also contains the polarisation current.

After every $Y''$ is eliminated with Legendre's equation, $Y''=-\cot\theta\,Y'-\lhat Y$, the angular structure of the sector is
\begin{align}
\propto\ \sin\theta\,Y'&:\ \mathcal E_{t\phi},\ \mathcal E_{r\phi},\ \mathcal M^{\theta};
\qquad
\propto\ \sin\theta\,Y:\ \mathcal M^{t},\ \mathcal M^{r};
\qquad
\propto\ \sin\theta\,(Y''-\cot\theta\,Y'):\ \mathcal E_{\theta\phi};\nonumber\\
\text{identically }0&:\ \mathcal E_{tt},\ \mathcal E_{tr},\ \mathcal E_{rr},\ \mathcal E_{t\theta},\ \mathcal E_{r\theta},\ \mathcal E_{\theta\theta},\ \mathcal E_{\phi\phi},\ \mathcal M^{\phi}.
\label{eq:angular}
\end{align}
The $\theta\phi$ component is proportional to the single axial tensor harmonic $\sin\theta\,(Y''-\cot\theta\,Y')=\sin\theta\,(-2\cot\theta\,Y'-\lhat\,Y)$. Its $Y$ and $Y'$ coefficients are assembled from different terms of the Einstein tensor. We extract them separately and find that they lock together in this fixed ratio, which checks the ansatz and the algebra in the same way as the traceless-angular check of \cite{Sadhukhan:2026czl}. The eight components in the last line of \eqref{eq:angular} vanish identically on both backgrounds. That is the parity selection rule at work. From here on the six symbols $\mathcal E_{t\phi},\mathcal E_{r\phi},\mathcal E_{\theta\phi},\mathcal M^t,\mathcal M^r,\mathcal M^\theta$ denote the radial coefficients so extracted. For $\mathcal E_{\theta\phi}$ we take the coefficient of $-\sin\theta\,(Y''-\cot\theta\,Y')$. The Maxwell residuals are divided by $q$, which the rescaling $a_i=qw_i/4$ leaves as an overall factor. The definition of the electric field, $E=p\,w_1-w_0'$ in the frequency domain, is carried as a seventh equation.

\subsection{Consistency conditions and the reduction algorithm}
\label{sec:consistency}

Six field equations plus the definition of $E$ constrain the five unknowns $h_0,h_1,w_0,w_1$ (in the gauge $a_2=0$) and $E$. With the bosonised scalar $X$ of Sec.~\ref{sec:mouthS} the $\chi$ equation joins them, and eight equations constrain six unknowns. In both cases there are exactly two redundancies,
\begin{equation}
\nabla_\mu\big(G^{\mu\phi}-\kappa T^{\mu\phi}\big)\equiv0
\qquad\text{and}\qquad
\partial_\nu\partial_\mu\big(\sqrt{-g}\,F^{\mu\nu}\big)\equiv0 .
\label{eq:redundancies}
\end{equation}
The first is the $\phi$ component of the linearised contracted Bianchi identity. The second is charge conservation. Both become nontrivial when the matter sector is switched on. In $\delta(\nabla_\mu T_Q^{\mu\phi})$ the perturbed connection multiplies the background fermion stress tensor \cite{Sadhukhan:2026czl}, and these terms do not vanish in the throat, or in the mouth at $O(\alpha)$, when the fermion response is set to zero. Once a fermion current is inserted into the Maxwell equations, the second identity in \eqref{eq:redundancies} requires $\nabla_\nu J^\nu=0$. The computational form of both statements is Step~4 of the algorithm below. After all eliminations, the two unused equations must reduce to $0=0$, with no remaining freedom to arrange it. The two identities constrain the divergence of the response but not the covariant constitutive relation itself. We return to this point in Secs.~\ref{sec:lll} and \ref{sec:controls}.

The reduction is performed on linear forms as in \cite{Sadhukhan:2026czl}, with derivatives of basis elements replaced by the linear forms established in earlier steps. The fixed order is the following. The bracketed items apply when the bosonised LLL scalar $X$ of Sec.~\ref{sec:mouthS} is present.
\begin{enumerate}
\item The $\theta\phi$ Einstein equation is algebraic in $h_0$ and gives $h_0=h_0[h_1,h_1']$.
\item In $\mathcal M^t$ and $\mathcal M^r$ the amplitudes $w_0,w_1$ appear only algebraically once $w_0'$ is traded for $p\,w_1-E$. $\mathcal M^r$ is solved for $w_1$ and $\mathcal M^t$ for $w_0$.
\item The $r\phi$ Einstein equation gives $h_1''$, the definition of $E$ gives $E''$, [and the $\chi$ equation gives $X''$].
\item Consistency. The two unused equations, $\mathcal E_{t\phi}$ (which after Step~1 carries $h_1'''$) and $\mathcal M^\theta$, must reduce identically to zero once the expressions of Steps 1--3 are inserted. These are the two redundancies \eqref{eq:redundancies}. They hold only if the fermion source obeys force balance, $\nabla_\mu T_Q^{\mu\phi}=-F^{\phi\lambda}J_\lambda$, and charge conservation, $\nabla_\nu J^\nu=0$. A source that violates either at the order kept is inconsistent. One of the two residuals then survives as an extra algebraic relation among the remaining fields, which the system cannot satisfy except trivially. A violation proportional to a field equation is invisible at that order, and closure (c) of Sec.~\ref{sec:controls} is an example. Section~\ref{sec:controls} displays such nonvanishing residuals for truncated closures. By a closure we mean the prescription that expresses the fermionic sources $\delta T^Q_{\mu\nu}$ and $J^\nu$ in terms of the perturbation fields.

\item Eliminations give the second-order system $\bm Y''+P\,\bm Y'+Q\,\bm Y=0$ for $\bm Y=(h_1,E\,[,X])^{\!\top}$, or equivalently the first-order system $\bm y'=\mathsf M\bm y$ for $\bm y=(h_1,h_1',E,E'\,[,X,X'])^{\!\top}$. Master variables and the tortoise coordinate $\dd x=\sqrt{B/A}\,\dd r$ then bring it to $\bm Z_{xx}+\tilde P\bm Z_x+\tilde Q\bm Z=0$ and, when $\tilde P$ can be removed, to Schr\"odinger form with $V=-\tilde Q-p^2$.
\end{enumerate}
The master variables are built from the prefactors $S\equiv\sqrt{A/B}$ (so that $\dd x=\dd r/S$), $\sqrt{AB}$ and $R$. We define
\begin{equation}
\Psi_g=\frac{S\,h_1}{R},\quad
\Psi_e=\frac{p\,r_e}{\lhat}\,\frac{R^2E}{\sqrt{AB}},\quad
\Xi=\frac{p\,r_e}{\lhat}\,X,\quad
\Psi_D=\Psi_e-m_S^2\,\Xi .
\label{eq:mastervars}
\end{equation}
These are four symbols for at most three fields. The Einstein--Maxwell system has the two master fields $\Psi_g$ and $\Psi_e$, and $X$ is absent. With the Schwinger response the system has three master fields, in the basis $(\Psi_g,\Psi_e,\Xi)$ or in the basis $(\Psi_g,\Psi_D,\Xi)$. The two bases are related by a constant triangular matrix and coincide for $m_S^2=0$. $\Psi_g$ is the Regge--Wheeler variable of the axial sector. $\Psi_e$ is the electric flux per unit solid angle, $R^2E/\sqrt{AB}$, times a factor $p$. This prefactor is a time derivative, $\Psi_e=\frac{r_e}{\lhat}\,R^2\dot E/\sqrt{AB}$, chosen so that the mixing matrix comes out frequency independent and rational in $l$. $\Psi_D$ is the corresponding displacement flux. The metric couples to $\Psi_D$ and not to $\Psi_e$, as shown in Sec.~\ref{sec:linesum}. We use the basis $(\Psi_g,\Psi_D,\Xi)$ by default. In it the gravitational channel couples to $\Xi$ only through the $O(\lhat/q)$ terms of Sec.~\ref{sec:kk}, and the $(\Psi_g,\Psi_D)$ block is the Einstein--Maxwell matrix for every $m_S^2$. At $\alpha^0$ a constant diagonal rescaling makes the potential matrix symmetric, which the stability analysis of Sec.~\ref{sec:stability} requires. In the basis $(\Psi_g,\Psi_e,\Xi)$ the metric row couples to $\Xi$ with a strength $\propto m_S^2$ while the $\Xi$ row couples back only at $O(1/q)$, and no diagonal rescaling can symmetrise the matrix. We switch to $\Psi_e$ in the derivation of the master system in Secs.~\ref{sec:Xmaster} and \ref{sec:mouthmaster}, in the isolation of the light content at large $m_S^2$ in Sec.~\ref{sec:mouthneutral}, and for the exactly massless mode of Sec.~\ref{sec:alpha0}, which is $\Psi_e=0$. In the mouth $(S,\sqrt{AB},R)=(A,1,r)$, and in the throat $(S,\sqrt{AB},R)=(f+\alpha\gamma,\;r_e^2,\;r_e\sqrt{1+\alpha\varphi})$. The change of variables $\bm Y=\mathsf D(r)\bm Z$, with $\mathsf D$ the inverse of the diagonal matrix of these prefactors, gives $\bm Z''+P_Z\bm Z'+Q_Z\bm Z=0$ with $P_Z=\mathsf D^{-1}(2\mathsf D'+P\mathsf D)$ and $Q_Z=\mathsf D^{-1}(\mathsf D''+P\mathsf D'+Q\mathsf D)$. The tortoise coordinate then gives $\tilde P=S\,P_Z-I\,S'$ and $\tilde Q=S^2Q_Z$ with $I$ the identity matrix. Throughout, $\bm Z$ denotes the vector of master variables, $Z_i$ with $i=1,2$ Chandrasekhar's decoupled combinations in the mouth (Sec.~\ref{sec:chandra}), and $Z_\pm$ the decoupled combinations \eqref{eq:Zexplicit} of the throat.

\section{The Schwinger response of the lowest Landau level}
\label{sec:mouthS}

\subsection{Why the fermions do not decouple}
\label{sec:whynot}
The companion paper treats the stress tensor of the LLL fermions in the mouth \cite{Sadhukhan:2026czl}. Its Casimir part has no appreciable local density there. Its anomaly part is local and gives $\kappa\bar T^{Q}=O(\alpha)$, of the same order as in the throat. The drift lift and the polarisation currents of Sec.~\ref{sec:lll} are built from $\bar T^Q$ and from the sphere pressure $\mathcal P$ of \eqref{eq:lift}. They are therefore present in the mouth at $O(\alpha)$, and Sec.~\ref{sec:mouth} treats them. This section treats the response that is not small.

The fermion charge also responds in the mouth. In this sector the perturbation has components $a_t,a_r\propto Y$. They pull back to a two-dimensional gauge field on the worldsheet of every magnetic field line, and the LLL fermions run along those field lines straight through the mouth. That is the mechanism by which the Casimir energy is generated in the first place \cite{Maldacena:2018gjk,Alford:1992yx}. A massless two-dimensional charged fermion responds linearly to a worldsheet electric field, and the response is exact. It is the chiral anomaly \cite{Schwinger:1962tp}. It is a local response of the vacuum on each field line, independent of the conformal circumference $L_c$ of the closed line. It descends from the axial anomaly and not from the trace anomaly, and it does not vanish with the curvature. The current is identically conserved, and both $J=0$ and $J\neq0$ pass Step~4. Consistency cannot decide and physics must. In the companion sector no such response exists. The gauge perturbation there is purely $a_\phi\propto\sin\theta\,Y'$, with no $(\tau,\rho)$ legs, and the fermions see no worldsheet electric field. This is also why $\delta J^\tau=\delta J^\rho=0$ was found there.

The size of the response settles its fate. In $w$-units the stress-tensor responses enter the field equations at $O(\alpha)$. The Schwinger current enters with the coefficient
\begin{equation}
m_S^2\;\equiv\;\frac{g_c^2q}{4\pi^2}\;=\;\alpha\,q^2 ,
\label{eq:mS2}
\end{equation}
since all $q$ field lines contribute coherently. For the fiducial parameters of \cite{Mondal:2025tht}, $q=10^{33}$ and $g_c=1$, this is $\sim10^{31}$, and it dominates the electromagnetic dynamics of the sector. We therefore keep $m_S^2$ as an independent, unexpanded parameter throughout. The physical value $m_S^2=\alpha q^2$ is at least $\alpha$. The value $m_S^2=0$ at $\alpha\neq0$ therefore lies outside the parameter space. We use it as a formal point at which the Schwinger sector decouples. At $\alpha=0$ the system is then Einstein--Maxwell theory, and at $O(\alpha)$ the Einstein--Maxwell--Casimir block of Sec.~\ref{sec:schrodinger} closes. The expansion in $\alpha$ is then an expansion at fixed $m_S^2$. Every equation is linear in $m_S^2$, and the $O(\alpha)$ background corrections multiply $m_S^2$ at most linearly. Within this expansion the neglected terms are $O(\alpha^2)$ relative to the terms kept in every channel. The running of $g_c$, which comes from the one-loop action of the higher Landau levels, is a second expansion, treated in Sec.~\ref{sec:hll}. Virtual transitions to the higher levels lie outside both expansions and are listed in Sec.~\ref{sec:summary}. The Schwinger mass $m_S$ is measured in units of $1/R$, in which the Landau gap is $\sqrt q$, and their ratio is $m_S/\sqrt q=g_c/2\pi$. This is the parameter that keeps the LLL truncation valid in the presence of screening.

We reduce the LLL to two-dimensional fermions on the field lines, solve the response of a single line by bosonisation, sum over the lines, and derive the second-order equation for the bosonised scalar.

\subsection{Two-dimensional fermions}
\label{sec:lllreduction}
On the background \eqref{eq:background} the LLL of a four-dimensional massless Dirac fermion of unit charge is a set of $q$ massless two-dimensional Dirac fermions on the $(t,r)$ plane, one per flux quantum, with the higher Landau levels gapped by $\sqrt q/R$ \cite{Wu:1976ge,Atiyah:1963zz,Aharonov:1978gb,Sadhukhan:2026czl}. The same reduction underlies the $s$-wave scattering of fermions off magnetic monopoles \cite{Callan:1982ah,Rubakov:1982fp} and off magnetic black holes \cite{Alford:1992yx}, and it is the starting point of MMP \cite{Maldacena:2018gjk}. The $q$ states of the LLL can be organised into states localised near points of the sphere. Each such state occupies the coordinate area $2\pi/|\bar F_{\theta\phi}|$, and we call its worldsheet a field line. The density of field lines per unit $\dd\theta\,\dd\phi$ is
\begin{equation}
n_{\rm L}(\theta)=\frac{|\bar F_{\theta\phi}|}{2\pi}=\frac{q}{4\pi}\,\sin\theta,
\qquad
\int_{S^2}n_{\rm L}\,\dd\theta\,\dd\phi=q .
\label{eq:linedensity}
\end{equation}
For $l\ll\sqrt q$ each field line sees the local value of the perturbation, and to leading order the four-dimensional response is the sum of the responses of independent lines \cite{Sadhukhan:2026czl}. Section~\ref{sec:kk} computes the response on the exact LLL states and finds two corrections of relative order $\lhat/q$. By the Weyl invariance of the massless Dirac action, the fermion on a field line through $(\theta,\phi)$ propagates on the two-dimensional metric $\hat g_{ab}=g_{ab}/R^2$, conformal to $-\dd t^2+\dd x^2$ with $\dd x=\dd r/S$ (MMP eq.~(2.12)). The factor $R$ is absorbed by a rescaling of the spinor and drops out of everything below. The construction is exact in the background.

\subsection{Response of one field line}
\label{sec:lineresponse}
The pullback of the gauge perturbation \eqref{eq:aansatz} to the field line through $(\theta_{\rm L},\phi_{\rm L})$ is $a_b(t,r)\,Y(\theta_{\rm L})$ with $b\in\{t,r\}$. Its worldsheet field strength is
\begin{equation}
\tilde\epsilon^{\,bc}\,\partial_b\big(a_c\,Y\big)=\big(\dot a_1-a_0'\big)\,Y=E\,Y(\theta_{\rm L}),
\label{eq:wsfield}
\end{equation}
with $\tilde\epsilon^{\,ab}$ the numerical alternating symbol, $\tilde\epsilon^{\,tr}=+1$. The transverse component $a_\theta=a_2Y'=\partial_\theta(a_2Y)$ is pure gauge in the $\theta$ direction. On a fixed field line it is removed by the transformation $\Lambda=a_2Y$. Only the invariants $\mathsf b_b=a_b-\partial_ba_2$ can enter, as the residual $U(1)$ of Sec.~\ref{sec:ansatz} requires.

A massless Dirac fermion in two dimensions is equivalent to a free boson \cite{Coleman:1975pw,Mandelstam:1975hb}. In the bosonic description the vector current is the topological current of the boson, and the gauge field couples to it. In the sign convention of Sec.~\ref{sec:sector}, $J=-\delta\mathcal I_{\rm fermion}/\delta A$, the bosonised action of one field line of unit charge is
\begin{equation}
\mathcal I_{\rm L}=\int\dd t\,\dd r\,\Big[-\frac1{2\pi}\sqrt{-\hat g}\;\hat g^{bc}\,\partial_b\chi_{\rm L}\,\partial_c\chi_{\rm L}
-\frac1\pi\,\chi_{\rm L}\;\tilde\epsilon^{\,bc}\,\partial_b\big(a_c\,Y\big)\Big].
\label{eq:Iline}
\end{equation}
The field $\chi_{\rm L}$ is $\sqrt\pi$ times the canonically normalised boson of Refs.~\cite{Coleman:1975pw,Mandelstam:1975hb}. Both terms of \eqref{eq:Iline} are Weyl invariant. The first contains the density $\sqrt{-\hat g}\,\hat g^{ab}$, which is Weyl invariant in two dimensions, and the second contains no metric at all. The current density $\mathfrak j^{\,b}\equiv\sqrt{-\hat g}\,j^b=-\delta\mathcal I_{\rm L}/\delta a_b$ and the field equation of $\chi_{\rm L}$ are
\begin{equation}
\mathfrak j^{\,a}=\frac1\pi\,\tilde\epsilon^{\,ab}\,\partial_b\chi_{\rm L},
\qquad
\partial_a\big(\sqrt{-\hat g}\;\hat g^{ab}\,\partial_b\chi_{\rm L}\big)=E\,Y(\theta_{\rm L}).
\label{eq:lineresponse}
\end{equation}
The first relation is Coleman's bosonisation formula for the current, written in covariant form. It makes charge conservation, $\partial_a\mathfrak j^{\,a}=0$, an identity. The second relation is the two-dimensional axial anomaly. The dual current density $\mathfrak j_5^{\,a}=\frac1\pi\sqrt{-\hat g}\,\hat g^{ab}\partial_b\chi_{\rm L}$ is the axial current, and the field equation states $\partial_a\mathfrak j_5^{\,a}=\frac1\pi\,E\,Y$. This is the anomaly of the Schwinger model \cite{Schwinger:1962tp}, the two-dimensional form of the Adler--Bell--Jackiw anomaly \cite{Adler:1969gk,Bell:1969ts}. The action \eqref{eq:Iline} is quadratic in $\chi_{\rm L}$ and linear in the gauge field. The response \eqref{eq:lineresponse} is therefore exact, with no higher-order terms in the gauge field and no loop corrections \cite{Schwinger:1962tp,Lowenstein:1971fc}. The auxiliary scalar $\chi_{\rm L}$ keeps the response local. It plays here the role that the transport fields $\xi_a$ played in the companion sector \cite{Sadhukhan:2026czl}.

The meaning of the response is clearest for an isolated line in flat space with a dynamical gauge field. Take $\hat g_{ab}=\eta_{ab}$, add the two-dimensional Maxwell term $\frac1{2e_2^2}\int F_{tx}^2$ with coupling $e_2$ and field strength $F_{tx}=\partial_ta_x-\partial_xa_t$, and drop the angular factor. The Maxwell equations $\partial_aF^{ab}=e_2^2\,\mathfrak j^{\,b}$ with the current \eqref{eq:lineresponse} give $\partial_a\big(F_{tx}-\tfrac{e_2^2}{\pi}\chi_{\rm L}\big)=0$. Without a background electric field the bracket vanishes, and the anomaly equation becomes
\begin{equation}
\Big(-\partial_t^2+\partial_x^2-\frac{e_2^2}{\pi}\Big)\chi_{\rm L}=0 .
\label{eq:schwmass}
\end{equation}
The boson acquires the mass$^2$ $e_2^2/\pi$. This is Schwinger's result \cite{Schwinger:1962tp}. The gauge field is screened, and the gauge-invariant spectrum is a free massive boson \cite{Lowenstein:1971fc,ColemanJackiwSusskind:1975}. For $N_f$ flavours of equal charge only the sum of the bosons couples to the gauge field. Its mass$^2$ is $N_fe_2^2/\pi$, and the $N_f-1$ relative combinations stay massless. The mass$^2$ is positive.

\subsection{Sum over the field lines}
\label{sec:linesum}
The source in \eqref{eq:lineresponse} is proportional to $Y(\theta_{\rm L})$, and the equation is linear. The response of every line is therefore $\chi_{\rm L}=\chi(t,r)\,Y(\theta_{\rm L})$ with a single radial function $\chi$. For the metric $g_{ab}=\mathrm{diag}(-A,B)$ the Weyl-invariant density is $\sqrt{-\hat g}\,\hat g^{ab}=\mathrm{diag}\big(-1/S,\;S\big)$ with $S=\sqrt{A/B}$. The anomaly equation in \eqref{eq:lineresponse} then becomes
\begin{equation}
-\frac{\ddot\chi}{S}+\big(S\,\chi'\big)'\;=\;E ,
\qquad S=A\ \text{(mouth)},\quad S=f+\alpha\gamma\ \text{(throat)} .
\label{eq:chieq}
\end{equation}
The four-dimensional current density is the line density \eqref{eq:linedensity} times the worldsheet current density of \eqref{eq:lineresponse}, $\sqrt{-g}\,J^a=n_{\rm L}(\theta)\,\mathfrak j^{\,a}$. Multiplied by the gauge coupling it reads
\begin{equation}
g_c^2\sqrt{-g}\;J_S^{a}
=m_S^2\,\sin\theta\;\tilde\epsilon^{\,ab}\partial_b\chi\;Y(\theta),
\qquad\text{that is}\qquad
g_c^2\sqrt{-g}\,J_S^{t}=m_S^2\sin\theta\,\chi'\,Y,\quad
g_c^2\sqrt{-g}\,J_S^{r}=-m_S^2\sin\theta\,\dot\chi\,Y,
\label{eq:JS}
\end{equation}
with
\begin{equation}
m_S^2=g_c^2\times\frac{q}{4\pi}\times\frac1\pi=\frac{g_c^2q}{4\pi^2}=\alpha\,q^2 .
\end{equation}
The factor $q/4\pi$ is the line density, the factor $1/\pi$ is the normalisation of the bosonised current, and together with $g_c^2$ they give precisely \eqref{eq:mS2}. The Schwinger current has no components across the field lines, $J_S^\theta=J_S^\phi=0$, since the LLL fermions move only along the lines. The currents across the lines of Sec.~\ref{sec:lll} have a different origin. The same number $m_S^2$ follows from dimensional reduction. The Maxwell term $\int\sqrt{-g}\,F^2/4g_c^2$ integrated over a sphere of radius $R$ gives a two-dimensional coupling $e_2^2=g_c^2/4\pi R^2$. By \eqref{eq:schwmass} the Schwinger mass$^2$ of $q$ unit-charge flavours is $q\,e_2^2/\pi=m_S^2/R^2$. In the dilaton language of \cite{Freivogel:2026ujn} this is the Schwinger mass $m^2=e^2N_f/\pi f(\Phi)$ of their eq.~(4.82). There $N_f$ is the number of flavours, equal to the Landau degeneracy $q$, and their $f(\Phi)$ is the dilaton-dependent coefficient of the two-dimensional Maxwell term, not the $f$ of \eqref{eq:throatbg}. Before their Weyl rescaling of the two-dimensional metric this coefficient is $4\pi R^2$, and with $e=g_c$ their mass is $m_S^2/R^2$. In throat units this is the mass $m^2R_{AdS_2}^2\propto g_c^2q$ of the massive mode in MMP's Appendix~A, with the coefficient fixed by their eqs.~(A.2) and (A.3), $m^2=g_2^2q/\pi$ and $g_2^2=g_c^2/4\pi$.

The source \eqref{eq:JS} enters only the $t$ and $r$ components of the Maxwell equations. For the metric \eqref{eq:background} one has $\sqrt{-g}\,F^{rt}=\sin\theta\,Y\,\mathcal F$ with the electric flux per unit solid angle
\begin{equation}
\mathcal F\equiv\frac{R^2E}{\sqrt{AB}} ,
\qquad
\mathcal F_D\equiv\mathcal F-m_S^2\,\chi .
\label{eq:flux}
\end{equation}
The longitudinal parts of the two Maxwell equations are $\partial_r\big(\sqrt{-g}F^{rt}\big)-g_c^2\sqrt{-g}J_S^t=\sin\theta\,Y\,\partial_r\mathcal F_D$ and $\partial_t\big(\sqrt{-g}F^{tr}\big)-g_c^2\sqrt{-g}J_S^r=-\sin\theta\,Y\,\partial_t\mathcal F_D$. The Schwinger current is the curl of a longitudinal polarisation, and the Maxwell equations see the electric flux only through the displacement flux $\mathcal F_D$. The remaining terms of $\mathcal M^t$ and $\mathcal M^r$ come from $\delta F_{a\theta}$ and from the metric perturbation, and they are unchanged. After the rescaling $\chi=\frac q4X$, which removes $q$ together with $a_i=\frac q4w_i$, we keep the symbol $\mathcal F_D$ for the rescaled combination. The two Maxwell equations of the mouth then give
\begin{equation}
w_1=-\frac{2}{r^2}\,h_1-\frac{p\,r^2}{\lhat\,\Delta}\,\mathcal F_D,
\qquad
w_0=-\frac{2}{r^2}\,h_0-\frac{\Delta}{\lhat\,r^2}\,\mathcal F_D',
\qquad
\mathcal F_D=r^2E-m_S^2X .
\label{eq:wD}
\end{equation}
For $m_S^2=0$ these reduce to the Einstein--Maxwell amplitudes of Sec.~\ref{sec:mouthmaster}. The Einstein equations contain the gauge field only through $w_0$ and $w_1$. Gravity therefore couples to the displacement flux and not to the electric flux. The coupling of Sec.~\ref{sec:kk} adds a direct coupling to $\chi$. This is the origin of the master variable $\Psi_D$ of \eqref{eq:mastervars}, since $\Psi_D=\frac{p\,r_e}{\lhat}\,\mathcal F_D$ in all regions.

Three checks fix the construction. First, $J_S$ is identically conserved, $\partial_a\big(g_c^2\sqrt{-g}J_S^a\big)=m_S^2\sin\theta\,Y\,\tilde\epsilon^{\,ab}\partial_a\partial_b\chi=0$. Second, in the flat region $r\gg r_e$ the electric channel of the mouth master system \eqref{eq:gexi} of Sec.~\ref{sec:mouthmaster} carries the potential $(\lhat+m_S^2)/r^2$. The first term is the centrifugal barrier. The second is an effective photon mass$^2$, $m_{\rm eff}^2=m_S^2/r^2=g_c^2q/4\pi^2r^2=g_c^2|\bm B|/2\pi^2$ with $|\bm B|=q/2r^2$. This is exactly the screening term of MMP eq.~(G.35), derived there for the ambient region, and the photon mass that the lowest Landau level induces in a strong magnetic field \cite{Gusynin:1999pq,Fukushima:2011nu}. Third, the sign is the screening one. The induced mass$^2$ in the master systems of both regions is $+m_S^2$ and not tachyonic.

\subsection{The equation for \texorpdfstring{$X$}{X}}
\label{sec:Xmaster}
In the frequency domain and after the rescaling $\chi=\frac q4X$, the anomaly equation \eqref{eq:chieq} reads
\begin{equation}
\big(S\,X'\big)'-\frac{p^2}{S}\,X=E .
\label{eq:Xeq}
\end{equation}
We call the residual of this equation $\mathcal X\equiv(S\,X')'-p^2X/S-E$, in the same way as $\mathcal E_{\mu\nu}$ and $\mathcal M^\nu$ of Sec.~\ref{sec:algorithm}. It appears under this name in the appendices. In the throat at $\alpha=0$ one has $S=f$ and $X''=E/f+p^2X/f^2-2\rho X'/f$, as in Appendix~\ref{app:throat0}.

The $X$ equation takes its simplest form in the tortoise coordinate. Since $\partial_x=S\,\partial_r$, one has $(S\,X')'=S^{-1}\partial_x^2X$, and \eqref{eq:Xeq} becomes a flat two-dimensional wave equation with a source,
\begin{equation}
\partial_x^2X-p^2X=S\,E .
\label{eq:Xflat}
\end{equation}
This is exact for every $S$. It is the anomaly equation written in the conformally flat coordinates of the worldsheet, and no first-derivative term appears, whatever the background. With the master variables \eqref{eq:mastervars}, $\Xi=\frac{p\,r_e}{\lhat}X$ and $\Psi_e=\frac{p\,r_e}{\lhat}\mathcal F$, the source is $\frac{p\,r_e}{\lhat}\,S\,E=\sigma\,\Psi_e$ with
\begin{equation}
\sigma(r)\equiv\frac{S\sqrt{AB}}{R^2} .
\end{equation}
Replacing $\Psi_e$ by $\Psi_D+m_S^2\,\Xi$ and $p^2$ by $-\omega^2$ gives the $\Xi$ row of the master system,
\begin{equation}
\partial_x^2\Xi+\big(\omega^2-\sigma\,m_S^2\big)\,\Xi-\sigma\,\Psi_D=0 .
\label{eq:Xirow}
\end{equation}
The $\Xi$ row of the potential matrix is therefore $\sigma\,(0,\,1,\,m_S^2)$ in the basis $(\Psi_g,\Psi_D,\Xi)$, exactly and in every region, apart from the coupling of Sec.~\ref{sec:kk}. In the mouth $\sigma=\Delta/r^4$, and \eqref{eq:Xirow} is the third row $\frac{\Delta}{r^5}(0,\,r,\,m_S^2r)$ of \eqref{eq:VmouthS}. In the throat at $\alpha=0$ one has $\sigma=f$, and \eqref{eq:Xirow} is the third row of the mass matrix $\mathcal N$ of \eqref{eq:Nfull}. At $O(\alpha)$ one has $S=f+\alpha\gamma$, $\sqrt{AB}=r_e^2$ and $R^2=r_e^2(1+\alpha\varphi)$, and
\begin{equation}
\sigma=f+\alpha\,(\gamma-f\varphi)+O(\alpha^2),
\qquad
f\varphi-\gamma=2\rho\,(\rho^2+2)\arctan\rho+\frac{8\rho^2}{3}+\frac43-\log f .
\end{equation}
This gives $\tilde Q_{1,\Xi D}=f\varphi-\gamma$, $\tilde Q_{1,\Xi\Xi}=m_S^2\,(f\varphi-\gamma)$, $\tilde Q_{1,\Xi g}=0$ and $\tilde P_{1,\Xi\Xi}=0$, which are the entries listed in Appendix~\ref{app:matrices}. Apart from the coupling of Sec.~\ref{sec:kk}, the whole $O(\alpha)$ content of the $\Xi$ row of $\tilde Q$ is the correction of $\sigma$ by the corrected background. The removal of the friction term in Sec.~\ref{sec:schrodinger} adds a commutator to the $\Xi D$ entry of $V_{1a}$.

\subsection{The exact LLL response and the coupling to \texorpdfstring{$h_{a\phi}$}{haphi}}
\label{sec:kk}
The sum over field lines treats every LLL state as a point of the sphere. The LLL states have a width, and their azimuthal momentum and spin couple them to $h_{a\phi}$. Both effects follow from the exact zero modes. In the frame $e^{\hat\theta}=R\,\dd\theta$, $e^{\hat\phi}=R\sin\theta\,\dd\phi$, with two Pauli matrices as sphere Dirac matrices and the potential $\bar A$ of \eqref{eq:background}, the sphere Dirac operator has the $q$ zero modes \cite{Wu:1976ge,Atiyah:1963zz}
\begin{equation}
\mathsf u_{m_{\rm L}}=\Big(0,\;\sin^{\frac{q-1}{2}}\theta\,\tan^{-m_{\rm L}}\frac\theta2\;e^{im_{\rm L}\phi}\Big)^{\!\top},\qquad m_{\rm L}=-\frac{q-1}{2},\dots,\frac{q-1}{2} .
\label{eq:zeromodes}
\end{equation}
The label $m_{\rm L}$ is the canonical angular momentum of Sec.~\ref{sec:polcurrent}, and the LLL field is a sum over $m_{\rm L}$ of two-dimensional Dirac spinors times $\mathsf u_{m_{\rm L}}$. We insert the perturbation \eqref{eq:hansatz}--\eqref{eq:aansatz} into the four-dimensional Dirac operator, with the vielbein $e^{\hat\phi}=R\sin\theta\,\dd\phi+\epsilon\,h_aY'\dd x^a/R$ and its spin connection, and project it on the zero modes. At $m=0$ the projection is diagonal in $m_{\rm L}$. The spinor with label $m_{\rm L}$ sees the two-dimensional gauge field $a_b\langle Y\rangle_{m_{\rm L}}+h_b\langle\mathcal V\rangle_{m_{\rm L}}$, where $\langle\cdot\rangle_{m_{\rm L}}$ is the average over the sphere with weight $|\mathsf u_{m_{\rm L}}|^2$ and
\begin{equation}
\mathcal V=\frac1{R^2}\Big[\big(m_{\rm L}-\tfrac q2\cos\theta\big)\frac{Y'}{\sin\theta}-\frac14\big(Y''-\cot\theta\,Y'\big)\Big].
\label{eq:Vvertex}
\end{equation}
The term with $m_{\rm L}-\tfrac q2\cos\theta=m_{\rm L}-\bar A_\phi$ is the kinetic azimuthal momentum $\Pi_\phi$ of the state, and the $Y''$ term comes from the spin connection. Legendre's equation gives the identity
\begin{equation}
\sin\theta\,|\mathsf u_{m_{\rm L}}|^2\Big(R^2\mathcal V+\frac{\lhat}{4}\,Y\Big)=-\frac12\,\partial_\theta\big(|\mathsf u_{m_{\rm L}}|^2\sin\theta\,Y'\big),
\end{equation}
and $|\mathsf u_{m_{\rm L}}|^2\sin\theta$ vanishes at both poles. The right side therefore integrates to zero, and
\begin{equation}
\langle\mathcal V\rangle_{m_{\rm L}}=-\frac{\lhat}{4R^2}\,\langle Y\rangle_{m_{\rm L}}
\label{eq:Vavg}
\end{equation}
for every zero mode, every two-dimensional metric and every $R$. The metric perturbation therefore acts on every LLL state as the shift $a_b\to a_b-\lhat h_b/4R^2$ of its worldsheet gauge field, or in $w$-units
\begin{equation}
w_b\;\longrightarrow\;w_b-\frac{\lhat}{q\,R^2}\,h_b .
\label{eq:kkshift}
\end{equation}
This is a relative correction of order $\lhat/q$. With $1/q=4\pi^2\alpha/g_c^2$ it is $\lhat/q\simeq237\,\alpha$ at $l=2$ and $g_c=1$, the order of the $O(\alpha)$ terms, and we keep it.

The width of the states changes the strength of the response. The zero modes form the multiplet of angular momentum $\mathsf j=(q-1)/2$. The Wigner--Eckart theorem and the explicit average in the state $m_{\rm L}=\mathsf j$ give $\langle Y\rangle_{m_{\rm L}}=\pm\langle\mathsf j\,m_{\rm L};l\,0|\mathsf j\,m_{\rm L}\rangle\langle\mathsf j\,\mathsf j;l\,0|\mathsf j\,\mathsf j\rangle$. The Schwinger source projected on $Y$ is $\sum_{m_{\rm L}}\langle Y\rangle_{m_{\rm L}}^2$ times the response of one state, against $q/(2l+1)$ in the sum over lines of Sec.~\ref{sec:linesum}. The orthogonality of the Clebsch--Gordan coefficients gives the ratio
\begin{equation}
\frac{2l+1}{q}\sum_{m_{\rm L}}\langle Y\rangle_{m_{\rm L}}^2=\langle\mathsf j\,\mathsf j;l\,0|\mathsf j\,\mathsf j\rangle^2=\frac{(q-1)!\,q!}{(q-1-l)!\,(q+l)!}=1-\frac{\lhat}{q}+O(q^{-2}).
\label{eq:widthfactor}
\end{equation}
This constant multiplies $m_S^2$, and the shift \eqref{eq:kkshift} enters through the same sum. Since $m_S^2$ is kept as a free parameter, we absorb the factor into it. It changes the physical value $\alpha q^2$ by a relative $\lhat/q$, and we quote the leading value $\alpha q^2$ throughout.

The shift \eqref{eq:kkshift} enters the bosonised action \eqref{eq:Iline} through $a_c$. The anomaly equation \eqref{eq:chieq} becomes
\begin{equation}
-\frac{\ddot\chi}{S}+\big(S\,\chi'\big)'=E-\partial_t\Big(\frac{\lhat\,h_1}{4R^2}\Big)+\partial_r\Big(\frac{\lhat\,h_0}{4R^2}\Big),
\label{eq:chieqkk}
\end{equation}
and the variation of the same action with respect to $h_{a\phi}$ gives the reciprocal Einstein source
\begin{equation}
\sqrt{-g}\,T_\chi^{a\phi}=\frac{m_S^2}{4g_c^2R^2}\,\tilde\epsilon^{\,ab}\partial_b\chi\;Y' .
\label{eq:Tchi}
\end{equation}
With both terms Step~4 closes in both regions. In the master variables they add matrices of order $1/q$ to the potentials, $\delta\mathcal N$ of \eqref{eq:dN} in the throat and $\delta\hat V$ of \eqref{eq:dVmouth} in the mouth. Their entries in the $\Psi_g$ and $\Psi_D$ rows are proportional to $m_S^2$. At $m_S^2=0$ only the $\Xi$ row changes, and $\Xi$ is driven without acting back. The products of the coupling with the $O(\alpha)$ terms are $O(\alpha/q)$, the order of the neglected $O(\alpha^2)$ terms. We keep the coupling exactly at $\alpha^0$. We have also computed its $O(\alpha/q)$ terms in the full $O(\alpha)$ system, with the coupling entering through the Schwinger action alone. They are listed in Appendix~\ref{app:matrices}, and we use them only as checks. The projection keeps the LLL. Virtual transitions to the higher Landau levels are suppressed by the gap $\sqrt q/R$. We have not estimated them. They could also contribute at order $1/q$, and we neglect them.

\section{The lowest-Landau-level sector in the throat}
\label{sec:lll}

\subsection{The background source and the Schwinger response}
\label{sec:notrespond}
The background fermion source is the LLL stress tensor of \cite{Sadhukhan:2026czl} with its trace retained, lifted with
\begin{equation}
T^Q_{ab}=\frac{t_{ab}}{4\pi R^2},\qquad T^Q_{AB}=\mathcal P\,g_{AB},\qquad
\mathcal P=-\frac{\hat g^{ab}t_{ab}}{8\pi R^4},\qquad T^Q_{aA}=0 ,
\label{eq:lift}
\end{equation}
where $\mathcal P$ is the sphere pressure. On the exact AdS$_2\times S^2$ background,
\begin{equation}
\bar T^{Q}_{\mu\nu}=\frac{q}{96\pi^2r_e^2}\,
\mathrm{diag}\Big(\rho^2-2,\;-\frac{\rho^2+4}{f^2},\;1,\;\sin^2\theta\Big),
\qquad
\mathcal P=\frac{q}{96\pi^2r_e^4},
\label{eq:TQ0}
\end{equation}
with $\hat g^{ab}\bar t_{ab}=-q/12\pi$ the anomalous trace on unit AdS$_2$, since the central charge is $q$. Below $\mathcal P$ denotes the background sphere pressure, in the throat and in the mouth, which the companion paper writes $\bar{\mathcal P}$. This tensor sources the $O(\alpha)$ background \eqref{eq:gamphi}. What is new in this sector is the response. In this sector the perturbation \eqref{eq:hansatz} has no $(\tau,\rho)$ block and does not touch the sphere radius. The two-dimensional metric $\hat g_{ab}=g_{ab}/R^2$ on each field line is unperturbed at $O(\epsilon)$. This holds even on the tilted worldsheet of Sec.~\ref{sec:drift}, since $g(e_a,e_b)=g_{ab}+O(\epsilon^2)$. There is no Weyl factor, no transport diffeomorphism, and the conformal-Ward-identity machinery of \cite{Sadhukhan:2026czl} is empty here. Nor does the worldsheet gauge field induce a stress-tensor response at linear order. The correlator of the stress tensor with the current vanishes in a state invariant under charge conjugation (Furry's theorem). We take the state of each closed line to be such a state, the antiperiodic vacuum with vanishing holonomy, and then
\begin{equation}
\delta t_{ab}=0 \qquad\text{at }O(\epsilon).
\end{equation}
The fermions respond only through their charge. The construction of Sec.~\ref{sec:mouthS}, including the coupling of Sec.~\ref{sec:kk}, applies verbatim with $S=f+\alpha\gamma$. The sources \eqref{eq:JS} enter the $\tau$ and $\rho$ Maxwell equations and the auxiliary equation \eqref{eq:chieq} joins the system. The current and the $\chi$ equation are Weyl invariant and independent of $R$. The response is exact in the background.

\subsection{The tilted lift, drift and twist}
\label{sec:drift}
In the background each LLL state moves only along its own magnetic field line. Its worldsheet is the $(\tau,\rho)$ plane at fixed $(\theta,\phi)$, and the Casimir stress tensor of \eqref{eq:lift} lives on these worldsheets. The perturbation of this sector adds an electric field across the lines and twists them. A charged particle in crossed electric and magnetic fields drifts perpendicular to both, with the velocity $\bm E\times\bm B/\bm B^2$ \cite{Jacksonelec}. The guiding centres of LLL states drift in the same way \cite{Girvin:1999}. Here the drift points along $\phi$, and the worldsheet of every LLL state is tilted toward the $\phi$ direction. The drift lift is the statement that the Casimir stress tensor follows this tilt. Its stress tensor is the background one, carried onto the tilted worldsheet. This produces the mixed components $\delta T^Q_{a\phi}$ that are new in this sector.

The direction of the worldsheet is fixed as follows. The transverse field strength $\delta F_{a\theta}=-\mathsf b_a Y'$, with $\mathsf b_a$ from \eqref{eq:beta}, is an electric field across the field lines for $a=\tau$ and a twist of them for $a=\rho$. The metric perturbation $h_{a\phi}$ displaces the lines azimuthally. A non-null field strength $F^\mu{}_\nu$ has two invariant two-planes, one timelike and one spacelike, and $F$ maps each of them into itself \cite{Synge:1956,Landau:1975}. In a local frame where the electric and magnetic fields are parallel, the timelike plane is spanned by the time direction and the common direction of the two fields, and the spacelike plane is orthogonal to it. In the background the spacelike plane is the sphere and the timelike plane is the $(\tau,\rho)$ plane. An LLL state is localised in the spacelike plane and moves freely in the timelike one. Its worldsheet is therefore the timelike invariant plane of $F$. The same field-sheet picture underlies force-free magnetospheres \cite{Gralla:2014yja}. At $O(\epsilon)$ the plane tilts. We describe it by two tangent vectors $e_a$, $a\in\{\tau,\rho\}$. At zeroth order they are the coordinate vectors, and at first order each acquires a component along $\partial_\phi$,
\begin{equation}
e_a=\partial_a+\epsilon\,v_a\,\partial_\phi,
\qquad
v_a=\frac{\delta F_{a\theta}}{\bar F_{\theta\phi}}
=\frac{2\,\mathsf b_a\,Y'}{q\,\sin\theta},
\qquad F_{A\nu}\,e_a^\nu=O(\epsilon^2)\quad(A=\theta,\phi) .
\label{eq:tilt}
\end{equation}
The last condition says that the transverse components of $F$ do not see the worldsheet. For $A=\theta$ it reads $F_{\theta a}+\epsilon\,v_a\bar F_{\theta\phi}=0$. With $F_{\theta a}=-\epsilon\,\delta F_{a\theta}$ this gives $v_a=\delta F_{a\theta}/\bar F_{\theta\phi}$, and $\bar F_{\theta\phi}=-\frac q2\sin\theta$ gives the second form. For $A=\phi$ the condition holds identically, since $\delta F_{a\phi}=0$ and $F_{\phi\phi}=0$. These are the transverse components of the Lorentz force on the plane, and their vanishing fixes $v_a$ uniquely. The longitudinal components $F^b{}_\nu e_a^\nu$ do not vanish. They carry the field $E$, which stays inside the plane and drives the Schwinger response. The full contraction $F_{\mu\nu}e_a^\nu$ is therefore not $O(\epsilon^2)$ when $E\neq0$, and $F$ has no kernel at all when $\bm E\cdot\bm B\neq0$. The invariant plane is what is meant. $v_\tau$ is the $\bm E\times\bm B$ drift velocity and $v_\rho$ the local twist of the field line. We adopt the corresponding covariant lift of the Casimir stress tensor,
\begin{equation}
T_Q^{\mu\nu}=\varrho^{ab}\,e_a^{\mu}e_b^{\nu}+\mathcal P\,\perp^{\mu\nu},
\label{eq:fluidlift}
\end{equation}
with $\varrho^{ab}$ the worldsheet energy-momentum density, equal to $\bar T^{ab}$ at zeroth order, and $\perp^{\mu\nu}$ the projector orthogonal to the worldsheet. Expanding to $O(\epsilon)$, the only new components are the mixed (axial) ones,
\begin{equation}
\delta T^{Q}_{a\phi}
=\big(\bar T_a{}^{c}-\mathcal P\,\delta_a^{c}\big)\,u_c+\mathcal P\,h_{a\phi},
\qquad
u_c=h_{c\phi}+\bar g_{\phi\phi}\,\frac{\delta F_{c\theta}}{\bar F_{\theta\phi}}
=\Big[h_c+\frac{2r_e^2}{q}\,\mathsf b_c\Big]\sin\theta\,Y' ,
\label{eq:driftlift}
\end{equation}
or equivalently $\delta(T^{a}{}_{\phi})=(\bar T^{ac}-\mathcal P\bar g^{ac})\,u_c$, with $h_\tau\equiv h_0$, $h_\rho\equiv h_1$ and likewise for $\mathsf b_c$. The tilt one-form $u_c=g_{\phi\mu}e_c^\mu$ is the covariant azimuthal component of the worldsheet tangent. It is invariant under both the residual $U(1)$ and axial diffeomorphisms. Componentwise, with $\tilde u_a\equiv h_a+\frac{2r_e^2}{q}\mathsf b_a$,
\begin{align}
\frac{\delta T^{Q}_{\tau\phi}}{\sin\theta\,Y'}
&=-\frac{q(\rho^2-2)}{96\pi^2r_e^4 f}\,h_0
-\frac{2\rho^2-1}{48\pi^2r_e^2 f}\,\mathsf b_0,\\
\frac{\delta T^{Q}_{\rho\phi}}{\sin\theta\,Y'}
&=-\frac{q(\rho^2+4)}{96\pi^2r_e^4 f}\,h_1
-\frac{2\rho^2+5}{48\pi^2r_e^2 f}\,\mathsf b_1,
\label{eq:dTcomp}
\end{align}
and the mixed background combination that recurs below is
\begin{equation}
\bar T^{ac}-\mathcal P\,\bar g^{ac}
=\frac{q}{96\pi^2r_e^6}\,
\mathrm{diag}\Big(\frac{2\rho^2-1}{f^2},\;-(2\rho^2+5)\Big).
\label{eq:TbarP}
\end{equation}
In the gauge $a_2=0$ of the reduction, $\mathsf b_0=a_0$ and $\mathsf b_1=a_1$. We keep the gauge-covariant form here because the covariance test uses a pure axial diffeomorphism, which generates a nonzero $a_2$.

Covariance under axial diffeomorphisms constrains the lift. A change of coordinates does not change the physics, and any constitutive relation for $\delta T^Q_{a\phi}$ must treat a pure coordinate change correctly. Consider the infinitesimal axial diffeomorphism
\begin{equation}
\xi=\epsilon\,\xi^\phi\,\partial_\phi,\qquad \xi^\phi=\zeta(\tau,\rho)\,\frac{Y'}{\sin\theta},
\label{eq:axdiffeo}
\end{equation}
with $\zeta$ an arbitrary function. It shifts the azimuthal coordinate of every point by an amount that depends on $(\tau,\rho,\theta)$. Under it every background field changes by its Lie derivative. The background is axisymmetric and $\xi^\phi$ does not depend on $\phi$. The Lie derivative of a background tensor $\bar{\mathsf X}_{\mu\nu}$ therefore reduces to
\begin{equation}
(\mathcal L_\xi\bar{\mathsf X})_{\mu\nu}=\bar{\mathsf X}_{\phi\nu}\,\partial_\mu\xi^\phi+\bar{\mathsf X}_{\mu\phi}\,\partial_\nu\xi^\phi .
\label{eq:lieaxial}
\end{equation}

For the metric only $\bar g_{\phi\phi}=r_e^2\sin^2\theta$ has a $\phi$ index, and \eqref{eq:lieaxial} gives $h_{\mu\phi}=\bar g_{\phi\phi}\,\partial_\mu\xi^\phi$. Explicitly,
\begin{equation}
h_{\tau\phi}=r_e^2\,\dot\zeta\,\sin\theta\,Y',\qquad
h_{\rho\phi}=r_e^2\,\zeta'\,\sin\theta\,Y',\qquad
h_{\theta\phi}=r_e^2\,\zeta\,\sin\theta\,\big(Y''-\cot\theta\,Y'\big).
\label{eq:diffeoh}
\end{equation}
The first two have the form of the ansatz \eqref{eq:hansatz} with $h_0=r_e^2\dot\zeta$ and $h_1=r_e^2\zeta'$. The third is the odd tensor harmonic with amplitude $h_2=r_e^2\zeta$, which Regge--Wheeler gauge sets to zero. The test is therefore carried out before the gauge is fixed. This costs nothing, since the constitutive relation is a local statement about $\delta T_{a\phi}$ and does not refer to the gauge chosen for $h_2$.

For the gauge field Cartan's formula gives $\mathcal L_\xi\bar A=i_\xi\bar F+\dd\big(i_\xi\bar A\big)$. The second term is a $U(1)$ transformation with parameter $i_\xi\bar A=\frac q2\,\zeta\cot\theta\,Y'$, which we remove by a compensating $U(1)$ transformation. The first term has the components $(i_\xi\bar F)_\mu=\xi^\phi\bar F_{\phi\mu}$. With $\bar F_{\phi\theta}=\frac q2\sin\theta$ only the $\theta$ component survives,
\begin{equation}
a_\theta=\xi^\phi\,\bar F_{\phi\theta}=\frac{\zeta\,Y'}{\sin\theta}\,\frac q2\sin\theta=\frac q2\,\zeta\,Y',
\qquad\text{that is}\qquad
a_0=a_1=0,\quad a_2=\frac q2\,\zeta,\quad
\mathsf b_0=-\frac q2\,\dot\zeta,\quad \mathsf b_1=-\frac q2\,\zeta' .
\label{eq:diffeoa}
\end{equation}
The gauge field must be dragged along with the metric in this way. Otherwise the configuration is not pure gauge, and it describes field lines that are physically tilted relative to the coordinates. The dragged field has $E=0$ but a nonzero transverse field strength $\delta F_{a\theta}=-\mathsf b_aY'=\frac q2\,\partial_a\zeta\,Y'$. Neither $h_{a\phi}$ nor $\delta F_{a\theta}$ alone is therefore a measure of a physical tilt.

The background stress tensor has worldsheet components $\bar T_{ab}$, sphere components $\bar T_{AB}=\mathcal P\,\bar g_{AB}$ and no mixed components. Then \eqref{eq:lieaxial} gives
\begin{equation}
(\mathcal L_\xi\bar T)_{a\phi}=\bar T_{\phi\phi}\,\partial_a\xi^\phi=\mathcal P\,\bar g_{\phi\phi}\,\partial_a\xi^\phi=\mathcal P\,h_{a\phi},
\qquad
(\mathcal L_\xi\bar T)_{\theta\phi}=\mathcal P\,h_{\theta\phi} .
\label{eq:lieT}
\end{equation}
The worldsheet components $\bar T_{ab}$ drop out, since $\bar T_{a\phi}=0$. Covariance therefore demands that on the pure-gauge data \eqref{eq:diffeoh}--\eqref{eq:diffeoa} the response reduces to $\delta T^Q_{a\phi}=\mathcal P\,h_{a\phi}$. The $\theta\phi$ component is reproduced by the pressure term of any lift of the form \eqref{eq:fluidlift} and gives no further condition.

We now impose this on a general closure. We restrict to local linear relations without derivatives of the fields. This is an assumption, since in the throat every length is of order $r_e$ and derivative terms are not suppressed. The most general such relation is
\begin{equation}
\delta T^Q_{a\phi}=\big(\mathsf A_a{}^c\,h_c+\mathsf B_a{}^c\,\mathsf b_c\big)\sin\theta\,Y' ,
\label{eq:closure}
\end{equation}
with $\mathsf A$ and $\mathsf B$ two $2\times2$ matrices of functions of $\rho$. $U(1)$ invariance allows the gauge field to enter only through $\mathsf b_c$ and $E$. The field $E$ carries the angular profile $Y$ and cannot appear in a component proportional to $\sin\theta\,Y'$ without an angular derivative. Inserting $h_c=r_e^2\partial_c\zeta$ and $\mathsf b_c=-\frac q2\partial_c\zeta$ into \eqref{eq:closure} and requiring the result to equal $\mathcal P\,h_{a\phi}=\mathcal P\,r_e^2\,\partial_a\zeta\,\sin\theta\,Y'$ gives
\begin{equation}
\Big(r_e^2\mathsf A_a{}^c-\frac q2\,\mathsf B_a{}^c\Big)\,\partial_c\zeta=\mathcal P\,r_e^2\,\partial_a\zeta .
\end{equation}
The function $\zeta$ is arbitrary, and its derivatives $\partial_\tau\zeta$ and $\partial_\rho\zeta$ are independent at every point. The matrices must therefore satisfy $r_e^2\mathsf A-\frac q2\mathsf B=\mathcal P\,r_e^2\,I_2$. Writing $\mathsf A=\mathcal P\,I_2+\mathsf K$, this gives $\mathsf B=\frac{2r_e^2}{q}\,\mathsf K$. Substituting back into \eqref{eq:closure},
\begin{equation}
\delta T^{Q}_{a\phi}=\mathcal P\,h_{a\phi}+\mathsf K_a{}^{c}\,u_c,
\qquad
u_c=\Big[h_c+\frac{2r_e^2}{q}\,\mathsf b_c\Big]\sin\theta\,Y' ,
\label{eq:family}
\end{equation}
with $u_c$ the tilt one-form of \eqref{eq:driftlift}. Covariance under axial diffeomorphisms fixes the structure of the response up to one matrix $\mathsf K$, the response kernel.

Equation \eqref{eq:family} is a family of closures labelled by the kernel $\mathsf K_a{}^c$. Its members differ in the azimuthal momentum the fermions acquire, $\delta(T^a{}_\phi)=\bar g^{ab}\mathsf K_b{}^c u_c$. The response-free closure $\mathsf K_a{}^c=0$ keeps the mixed components at their background values. Its term $\mathcal P\,h_{a\phi}$ arises only from lowering an index with the perturbed metric. The drift lift is $\mathsf K_a{}^c=\bar T_a{}^c-\mathcal P\,\delta_a^c$, with $\bar T_a{}^c$ the background worldsheet stress tensor. In a local orthonormal frame with energy density $\mathsf e$, pressure $\mathsf p_\parallel$ along the line and $u_c=v_c$, this reads $\delta T^{\tau\phi}=(\mathsf e+\mathcal P)\,v_\tau$ and $\delta T^{\rho\phi}=(\mathsf p_\parallel-\mathcal P)\,v_\rho$. The first is the momentum density of the Casimir energy boosted with the drift velocity, with the transverse pressure in the inertia $\mathsf e+\mathcal P$. The second is the background anisotropic stress rotated with the twisted field line. To first order in $\epsilon$ the lift \eqref{eq:fluidlift} is the stress tensor $(\mathsf e+\mathcal P)\,U^\mu U^\nu+\mathcal P\,g^{\mu\nu}+(\mathsf p_\parallel-\mathcal P)\,b^\mu b^\nu$ of a gyrotropic fluid whose velocity $U$ is along $e_\tau$ and whose field direction $b$ is along $e_\rho$. This is the guiding-centre form of Chew, Goldberger and Low \cite{Chew:1956cgl}. The drift lift treats the Casimir medium as frozen into the field lines.

The tilt one-form is the physical variable in this result. It is built from $\mathsf b_c$ and is therefore $U(1)$ invariant. On the pure-gauge data it vanishes, $h_c+\frac{2r_e^2}{q}\mathsf b_c=r_e^2\partial_c\zeta-r_e^2\partial_c\zeta=0$. It measures the tilt of the field lines relative to the coordinates, which a coordinate change cannot produce, since it moves the lines and the coordinates together. The mixed component that drives the current, $\delta(T^a{}_\phi)=\bar g^{ab}\big(\delta T_{b\phi}-\mathcal P\,h_{b\phi}\big)=\bar g^{ab}\mathsf K_b{}^c u_c$, vanishes on pure gauge as well. A coordinate change therefore induces no current through the force balance of Sec.~\ref{sec:polcurrent}.

The test separates the candidate closures as follows. The naive closure obtained by index gymnastics with rigid field lines, $\delta T_{a\phi}=\bar T_a{}^{c}h_{c\phi}$, has $\mathsf A=\bar T$ and $\mathsf B=0$. The condition $r_e^2\mathsf A-\frac q2\mathsf B=\mathcal P\,r_e^2I_2$ would then require $\bar T_a{}^c=\mathcal P\,\delta_a^c$. In the throat, however, \eqref{eq:TbarP} gives $\bar T_a{}^c-\mathcal P\,\delta_a^c=-\frac{q}{96\pi^2r_e^4f}\,\mathrm{diag}\big(2\rho^2-1,\;2\rho^2+5\big)\neq0$. The naive closure is not covariant. It moves the stress tensor with the metric but keeps the field lines fixed. The response-free closure $\mathsf K=0$, which carries the Casimir stress tensor rigidly with the coordinates while the field lines tilt, and the drift lift $\mathsf K=\bar T-\mathcal P$ both pass. Covariance cannot decide between them. Section~\ref{sec:controls} shows that the Bianchi identities cannot select the kernel either. The next subsection states the assumption that fixes it.

\subsection{The response kernel}
\label{sec:driftderivation}
Two statements bear on the kernel, and an assumption fixes it. The first statement is kinematic, and the second rests on the local state of the LLL fermions. The first is uniqueness of the plane. Within the ansatz, the vectors $e_a=\partial_a+\epsilon v_a\partial_\phi$ that span an invariant two-plane of $F=\bar F+\epsilon\,\delta F$ are determined by the vanishing of the transverse components $F^\theta{}_\nu e_a^\nu$ and $F^\phi{}_\nu e_a^\nu$ at $O(\epsilon)$. The first condition gives $v_a=\delta F_{a\theta}/\bar F_{\theta\phi}$ and the second is empty since $\delta F_{a\phi}=0$. A tilt in the $\theta$ direction cannot appear since $\delta F_{\theta\phi}=0$, and adding components along $\partial_b$ only reparametrises the plane. The timelike invariant plane of $F$ is therefore unique to this order. Other covariant constructions can single out other planes. The principal planes of the curvature, for instance, tilt with $h_{a\phi}$ and its derivatives. That the fermions follow the plane of $F$ is part of the assumption below. The coordinate plane spanned by $\partial_\tau,\partial_\rho$ is the invariant plane of the background field $\bar F$, and a closure built on it depends on $\bar F$ separately from the dynamical field $F$. It is covariant under axial diffeomorphisms, which move $\bar F$ and $F$ together, but it is not a functional of the fields.

The second statement is the local Lorentz transformation. Take flat space with coordinates $(t,z,x_1,x_2)$, field lines along $z$, and a frame in which the field is purely magnetic, $F'_{x_1x_2}=(\bm B')_z$, and the LLL fermions are at rest with $T'^{ab}=\mathrm{diag}(\mathsf e,\mathsf p_\parallel,\mathcal P,\mathcal P)$ on the sheet $(t,z)$ and transverse pressure $\mathcal P$. A boost along $x_2$ with velocity $v$ produces $(\bm E)_{x_1}/(\bm B)_z=v$, which identifies $v$ with the $\bm E\times\bm B$ drift velocity. The transformed stress tensor to first order in $v$ is
\begin{equation}
T^{\mu\nu}=\begin{pmatrix}\mathsf e&0&0&(\mathsf e+\mathcal P)v\\0&\mathsf p_\parallel&0&0\\0&0&\mathcal P&0\\(\mathsf e+\mathcal P)v&0&0&\mathcal P\end{pmatrix}
=\varrho^{ab}e_a^\mu e_b^\nu+\mathcal P\perp^{\mu\nu}\Big|_{e_t=\partial_t+v\partial_{x_2}}+O(v^2),
\end{equation}
which is \eqref{eq:fluidlift}. Lowering an index, $\delta T_{tx_2}=-(\mathsf e+\mathcal P)v=(\bar T_t{}^t-\mathcal P)\,u_t$ with $u_t=g_{x_2x_2}v$, which is \eqref{eq:driftlift} with $\mathsf K=\bar T-\mathcal P$ and nothing else. For the twist $v_\rho$ we assume the analogous statement for a local rotation of the field-line direction. Both steps assume that the LLL state in the frame where the transverse field vanishes is the unperturbed Casimir state, and that its stress responds locally. The normal frequencies of the throat are of order one in $\Om$, as are the excitation energies of the CFT on a closed line of conformal length $\pi$, and a local response is not justified by a separation of scales. The lift ignores the finite width of the LLL states. For the charge response this width gives the corrections of Sec.~\ref{sec:kk}. For the stress response it is not computed here.

We assume that the Casimir stress tensor of the LLL follows the tilted worldsheet, that is, $\mathsf K=\bar T-\mathcal P$. The index theorem fixes the number of modes per unit flux, the per-line state is the two-dimensional Casimir vacuum, and the density of lines is unchanged in this sector since $\delta F_{\theta\phi}=0$. None of this fixes the $O(\epsilon)$ stress response of an LLL state to a transverse electric field or to a tilt of the field, and the step to $\mathsf K=\bar T-\mathcal P$ is an assumption.

\subsection{Force balance and the polarisation current}
\label{sec:polcurrent}
We now impose the $O(\epsilon)$ force balance $\nabla_\mu T_Q^{\mu\nu}=-F^{\nu\lambda}J_\lambda$ on the lifted tensor. The background divergence vanishes, and at $O(\epsilon)$
\begin{equation}
\nabla_\mu T_Q^{\mu\tau}\big|_{\epsilon}=
\nabla_\mu T_Q^{\mu\rho}\big|_{\epsilon}=
\nabla_\mu T_Q^{\mu\theta}\big|_{\epsilon}=0
\qquad\text{identically},
\end{equation}
for arbitrary $h_a$, $a_i$. The $\phi$ component does not vanish. Since $\bar F_{\phi\lambda}$ has only the $\lambda=\theta$ entry, the $\phi$ force balance determines a current across the field lines,
\begin{equation}
\nabla_\mu T_Q^{\mu\phi}\big|_{\epsilon}=-\bar F^{\phi\theta}\,\delta J_\theta
\qquad\Longrightarrow\qquad
J^{\theta}=\frac{1}{\sqrt{-g}}\,\partial_a\big(\sqrt{-g}\,\mathcal D^{a}\big),
\qquad
\mathcal D^{a}\equiv\frac{\delta(T^{a}{}_{\phi})}{\bar F_{\theta\phi}} .
\label{eq:Jtheta}
\end{equation}
The second equality holds because $\partial_\phi$ is a Killing vector of the perturbed metric at $m=0$. For a symmetric tensor this gives $\nabla_\mu T^\mu{}_\phi=(-g)^{-1/2}\partial_\mu\big(\sqrt{-g}\,T^\mu{}_\phi\big)$. At $O(\epsilon)$ only the $\tau$ and $\rho$ derivatives survive, since $T^\theta{}_\phi$ vanishes at this order and nothing depends on $\phi$, and $\bar F_{\theta\phi}$ does not depend on $\tau$ and $\rho$. The required current is the divergence, along the worldsheet, of $\mathcal D^a$. Explicitly,
\begin{equation}
\mathcal D^{\tau}=-\frac{(2\rho^2-1)\,\tilde u_\tau}{48\pi^2r_e^6 f^2}\,Y',
\qquad
\mathcal D^{\rho}=+\frac{(2\rho^2+5)\,\tilde u_\rho}{48\pi^2r_e^6}\,Y',
\qquad
J^{\theta}=\partial_\tau \mathcal D^{\tau}+\partial_\rho \mathcal D^{\rho}.
\label{eq:Nclosed}
\end{equation}
A current with a $\theta$ leg alone would violate charge conservation. We complete it with the polarisation form
\begin{equation}
J^{\mu}=\nabla_\nu\,\mathcal D^{\mu\nu},\qquad
\mathcal D^{\theta a}=-\mathcal D^{a\theta}=\mathcal D^{a},\ \text{others }0,
\qquad\Longrightarrow\qquad
J^{a}_{\rm pol}=-\frac{1}{\sin\theta}\,\partial_\theta\big(\sin\theta\,\mathcal D^{a}\big)
=\lhat\,\frac{\mathcal D^{a}}{Y'}\;Y ,
\label{eq:Jpol}
\end{equation}
which is identically conserved by antisymmetry. Charge conservation fixes the completion only up to currents that are conserved by themselves. These are the Schwinger current, which the anomaly fixes independently, and other worldsheet curls of the same form, which we set to zero as part of the closure. The full current \eqref{eq:Jtheta} plus \eqref{eq:Jpol} satisfies $\nabla_\mu J^\mu=0$. The perturbation is axisymmetric, the canonical angular momentum $L_z=m_{\rm L}$ of every LLL state is conserved, and its kinetic part $\Pi_\phi=m_{\rm L}-A_\phi(\theta)$ changes only if the guiding centre moves, $\delta\theta=-\delta\Pi_\phi/\bar F_{\theta\phi}$. For the current $-J^\mu$ that couples to $A_\mu$, $-\mathcal D^\tau=-\delta T^\tau{}_\phi/\bar F_{\theta\phi}$ is the resulting dipole density, its $\tau$ derivative is the polarisation current and minus its $\theta$ divergence is the bound charge. The displacement of one guiding centre is $O(1/q)$, and summed over the $q$ species the current is $O(1)$. In $w$-units \eqref{eq:Nclosed}--\eqref{eq:Jpol} are independent of $q$. The index theorem sets the central charge equal to $q$, and the resulting cancellation of $q$ operates exactly as it did for the Hall current of \cite{Sadhukhan:2026czl}. The currents enter the Maxwell equations multiplied by $g_c^2=4\pi^2\alpha q$, at $O(\alpha)$, through the sources ($\nu=\tau,\rho,\theta$)
\begin{align}
g_c^2\sqrt{-g}\,J^{\tau}_{\rm pol}
&=-\,\frac{g_c^2\,\lhat\,(2\rho^2-1)}{48\pi^2 q\,r_e^2\,f^2}\,
\big(q\,h_0+2r_e^2(a_0-\dot a_2)\big)\,\sin\theta\;Y ,\nonumber\\
g_c^2\sqrt{-g}\,J^{\rho}_{\rm pol}
&=+\,\frac{g_c^2\,\lhat\,(2\rho^2+5)}{48\pi^2 q\,r_e^2}\,
\big(q\,h_1+2r_e^2(a_1-a_2')\big)\,\sin\theta\;Y ,
\label{eq:sources}
\end{align}
while $g_c^2\sqrt{-g}\,J^{\theta}=g_c^2 r_e^4\sin\theta\,(\partial_\tau \mathcal D^\tau+\partial_\rho \mathcal D^\rho)$ at $\alpha^0$. The full radial equations with these sources are in Appendix~\ref{app:alpha}. What remains open at the level of first principles is a direct Landau-level computation of $\langle T_{a\phi}\rangle$ in crossed electric and magnetic fields. A variational derivation analogous to the effective-action derivation of \cite{Sadhukhan:2026czl}, in the spirit of the Wen-Zee analysis \cite{Wen:1992ej,Abanov:2014ula,Gromov:2014gta,Bradlyn:2014wla}, also remains open. Both would test the assumption of Sec.~\ref{sec:driftderivation}.
\section{Master equations of the throat at order \texorpdfstring{$\alpha$}{alpha}}
\label{sec:alpha}

We assemble the throat system to first order in $\alpha$ with $m_S^2$ unexpanded, reduce it and bring it to Schr\"odinger form. The matching to the mouth is in Sec.~\ref{sec:matching}. The $\alpha\to0$ limit is read off in Sec.~\ref{sec:alpha0}.

\subsection{Ingredients and tortoise form}
\label{sec:alphaingredients}
The background is \eqref{eq:throatbg}--\eqref{eq:gamphi}. The function $S=\sqrt{A/B}=f+\alpha\gamma$ controls the tortoise coordinate and the Regge--Wheeler variable, $\sqrt{AB}=r_e^2$ exactly, and $R=r_e(1+\frac\alpha2\varphi)+O(\alpha^2)$ enters the normalisation of $\Psi_g$ and $\Psi_e$. The tortoise coordinate is
\begin{equation}
x=\int\frac{\dd\rho}{S}=\arctan\rho-\alpha\,\iota(\rho)+O(\alpha^2),
\qquad
\iota(\rho)=\int_0^\rho\frac{\gamma}{f^2}\,\dd\rho' .
\label{eq:iota}
\end{equation}
Every function of $\alpha$ is truncated after the linear term at each step. All coefficients are then rational functions of $\rho$, $\arctan\rho$ and $\log f$. In throat units $\kappa=8g_c^2r_e^2/q^2$ and $g_c^2=4\pi^2\alpha q$. The fermionic inputs are the Schwinger sources \eqref{eq:JS} with $S=f+\alpha\gamma$ in the $\chi$ equation \eqref{eq:chieq}, the drift lift \eqref{eq:driftlift} evaluated on the AdS$_2\times S^2$ background, and the currents \eqref{eq:Jtheta}--\eqref{eq:Jpol}. The lift and the currents are already $O(\alpha)$ in $w$-units once multiplied by $\kappa$ or $g_c^2$, and their own $O(\alpha)$ corrections would be $O(\alpha^2)$ in the field equations. The Schwinger sector is exact and its $O(\alpha)$ content enters only through $S$. After the rescaling $a_i=qw_i/4$, $\chi=qX/4$, the charge $q$ cancels from every equation.

The six field equations and the $\chi$ equation \eqref{eq:chieq} are listed in Appendix~\ref{app:alpha}, and the definition of $E$ completes the set of eight. Each is multiplied by the factor that clears the denominators of its $\alpha^0$ part. That part is the corresponding equation of Appendix~\ref{app:throat0}, and the $O(\alpha)$ part carries $\gamma$, $\varphi$, $\delta T^Q$ and the currents. The reduction of Sec.~\ref{sec:consistency} runs order by order. The eliminated amplitudes and the three second-order equations are also in Appendix~\ref{app:alpha}. Step 4 gives
\begin{equation}
\text{Res}\,\mathcal E_{\tau\phi}=\text{Res}\,\mathcal M^{\theta}=0
\qquad\text{at }O(\alpha^0)\ \text{and}\ O(\alpha^1),
\end{equation}
with the full response inserted. The $\phi$-Bianchi identity and charge conservation close exactly at this order. Section~\ref{sec:controls} tests which ingredients this fixes.

With the master variables \eqref{eq:mastervars}, $S=f+\alpha\gamma$, $R=r_e(1+\tfrac\alpha2\varphi)$, $\sqrt{AB}=r_e^2$ and $\dd x=\dd\rho/S$, the system reads
\begin{equation}
\bm Z_{xx}+\tilde P\,\bm Z_x+\tilde Q\,\bm Z=0,\qquad \bm Z=(\Psi_g,\Psi_D,\Xi)^{\!\top},
\qquad
\tilde P=\alpha\,\tilde P_1,\qquad
\tilde Q=-p^2\,I_3-f\,\mathcal N+\alpha\,\tilde Q_1,
\label{eq:tortoiseform}
\end{equation}
with $I_3$ the $3\times3$ identity, $\mu^2\equiv(l-1)(l+2)$ and
\begin{equation}
\mathcal N=\begin{pmatrix}
\mu^2 & -2 & 0\\
-2\mu^2 & \lhat+4 & \lhat\,m_S^2\\
0 & 1 & m_S^2
\end{pmatrix}.
\label{eq:Nfull}
\end{equation}
The coupling of Sec.~\ref{sec:kk} adds $-f\,\delta\mathcal N$ to $\tilde Q$ at $\alpha^0$, with
\begin{equation}
\delta\mathcal N=\frac1q\begin{pmatrix}
0 & 0 & \lhat\,m_S^2\\
0 & 0 & -2\lhat\,m_S^2\\
\mu^2 & -2 & \lhat\,m_S^2/q
\end{pmatrix}.
\label{eq:dN}
\end{equation}
At $m_S^2=0$ only its $\Xi$ row survives. We analyse the system without $\delta\mathcal N$ first and give its effect on the neutral channel in Secs.~\ref{sec:zmalpha} and \ref{sec:alpha0} and on the light massive channel in Sec.~\ref{sec:lightchannel}. The first-derivative matrix is $O(\alpha)$ and sparse,
\begin{equation}
\tilde P_1=\mathrm{diag}\Big(0,\;-\frac{\rho}{f},\;0\Big).
\end{equation}
Only the electric channel acquires a friction term. The entries of $\tilde Q_1$ carry $\arctan\rho$ and $\log f$ from \eqref{eq:gamphi} and are listed in Appendix~\ref{app:matrices}.

\subsection{Schr\"odinger form}
\label{sec:schrodinger}
The first-derivative term is removed by $\bm Z=(I_3-\tfrac\alpha2\mathcal G)\bm W$ with $\partial_x\mathcal G=\tilde P_1$ and $\mathcal G(0)=0$,
\begin{equation}
\mathcal G=\mathrm{diag}\Big(0,\;-\frac12+\frac{1}{2f},\;0\Big).
\end{equation}
Multiplying from the left by $(I_3+\tfrac\alpha2\mathcal G)$ restores the unit coefficient of $\bm W_{xx}$. The first-derivative terms $-\alpha \mathcal G_x\bm W_x+\alpha\tilde P_1\bm W_x$ cancel since $\partial_x\mathcal G=\tilde P_1$, and the potential acquires the term $-\tfrac\alpha2[\mathcal G,\tilde Q_0]$ with $\tilde Q_0=-p^2I_3-f\mathcal N$. The commutator does not vanish since $\mathcal G$ is not proportional to the identity. The result is
\begin{equation}
V=-\tilde Q-p^2\,I_3+\tfrac12\,\partial_x\tilde P+\tfrac14\,\tilde P^2-\tfrac\alpha2\,[\mathcal G,\tilde Q_0],
\qquad
\partial_x^2\bm W+\Big[\Om^2\big(I_3+\alpha V_{1b}\big)-V_0-\alpha V_{1a}\Big]\bm W=0 ,
\label{eq:schrodinger}
\end{equation}
with $V_0=f\,\mathcal N$ and the $O(\alpha)$ potential $V_{1a}+p^2V_{1b}$ exactly quadratic in the frequency. The frequency-dependent part is confined to the electric channel,
\begin{equation}
V_{1b}=\mathrm{diag}\Big(0,\;-\frac1f,\;0\Big)
\qquad\text{in }(W_g,W_D,W_\Xi),
\end{equation}
and $V_{1a}$ is a non-symmetric matrix whose entries are listed in Appendix~\ref{app:matrices}. Outside this subsection we call the transformed variable $\bm Z$ again.

The gravitational row of $V_{1a}$ does not couple to $W_\Xi$, and the coupling of the $W_D$ row to $W_\Xi$ is proportional to $m_S^2$, Appendix~\ref{app:matrices}. At $m_S^2=0$ the $(W_g,W_D)$ block of \eqref{eq:schrodinger} is therefore closed. This holds with $\delta\mathcal N$ as well, whose $\Psi_g$ and $\Psi_D$ rows are proportional to $m_S^2$. $W_\Xi$ is still sourced by $W_D$, through $f\mathcal N_{\Xi D}=f$ at $\alpha^0$ and through $(V_{1a})_{\Xi D}$ at $O(\alpha)$, and with $\delta\mathcal N$ also by $W_g$, but it does not feed back. The three-channel system is therefore block triangular. Its spectrum is the spectrum of the $2\times2$ block together with that of the driven massless channel, which is controlled by the mouths (Sec.~\ref{sec:zmalpha}). We call the block the Einstein--Maxwell--Casimir block and write it in the basis
\begin{equation}
Z_+=(l-1)\,W_g-W_D,\qquad Z_-=(l+2)\,W_g+W_D .
\label{eq:Zexplicit}
\end{equation}
The rows $(l-1,-1)$ and $(l+2,1)$ are the left eigenvectors of the $2\times2$ block of $\mathcal N$ with eigenvalues $m_+^2=(l+1)(l+2)$ and $m_-^2=l(l-1)$, and $V_0$ is diagonal in this basis. The basis is fixed at $\alpha=0$ and not rediagonalised at $O(\alpha)$, which is why $V_{1a}$ and $V_{1b}$ have off-diagonal entries in it. In terms of the master variables $W_g=\Psi_g$ and $W_D=\big(1-\tfrac{\alpha\rho^2}{4f}\big)\Psi_D+O(\alpha^2)$. In terms of the radial fields, with \eqref{eq:mastervars} and \eqref{eq:throatbg},
\begin{equation}
\begin{aligned}
Z_+&=(l-1)\,\frac{h_1}{r_e}\Big[f-\alpha\Big(2\rho^2+\tfrac23-\log f+\tfrac{\rho}{2}(3\rho^2+7)\arctan\rho\Big)\Big]
-\frac{p\,r_e}{\lhat}\Big[1+\alpha\Big(\rho\arctan\rho+\tfrac43-\frac{\rho^2}{4f}\Big)\Big]E ,\\
Z_-&=(l+2)\,\frac{h_1}{r_e}\Big[f-\alpha\Big(2\rho^2+\tfrac23-\log f+\tfrac{\rho}{2}(3\rho^2+7)\arctan\rho\Big)\Big]
+\frac{p\,r_e}{\lhat}\Big[1+\alpha\Big(\rho\arctan\rho+\tfrac43-\frac{\rho^2}{4f}\Big)\Big]E ,
\end{aligned}
\label{eq:Zexplicitfields}
\end{equation}
up to $O(\alpha^2)$. The first bracket is $(f+\alpha\gamma)(1-\tfrac\alpha2\varphi)$ from $\Psi_g=Sh_1/R$. The second is $(1+\alpha\varphi)(1-\tfrac{\alpha\rho^2}{4f})$ from $\Psi_e$ and the removal of the first-derivative term. At finite $m_S^2$ the same combinations carry the extra term $\pm\frac{p\,r_e}{\lhat}\,m_S^2\big(1-\tfrac{\alpha\rho^2}{4f}\big)X$, which comes from $-m_S^2\Xi$ in $\Psi_D$, but the block is then not closed and we do not use them. At $\alpha=0$ they reduce to \eqref{eq:Zpm}. At extremality they are the near-horizon limits of Chandrasekhar's mouth combinations $Z_i=\beta_i\Psi_g+2r_e\Psi_e$ of Sec.~\ref{sec:chandra}, $Z_1\to2r_e\,Z_-$ and $Z_2\to-2r_e\,Z_+$, since the roots $\beta_{1,2}$ defined there tend to $2r_e(l+2),\,-2r_e(l-1)$ at $GM=r_e$. In this basis
\begin{equation}
V_{1b}\Big|_{Z_\pm}=\frac{1}{(2l+1)f}
\begin{pmatrix}-(l+2)&l-1\\ l+2&-(l-1)\end{pmatrix},
\end{equation}
with $V_{1a}|_{Z_\pm}$ in Appendix~\ref{app:matrices}. At the throat centre, for $l=2$,
\begin{equation}
V_0\Big|_{\rho=0}=\begin{pmatrix}12&0\\0&2\end{pmatrix},\qquad
V_{1a}\Big|_{\rho=0}=\begin{pmatrix}-\tfrac{123}{5}&\tfrac{7}{30}\\[2pt]\tfrac{34}{15}&-\tfrac{77}{30}\end{pmatrix},\qquad
V_{1b}\Big|_{\rho=0}=\begin{pmatrix}-\tfrac45&\tfrac15\\[2pt]\tfrac45&-\tfrac15\end{pmatrix},
\end{equation}
and at large $\rho$ the diagonal entries grow as
\begin{equation}
V_{1a,++}\sim-\frac{\pi(2l^2+7l+7)}{2}\,\rho^3,
\qquad
V_{1a,--}\sim-\frac{\pi(2l^2-3l+2)}{2}\,\rho^3 .
\label{eq:rho3}
\end{equation}
The $\rho^3$ growth is the hand-over to the mouth, as in \cite{Sadhukhan:2026czl}. The throat expansion is valid for $\alpha\rho\ll1$, and in the overlap $1\ll\rho\ll1/\alpha$ both descriptions apply.

\subsection{Control experiments}
\label{sec:controls}
The closure used throughout is the drift lift \eqref{eq:driftlift}, the currents \eqref{eq:Jtheta}--\eqref{eq:Jpol} and the Schwinger current \eqref{eq:JS}. To see what the Bianchi identities fix, we recomputed the Step-4 residuals with parts of the response removed (Table~\ref{tab:controls}).
\begin{table}[htbp]
\caption{Step-4 residuals of the throat at $O(\alpha)$ for the closure of this paper with parts of the response removed, and for the response-free covariant closure.}
\label{tab:controls}
\centering
\begin{tabular}{llll}
\hline\hline
& closure & $\text{Res}\,\mathcal E_{\tau\phi}$ & $\text{Res}\,\mathcal M^{\theta}$\\
\hline
(a)& $J^\theta=J_{\rm pol}=0$, $\delta T^Q$ kept & nonzero & zero\\
(b)& $\delta T^Q=0$, $J^\theta$ and $J_{\rm pol}$ kept & nonzero & zero\\
(c)& $\delta T^Q=0$ and $J=0$ on the corrected background & zero & zero\\
(d)& Schwinger response omitted ($m_S^2=0$) & zero & zero\\
(e)& response-free covariant closure $\delta T^Q_{a\phi}=\mathcal P\,h_{a\phi}$ & zero & zero\\
\hline\hline
\end{tabular}
\end{table}
Closures (a) and (b) fail Step~4 with equal and opposite residuals,
\begin{equation}
\text{(a)}:\quad
\text{Res}\,\mathcal E_{\tau\phi}
=\frac{\alpha}{\lhat}\,\sqrt f\,\partial_\rho\!\Big(\frac{E-m_S^2X}{\sqrt f}\Big)
\;=\;-\,\text{(b)} .
\label{eq:ctrlres}
\end{equation}
The combination $E-m_S^2X$ is the displacement flux of the throat at $\alpha=0$, and the residual is the torque that the missing piece leaves unbalanced. In (a) the tilted Casimir stress tensor carries azimuthal momentum and no current crosses the field lines to supply the torque. In (b) the currents exert a Lorentz force and no fermion momentum balances it. The $\mathcal M^\theta$ residual vanishes in both cases, since $J^\theta$ and $J_{\rm pol}$ are conserved together. Step~4 fixes $\delta T^Q$ and the currents as a pair but does not decide whether the pair is present.

Closure (c) drops the whole response. The force balance $\nabla_\mu T_Q^{\mu\nu}=-F^{\nu\lambda}J_\lambda$ then requires $\delta\big(\nabla_\mu T_Q^{\mu\phi}\big)=0$. Instead
\begin{equation}
\delta\big(\nabla_\mu T_Q^{\mu\phi}\big)
=-\mathcal P\,\bar g^{\phi\phi}\,\nabla^a h_{a\phi}
=-\frac{q}{96\pi^2r_e^8\sin\theta}\Big[\partial_\rho\big(f\,h_1\big)-\frac{\dot h_0}{f}\Big]\,Y' ,
\label{eq:ctrlc}
\end{equation}
with $\nabla^a$ the covariant divergence on the AdS$_2$ factor. The $\tau$, $\rho$ and $\theta$ components vanish identically. The violation comes from the connection terms of $\delta(\nabla_\mu T_Q^{\mu\phi})$, in which the perturbed metric meets the background Casimir tensor. The bracket in \eqref{eq:ctrlc} equals $-2r_e^2\,\mathcal E_{\theta\phi}$ at $\alpha^0$ and vanishes on shell up to $O(\alpha)$. The source enters the field equations multiplied by $\kappa$, at $O(\alpha)$. The violation is $O(\alpha^2)$, beyond the order kept, and Step~4 passes. Closure (c) is nevertheless wrong. A pure axial diffeomorphism requires $\delta T^Q_{a\phi}=\mathcal P\,h_{a\phi}$ (Sec.~\ref{sec:drift}) and (c) gives zero. Closure (d) omits the Schwinger response and passes because the Schwinger current is identically conserved and exerts no force at $O(\epsilon)$. Its presence is a physical input. Closure (e) is the response-free covariant closure $\delta T^Q_{a\phi}=\mathcal P\,h_{a\phi}$, the member $\mathsf K=0$ of the family \eqref{eq:family}. It passes Step~4 and the covariance test. The Bianchi identities reject an inconsistent closure but cannot choose between the drift lift and (e). That choice rests on the assumption of Sec.~\ref{sec:driftderivation}.

The $\rho^3$ coefficients of $V_{1a}$ are identical for the full response, for (e) and for (c), since the $\rho^3$ growth is the background deformation $\gamma\sim-\tfrac\pi2\rho^3$. The difference between the full response and (e), in the $(\Psi_g,\Psi_e)$ block at $m_S^2=0$, is
\begin{equation}
V_{1a}^{\rm drift}-V_{1a}^{\rm (e)}=
\begin{pmatrix}
0 & -\dfrac{\rho^2}{2}\\[6pt]
-\dfrac{(l-1)(l+2)(\rho^2-2)}{6} & \dfrac{2l^2\rho^4+l^2\rho^2-l^2+2l\rho^4+l\rho^2-l+8\rho^4+7\rho^2-7}{6f}
\end{pmatrix}.
\label{eq:driftdiff}
\end{equation}
It grows as $\alpha\rho^2$ with coefficient matrix $\big(\begin{smallmatrix}0&-1/2\\ -\mu^2/6&(l^2+l+4)/3\end{smallmatrix}\big)$, subleading to $\rho^3$ in the overlap. Closure (e) has $\tilde P_1=0$ and $V_{1b}=0$. The drift lift produces both, and the entry $-\rho^2/2$ of \eqref{eq:driftdiff} is the commutator term $-\tfrac12[\mathcal G,\tilde Q_0]$ left by the removal of its friction. At $m_S^2=0$ the matrix \eqref{eq:driftdiff} together with the frequency-dependent term $\alpha\Om^2V_{1b}$ is therefore the entire imprint of the guiding-centre drift on the Einstein--Maxwell--Casimir block of the throat at this order. Closure (e) does not mix the modes of a level shared by $Z_+$ and $Z_-$ at first order. In the basis $Z_\pm$ its off-diagonal entries are constant multiples of $f=\sec^2x$, whose matrix element between the two modes vanishes (Sec.~\ref{sec:shifts}). The mixing of Sec.~\ref{sec:shifts} therefore comes entirely from the drift. It exists only at the formal point $m_S^2=0$. At the physical $m_S^2$ the imprint of the drift on the spectrum of the throat is the shift of the light levels that \eqref{eq:lam1minusdrift} and $\lambda^-_{1b}$ produce (Sec.~\ref{sec:lightchannel}). A first-principles Landau-level computation would have to reproduce this shift. At finite $m_S^2$ the drift also changes the $D\Xi$ and $\Xi D$ entries of $V_{1a}$, by $\lhat m_S^2(\rho^2-2)/12$ and $\rho^2/4$, and the neutral channel of Sec.~\ref{sec:zmalpha} depends on the closure. Section~\ref{sec:matching} shows that the mouth carries the same imprint and that the matching cannot separate the closures either.

\subsection{Shifts of the normal frequencies}
\label{sec:shifts}
In this subsection $m_S^2=0$, the formal point of Sec.~\ref{sec:whynot} at which the Einstein--Maxwell--Casimir block closes. The light massive channel at the physical $m_S^2$ is treated in Sec.~\ref{sec:lightchannel}. The $\alpha=0$ normal modes of the massive channels are the Poschl-Teller modes $\psi_n=\cos^sx\,C_n^{(s)}(\sin x)$, with $C_n^{(s)}$ the Gegenbauer polynomials. They have $V_0=m^2\sec^2x$, $m^2=s(s-1)$ and $\Om_0=n+s$, where $s=l$ for $Z_-$ and $s=l+2$ for $Z_+$. The first-order shifts are computed as for the polar sector in \cite{Sadhukhan:2026czl}. The potentials in \eqref{eq:schrodinger} are functions of $\rho$, and inverting \eqref{eq:iota} to first order gives
\begin{equation}
\rho=\tan x+\alpha\,\iota\sec^2x,\qquad V_0=f\mathcal N=\sec^2x\,(1+2\alpha\rho\,\iota)\,\mathcal N .
\label{eq:remap}
\end{equation}
In $V_{1a}$ and $V_{1b}$ one may set $\rho=\tan x$ and in $V_0$ one may not. On a non-degenerate level
\begin{equation}
\Om^2=\Om_0^2+\alpha\,\varsigma,\qquad
\varsigma=\frac{\int\dd x\,\psi_n\big(V_{1a}-\Om_0^2V_{1b}+2\rho\,\iota\,V_0\big)\psi_n}{\int\dd x\,\psi_n^2},
\qquad \rho=\tan x,
\label{eq:shift}
\end{equation}
with the diagonal entries of the $Z_\pm$ matrices and $V_0=m^2\sec^2x$. The term $2\rho\,\iota\,V_0$ is the remap term of \cite{Sadhukhan:2026czl}, which comes from the difference between $x$ and $\arctan\rho$. It supplies between 4 and 50 per cent of a shift for the modes of Table~\ref{tab:shifts}, and its share grows with $n$. The integrals converge, since $V_{1a}\sim\sec^3x$ and $\rho\,\iota\,V_0\sim\sec^3x\,\log\sec x$ at the ends while $\psi_n^2\sim\cos^{2s}x$ with $s\ge2$. We evaluate them by quadrature and in closed form (Appendix~\ref{app:numerics}). On every level the diagonal shift is $2\Om_0^2\log2$ plus a rational number. The logarithm comes from the $\log f=-2\log\cos x$ terms of the corrected background. The part $2m^2\log2$ comes from the term $m^2\log f$ of $V_{1a}$, and the remap term adds $2(\Om_0^2-m^2)\log2$ by the virial identity of the Poschl-Teller modes. The axial column of Table~\ref{tab:shifts} in Sec.~\ref{sec:isothroat0} lists them for $n=0,1,2$ and $l=2,3,4$. On the levels shared by $Z_+$ and $Z_-$, which are all $Z_+$ levels and the $Z_-$ levels with $n\ge2$, the listed values are the diagonal entries of the matrix \eqref{eq:Smatrix} below. At $\eta_{\rm H}=0$ the shifts on these levels are its eigenvalues. With the higher Landau levels the shared levels are split, and the shift is the diagonal entry (Sec.~\ref{sec:hll-throat}). All these shifts are negative. They grow with $n$ since the higher modes sample the region of larger $\rho$ where the corrected background lowers the potential. Their sign refers to the clock of Sec.~\ref{sec:sector}. A change of clock leaves the splittings, the off-diagonal entries of $\mathcal S$ and the reality of the shifts unchanged, and the term $2\Om_0^2\log2$ of every diagonal entry has the form $\gamma_0\Om_0^2$ of such a change (Sec.~\ref{sec:sector}). The levels $\Om_0=n+l+2$ are shared by $Z_+$ with quantum number $n$ and $Z_-$ with quantum number $n+2$. There the shifts are the eigenvalues, as in degenerate first-order perturbation theory \cite{Kato:1995}, of
\begin{equation}
\mathcal S_{ij}=\frac{\int\dd x\,\psi_i\Big[\big(V_{1a}-\Om_0^2V_{1b}\big)_{ij}+\delta_{ij}\,2\rho\,\iota\,V_{0,i}\Big]\psi_j}{\int\dd x\,\psi_i^2},
\label{eq:Smatrix}
\end{equation}
whose off-diagonal entries are rational. For $l=2$ and $\Om_0=4$
\begin{equation}
\mathcal S=\begin{pmatrix}-\tfrac{5188}{105}+32\log2 & \tfrac{72}{35}\\[2pt] \tfrac65 & -\tfrac{1316}{15}+32\log2\end{pmatrix},
\end{equation}
with eigenvalues $-27.16$ and $-65.62$. These contain the square root of a rational number that is not a perfect square, and they are not of the form rational plus $2\Om_0^2\log2$. For $\Om_0=5$ and $\Om_0=6$
\begin{equation}
\mathcal S=\begin{pmatrix}-\tfrac{7793}{84}+50\log2 & \tfrac{15}{7}\\[2pt] 5 & -\tfrac{299}{2}+50\log2\end{pmatrix},
\qquad
\mathcal S=\begin{pmatrix}-\tfrac{5269}{35}+72\log2 & \tfrac{11}{5}\\[2pt] \tfrac{66}{5} & -\tfrac{8032}{35}+72\log2\end{pmatrix},
\label{eq:Sl2ax}
\end{equation}
with eigenvalues $(-57.93,-115.03)$ and $(-100.27,-179.95)$. The product of the two off-diagonal entries does not depend on the normalisation of the Gegenbauer polynomials. It is $\tfrac{432}{175}$, $\tfrac{75}{7}$ and $\tfrac{726}{25}$ on the three levels. On every shared level the two entries are
\begin{equation}
\begin{aligned}
\mathcal S_{+-}&=\frac{2l(l^2-1)\,\Om_0(\Om_0+2l+1)}{(2l+1)(\Om_0+l)(\Om_0+l+1)},\\
\mathcal S_{-+}&=\frac{(l+2)\,\Om_0(\Om_0-l)(\Om_0-l-1)(\Om_0+2l+1)}{8l(l+1)(2l+1)},
\end{aligned}
\label{eq:Sentries}
\end{equation}
and their product is positive for every $\Om_0\ge l+2$. The off-diagonal entries of $V_{1a}-\Om_0^2V_{1b}$, with $V_{1a}$ of Appendix~\ref{app:matrices} and $V_{1b}$ of Sec.~\ref{sec:schrodinger}, are $\tfrac{l-1}{2(2l+1)}V_{\rm mix}$ and $\tfrac{l+2}{2(2l+1)}V_{\rm mix}$ plus constant multiples of $f=\sec^2x$, with
\begin{equation}
V_{\rm mix}=\lhat-2\big(\Om_0^2-1\big)\cos^2x .
\label{eq:Vmixax}
\end{equation}
The two diagonal entries of $V_0$ differ by $(4l+2)\sec^2x$, and both modes solve their equations at the same $\Om_0$. The multiples of $\sec^2x$ therefore have no matrix element between the two modes. The matrix element of $V_{\rm mix}$ follows from the Gegenbauer polynomials, as in the appendix on the off-diagonal entries of \cite{Sadhukhan:2026czl}, and gives \eqref{eq:Sentries}. On the levels computed, up to $\Om_0=20$ for $l=2$, up to $\Om_0=15$ for $l=3$ and the lowest three shared levels for $l=4,5,6$, the ratio $\max|\varsigma|/\Om_0^2$ grows slowly with $\Om_0$. For $l=2$ it rises from $4.1$ at $\Om_0=4$ to $7.5$ at $\Om_0=20$, roughly as $2\log\Om_0$, and for $l=3$ it stays below $5.4$ up to $\Om_0=15$. With a positive product the eigenvalues are real for any diagonal entries. On the levels examined the mixing moves each eigenvalue away from its diagonal entry by at most $1.13$ per cent of the eigenvalue, at $l=6$ and $\Om_0=10$. The two modes of a shared level repel by slightly more than their diagonal splitting. We also solved the $O(\alpha)$ eigenvalue problem directly for $l=2$ and $\Om_0\le6$ at $\alpha=10^{-3}$, $3\times10^{-4}$ and $10^{-4}$, with the throat closed by Dirichlet conditions at $\alpha\rho=\pm0.1$ (Appendix~\ref{app:numerics}). Its eigenvalues approach these values as $\alpha$ decreases, and none is complex. The massless channel has no finite first-order shift in this idealisation. The remap term vanishes on it, since $V_0$ annihilates it. For $m_S^2>0$ its $O(\alpha)$ potential grows as $\rho^3=\tan^3x$ (Sec.~\ref{sec:zmalpha}) while the squares of its free eigenfunctions fall as $\cos^2x$, and the integral in \eqref{eq:shift} diverges logarithmically at the ends of the throat. At $m_S^2=0$ its projected potential vanishes, but its levels depend on the tortoise length of the throat, which grows with the cutoff since $\iota\simeq-\frac\pi2\log\rho$ at large $\rho$. In both cases the mode is controlled by the mouths.

\subsection{The massless mode}
\label{sec:zmalpha}
The matrix $\mathcal N$ of \eqref{eq:Nfull} has $\det\mathcal N=0$ for every $l$ and $m_S^2$ (Sec.~\ref{sec:alpha0}). Its right and left null vectors are
\begin{equation}
\begin{gathered}
\bm v_0=\Big(-\frac{2m_S^2}{\mu^2},\;-m_S^2,\;1\Big)^{\!\top},\qquad
\bm L_0=\Big(-\frac{2}{\lhat},\;-\frac{1}{\lhat},\;1\Big),\\
\bm L_0\bm v_0=\frac{\mathcal H}{(l-1)l(l+1)(l+2)},
\qquad
\mathcal H\equiv(l-1)l(l+1)(l+2)+m_S^2(\lhat+2).
\end{gathered}
\label{eq:nullvectors}
\end{equation}
The two differ since $\mathcal N$ is not symmetric. At $\alpha=0$ the potential $f\mathcal N$ annihilates $\bm v_0$ and the mode $\bm Z=\psi_{\rm zm}(x)\,\bm v_0$ is free. At $O(\alpha)$ first-order perturbation theory with the two null vectors gives the potential of the neutral channel, $-\partial_x^2\psi_{\rm zm}+V_{\rm zm}\,\psi_{\rm zm}=\Om^2\psi_{\rm zm}$, with
\begin{equation}
V_{\rm zm}=\alpha\big(\lambda_{1a}(\rho)+p^2\lambda_{1b}(\rho)\big)+O(\alpha^2),
\qquad
\lambda_{1a,b}=\frac{\bm L_0V_{1a,b}\,\bm v_0}{\bm L_0\bm v_0},
\end{equation}
\begin{equation}
\lambda_{1b}=-\frac{m_S^2\,\mu^2}{f\,\mathcal H},
\qquad
\lambda_{1a}
=-\frac{m_S^2}{2f\,\mathcal H}
\Big[\,8\rho^5\arctan\rho+8\rho^4+16\rho^3\arctan\rho+18\rho^2+8\rho\arctan\rho+6
\;-\;\lhat\,(\rho^2-1)\,\Big].
\label{eq:lam1a}
\end{equation}
At the throat centre and at large $\rho$,
\begin{equation}
\lambda_{1a}(0)=-\frac{m_S^2\,(\lhat+6)}{2\,\mathcal H}<0,
\qquad
\lambda_{1a}\sim-\frac{2\pi\,m_S^2}{\mathcal H}\,\rho^3\ \ (\rho\to\infty).
\end{equation}
For $m_S^2\to\infty$, $\mathcal H\to m_S^2(\lhat+2)$ and $\lambda_{1a}$ saturates. Near the centre $\lambda_{1a}''(0)=2m_S^2(\lhat-10)/\mathcal H$, and for $l=2$ the centre is a maximum of $\lambda_{1a}$, which decreases outward. The negative values of $\lambda_{1a}$ take their meaning from the mouth. Under $\rho=\ell y/r_e^2$ the large-$\rho$ form becomes $\alpha\lambda_{1a}\to-8m_S^2y^3/(\mathcal H r_e^5)$ in mouth units. This is the near-horizon $O(y^3)$ coefficient of the neutral channel of the mouth potential \eqref{eq:VmouthS}. At $GM=r_e$ one has $\hat V=r_e\mathcal N+y\,\partial_r\hat V(r_e)+O(y^2)$, where $\partial_r\hat V(r_e)$ has the rows $(\lhat-4,0,0)$, $(0,\lhat-4,\lhat m_S^2)$ and $(0,1,m_S^2)$. The projection of $y\,\partial_r\hat V(r_e)$ with $\bm v_0$ and $\bm L_0$ moves the zero eigenvalue of $\hat V$ to $-8m_S^2y/\mathcal H$, and with $\Delta/r^5\simeq y^2/r_e^5$ this gives the coefficient above. For $m_S^2\to\infty$ it is the coefficient $-8y^3/((\lhat+2)r_e^5)$ that Sec.~\ref{sec:mouthneutral} finds in the light system. The neutral channel matches through $O(\alpha)$ like the massive ones. The $\rho^3$ tail of the throat is the inner end of the region of the mouth between $r_e$ and $2r_e$ where $\hat V$ has a negative eigenvalue. Integrated against $\dd r_*=r_e^2\dd y/y^2$ the near-horizon part contributes $\int V\,\dd r_*=-\tfrac{4m_S^2}{\mathcal H}(y/r_e)^2/r_e$, and the weight of the negative region sits at $y\sim r_e$, in the mouth. The interior of the throat, $x=O(1)$, contributes $O(\alpha)$ in throat units, which is $O(\alpha/\ell)\sim10^{-70}$ in mouth units. The global analysis is in Sec.~\ref{sec:neutralstab}.

The neutral channel depends on the closure. The response-free closure (e) of Sec.~\ref{sec:controls} has $V_{1b}=0$ and therefore $\lambda_{1b}^{\rm (e)}=0$, and $\lambda_{1a}-\lambda_{1a}^{\rm (e)}=m_S^2\mu^2(\rho^2-1)/(2f\mathcal H)$. At the centre, for $l=2$ and $m_S^2\to\infty$, this difference is $-\tfrac14$ out of $\lambda_{1a}(0)=-\tfrac34$. The $\rho^3$ tail is common to both closures.

The coupling of Sec.~\ref{sec:kk} keeps the channel massless. The matrix $\mathcal N+\delta\mathcal N$ has $\det(\mathcal N+\delta\mathcal N)=0$, with the null vectors
\begin{equation}
\bm v_0=\Big(-\Big(2+\frac{\lhat}{q}\Big)\frac{m_S^2}{\mu^2},\;-m_S^2,\;1\Big)^{\!\top},\qquad
\bm L_0=\Big(-\frac{2+\lhat/q}{\lhat},\;-\frac1{\lhat},\;1\Big),
\label{eq:nullvectorskk}
\end{equation}
exactly in $1/q$. They reduce to \eqref{eq:nullvectors} at $1/q\to0$. With these vectors and the $O(\alpha/q)$ terms of Appendix~\ref{app:matrices} the projection changes $\lambda_{1a}$ and $\lambda_{1b}$ by a relative amount of order $1/q$, which is $O(\alpha/q)$ in absolute terms. The correction to $\lambda_{1a}$ has no $\rho^2$ term, and its $\rho^3$ coefficient matches the near-horizon $y^3$ term of the mouth with the coupling to first order in $1/q$.

\subsection{The \texorpdfstring{$\alpha\to0$}{alpha->0} limit and the exact AdS\texorpdfstring{$_2\times S^2$}{2 x S2} throat}
\label{sec:alpha0}

Setting $\alpha=0$ while keeping $m_S^2$, which is not part of the $\alpha$ counting, collapses the system \eqref{eq:tortoiseform} onto the exact AdS$_2\times S^2$ throat. Unlike the polar-metric/axial-gauge sector of \cite{Sadhukhan:2026czl}, the throat is not pure Einstein--Maxwell. The polar gauge field is a worldsheet gauge field, and the LLL charge still responds through the Schwinger current, which on this background is MMP's Appendix~A result. The radial equations and the reduction, derived independently on the exact background as a check, are in Appendix~\ref{app:throat0}. Here $\tilde P\equiv0$ exactly and, with $x=\arctan\rho\in(-\tfrac\pi2,\tfrac\pi2)$,
\begin{equation}
\partial_x^2\bm Z+\big(\Om^2-f\,\mathcal N\big)\bm Z=0,
\qquad f=\sec^2x,
\label{eq:alpha0system}
\end{equation}
in the basis $(\Psi_g,\Psi_D,\Xi)$ with $\mathcal N$ of \eqref{eq:Nfull}. These are three coupled Poschl-Teller problems whose masses$^2$ are the eigenvalues of $\mathcal N$. The constant matrix $\Sigma$ of \eqref{eq:Ssym} in Sec.~\ref{sec:mouthneutral} symmetrises $\mathcal N$.

At $m_S^2=0$ the $\Xi$ channel is driven by $\Psi_D$ but does not act back on it, $\Psi_e=\Psi_D$, and the $2\times2$ block diagonalises exactly in the combinations \eqref{eq:Zexplicit}, which at $\alpha=0$ read
\begin{equation}
Z_+=(l-1)\,\Psi_g-\Psi_e,\qquad Z_-=(l+2)\,\Psi_g+\Psi_e ,
\label{eq:Zpm}
\end{equation}
\begin{equation}
\partial_x^2Z_\pm+\big(\Om^2-m_\pm^2\sec^2x\big)Z_\pm=0,
\qquad
m_+^2=(l+1)(l+2),\qquad m_-^2=l(l-1).
\label{eq:PTpm}
\end{equation}
These are the integer masses of the polar-metric sector of \cite{Sadhukhan:2026czl}. In the Einstein--Maxwell throat the two parity sectors are isospectral, as the near-horizon limit of Chandrasekhar's Darboux pair requires (Sec.~\ref{sec:isothroat0}), and the masses agree with the Kaluza--Klein spectrum of \cite{Michelson:1999zc}. The mouth reproduces the matrix \eqref{eq:Nfull} at $m_S^2=0$ in its near-horizon limit (Sec.~\ref{sec:matching}), which closes the triangle of mouth, throat and companion sector. The integer values make the potentials the trigonometric counterparts of reflectionless Poschl-Teller potentials \cite{Cooper:1994eh}, with the consequences for the echo structure discussed in \cite{Mondal:2025tht,Cardoso:2016rao,Cardoso:2017cqb,Bueno:2017hyj}.

At finite $m_S^2$ the spectrum of $\mathcal N$ follows from
\begin{equation}
\det(\mathcal N-\lambda I_3)=-\lambda\Big[\lambda^2-\big(2\lhat+2+m_S^2\big)\lambda
+(l-1)\,l\,(l+1)(l+2)+m_S^2\big(\lhat+2\big)\Big].
\label{eq:charpoly}
\end{equation}
Hence $\det\mathcal N=0$ for every $l$ and every $m_S^2$. The constant term of the quadratic factor is the polynomial $\mathcal H$ of \eqref{eq:nullvectors}. The two nonzero masses are
\begin{equation}
\lambda_\pm=\tfrac12\Big[2\lhat+2+m_S^2\pm\sqrt{(m_S^2-2)^2+16\lhat}\,\Big],
\label{eq:lampm}
\end{equation}
real for all $m_S^2\ge0$ by the discriminant identity
\begin{equation}
\text{disc}=\big(m_S^2-2\big)^2+16\,\lhat\;\ge\;0 ,
\label{eq:disc}
\end{equation}
and positive for $l\ge2$, since their sum and product are positive. At $m_S^2=0$ they reduce to $(l+1)(l+2)$ and $l(l-1)$. At large $m_S^2$, $\lambda_+\simeq m_S^2$ is the screened longitudinal electric field, the throat analogue of the massive Schwinger boson, and
\begin{equation}
\lambda_-=l^2+l+2-\frac{4\lhat}{m_S^2}-\frac{8\lhat}{m_S^4}+O\big(m_S^{-6}\big) .
\label{eq:lamseries}
\end{equation}
The limit $l^2+l+2$ is not of the form $k(k+1)$ with integer $k$. At the physical, large value of $m_S^2$ the reflectionless structure of the throat and the isospectrality between the parity sectors are properties of the Einstein--Maxwell limit only. With the coupling of Sec.~\ref{sec:kk} the mass matrix is $\mathcal N+\delta\mathcal N$. Its determinant still vanishes, and the quadratic factor of \eqref{eq:charpoly} becomes $\lambda^2-\big(2\lhat+2+m_S^2+\lhat m_S^2/q^2\big)\lambda+\mathcal H+\lhat m_S^2(4q+\lhat)/q^2$. Its coefficients stay positive, and the matrix $\Sigma$ of \eqref{eq:Ssym} symmetrises $\mathcal N+\delta\mathcal N$. The masses therefore stay real and positive. At large $m_S^2$ the light mass $\lambda_-$ moves up by $4\lhat/q$, a relative $4\lhat/((\lhat+2)q)$, which is $3\times10^{-33}$ at $l=2$. The term $-4\lhat/m_S^2$ of \eqref{eq:lamseries} is larger by the factor $q/m_S^2=4\pi^2/g_c^2$. The mass $\lambda_+$ moves down by nearly the same amount, and its relative change is smaller by the factor $m_S^2/(\lhat+2)$. Section~\ref{sec:lightchannel} compares these shifts with the $O(\alpha)$ shifts.

The zero eigenvalue has the right null vector $\bm v_0$ of \eqref{eq:nullvectors}, whose components give
\begin{equation}
\Psi_e=\Psi_D+m_S^2\,\Xi=0,\qquad E=0 .
\label{eq:zeromode}
\end{equation}
This is an exactly massless mode, for every $m_S^2$, in which the electric field along the field lines vanishes identically while the bosonised scalar $\chi$ and its gravitational dressing $\Psi_g$ do not. In the radial fields, take $X$ a free two-dimensional field, $X''=p^2X/f^2-2\rho X'/f$, set $E\equiv0$ and
\begin{equation}
h_1=-\frac{2m_S^2\,p\,r_e^2}{(l-1)l(l+1)(l+2)\,f}\,X,\quad
w_1=\frac{m_S^2\,p\,(l^2+l+2)}{(l-1)l(l+1)(l+2)\,f}\,X,\quad
w_0=\frac{m_S^2\,f\,(l^2+l+2)}{(l-1)l(l+1)(l+2)}\,X',
\label{eq:zmfields}
\end{equation}
with $h_0$ from Step 1. All eight radial equations are then satisfied identically. They are the six field equations, the definition of $E$ and the $\chi$ equation. With the coupling of Sec.~\ref{sec:kk} the null vector is \eqref{eq:nullvectorskk}, the factor $2$ in $h_1$ becomes $2+\lhat/q$, and $l^2+l+2$ in $w_0$ and $w_1$ becomes $l^2+l+2+2\lhat/q$. The eight equations are then again satisfied identically. The transverse field $\delta F_{a\theta}=-\tfrac q4w_aY'$ that connects neighbouring lines is nonzero, no longitudinal field is produced, and the metric is dressed at order $m_S^2X$. This is a massless wave of the bosonised charge and current of the field lines. Both vary over the sphere as $Y$, and the charge has no net value. In the time domain \eqref{eq:zmfields} holds with $pX\to\dot X$, and $X=X_1t$ with a constant $X_1$ solves the free equation. It gives $E=0$, $h_0=w_0=0$ and static $h_1$ and $w_1$. By \eqref{eq:JS} it is a zero-frequency persistent current along the field lines, $J^r_S\propto-\dot\chi$, with no charge density, $J^t_S\propto\chi'=0$. Its static fields are the magnetic field $\delta F_{r\theta}=-\frac q4w_1Y'$ of the current and the metric component $h_{r\phi}$. The same form solves the mouth equations of Appendix~\ref{app:mouth} for every $GM$, and it is the zero-energy solution of Sec.~\ref{sec:neutralstab}. The massless mode is the gravitationally dressed image of the massless sector of MMP's Appendix~A. Bosonising $q$ fermions, only the centre-of-mass boson acquires the Schwinger mass, and the relative $SU(q)_1$ WZW modes stay massless \cite{Witten:1983ar}. The analogue of the massless mode in chiral transport is the chiral magnetic wave \cite{Kharzeev:2010gd}. With a dynamical gauge field the wave of the total charge is gapped. Here that wave is the screened channel, and the relative charges stay gapless. The $l$-th sphere multipole of the line-resolved charge density has no net charge, and we identify it with these relative modes. Equation \eqref{eq:zeromode} shows it surviving the coupling to gravity and electromagnetism at every $m_S^2$. The mode is massless without gravity as well. With $h_0=h_1=0$ the Maxwell equations give $w_1=p\,(m_S^2X-E)/(\lhat f)$ and $w_0=-f\,(E-m_S^2X)'/\lhat$, and the test-field system has, in the basis $(\Psi_D,\Xi)$, the mass matrix
\begin{equation}
\mathcal N_{\rm test}=\begin{pmatrix}\lhat & \lhat\,m_S^2\\ 1 & m_S^2\end{pmatrix},\qquad \det\mathcal N_{\rm test}=0,
\end{equation}
with the null vector $(-m_S^2,1)^{\!\top}$. Gravity adds $4$ to the $DD$ entry and adds the $\Psi_g$ row and column. The null vector survives, dressed by $\Psi_g=-2m_S^2\,\Xi/\mu^2$. The fate of the mode at $O(\alpha)$ is treated in Sec.~\ref{sec:zmalpha}, and in the mouths in Sec.~\ref{sec:neutralstab}.

\subsection{The light massive channel at the physical Schwinger mass}
\label{sec:lightchannel}
The shifts of Sec.~\ref{sec:shifts} are computed at the formal point $m_S^2=0$. At the physical value $m_S^2=\alpha q^2$ the screened channel has the large mass $\lambda_+\simeq m_S^2$, and the light massive content of the throat is the single channel $\lambda_-$ of \eqref{eq:lampm}. Its first-order shifts follow from the left-right projection used for the neutral channel in Sec.~\ref{sec:zmalpha}. For an eigenvalue $\lambda\neq0$ of $\mathcal N$, with $\lambda\neq\mu^2$ and $\lambda\neq m_S^2$, the right and left eigenvectors are
\begin{equation}
\bm R_\lambda=\Big(2,\;\mu^2-\lambda,\;\frac{\lambda-\mu^2}{m_S^2-\lambda}\Big)^{\!\top},\qquad
\bm L_\lambda=\Big(\frac{2\mu^2}{\mu^2-\lambda}\,\frac{\lambda-m_S^2}{\lhat\,m_S^2},\;\frac{\lambda-m_S^2}{\lhat\,m_S^2},\;1\Big),
\label{eq:RLlam}
\end{equation}
and at $\lambda=\lambda_-$ the channel carries the potential $\lambda_-f+\alpha\big(\lambda^-_{1a}+p^2\lambda^-_{1b}\big)$ with $\lambda^-_{1a,b}=\bm L_\lambda V_{1a,b}\bm R_\lambda/(\bm L_\lambda\bm R_\lambda)$. Several entries of $V_{1a}$ grow like $m_S^2$, but the projection stays finite as $m_S^2\to\infty$. In this limit
\begin{equation}
\begin{aligned}
\lambda^-_{1a}&=(\lhat+2)\log f-\frac{\rho\big[(2\lhat^2+13\lhat+14)\rho^2+4\lhat^2+21\lhat+22\big]}{\lhat+2}\,\arctan\rho\\
&\quad-\frac{(8\lhat^2+47\lhat+50)\rho^4+6(2\lhat^2+14\lhat+15)\rho^2+4\lhat^2+37\lhat+52}{3(\lhat+2)f},
\qquad
\lambda^-_{1b}=-\frac{4}{(\lhat+2)f},
\end{aligned}
\label{eq:lam1minus}
\end{equation}
with corrections of relative order $1/m_S^2$, a few times $10^{-31}$ at the fiducial parameters. These corrections multiply $\alpha$ and are negligible, unlike the term $-4\lhat/m_S^2$ of \eqref{eq:lamseries} in $\lambda_-$ itself, which we treat at the end of this subsection. At the centre $\lambda^-_{1a}(0)=-(4\lhat^2+37\lhat+52)/[3(\lhat+2)]$, which is $-\tfrac{209}{12}$ for $l=2$, and at large $\rho$
\begin{equation}
\lambda^-_{1a}\sim-\frac{\pi\,(2\lhat^2+13\lhat+14)}{2(\lhat+2)}\,\rho^3 .
\label{eq:lam1minusrho3}
\end{equation}
This tail matches the mouth. At $GM=r_e$ and $r=r_e$ the matrix $r^5V_{\rm light}/\Delta$ of \eqref{eq:Vlight} has the eigenvalue $(\lhat+2)r_e$ with the right and left eigenvectors $(1,2)^{\!\top}$ and $(\mu^2,2)$. We expand this eigenvalue to first order in $y=r-r_e$, multiply by $\Delta/r^5=y^2(1-5y/r_e)/r_e^5+O(y^4)$ and insert the MMP dictionary $r=r_e+\rho r_e^2/\ell$. The expansion of $\Delta/r^5$ contributes, since the eigenvalue does not vanish at $y=0$. The result is $\ell^2V\to(\lhat+2)\rho^2+\alpha\rho^3$ times the coefficient of \eqref{eq:lam1minusrho3}. The channel therefore matches through $O(\alpha)$, like the neutral one. In $V_{1a}$ the Schwinger mass enters only the $D\Xi$ and $\Xi\Xi$ entries, while the $\Xi$ component of $\bm R_\lambda$ is $O(1/m_S^2)$. The product of the projected $O(\alpha)$ couplings of this channel with the neutral channel stays finite as $m_S^2\to\infty$, since the growth of $\bm v_0$ like $m_S^2$ is cancelled by $\bm L_0\bm v_0\propto m_S^2$. With the screened channel the product grows like $m_S^2$, but it enters at second order divided by $\lambda_+\simeq m_S^2$. The second-order terms are thus $O(\alpha^2)$ at every $m_S^2$.

The potential of the channel is positive on the validity window. For $l=2,3,4$ and $\alpha\le10^{-2}$ the ratio of $\lambda_-f+\alpha\lambda^-_{1a}$ to $\lambda_-$ is at least $0.978$ for $\alpha\rho\le0.1$. The potential first vanishes at $\alpha\rho=0.248$, $0.272$ and $0.287$ for $l=2,3,4$ at $\alpha=10^{-6}$, and about $0.002$ lower at $\alpha=10^{-2}$. The weight $1+\alpha\lambda^-_{1b}\ge1-4\alpha/(\lhat+2)$ is positive. At $\alpha=0$ the channel is a Poschl-Teller problem with the mass $\lambda_-$. Its normalisable modes are $\psi_n=\cos^{s_-}x\,C_n^{(s_-)}(\sin x)$ with $\Om_0=n+s_-$ and
\begin{equation}
s_-=\frac12+\frac12\sqrt{1+4\lambda_-}\;\longrightarrow\;\frac12+\frac12\sqrt{4\lhat+9}\qquad(m_S^2\to\infty).
\end{equation}
For $l\ge1$ the integer $4\lhat+9=(2l+1)^2+8$ is not a perfect square. The offset $s_-$ is irrational, and its fractional part is $0.372$, $0.275$ and $0.217$ for $l=2,3,4$. At the physical $m_S^2$ the offset differs from this limit by less than $10^{-30}$, far less than its distance from an integer. The levels of this channel therefore stay apart from the integer levels $\Om_0=n+1$ of the neutral channel. The levels of the screened channel lie above $\sqrt{\lambda_+}\simeq m_S$. Below them the first-order shift of each level is therefore the diagonal matrix element \eqref{eq:shift}, with $V_{1a}-\Om_0^2V_{1b}$ replaced by $\lambda^-_{1a}-\Om_0^2\lambda^-_{1b}$ and the remap term $2\rho\,\iota\,\lambda_-\sec^2x$, and it is real. Table~\ref{tab:lightshifts} lists it for $l=2,3,4$ and $n=0,1,2$, with $s_-$, $\lambda^-_{1a}$ and $\lambda^-_{1b}$ in the limit $m_S^2\to\infty$. All these shifts are negative, they grow with $n$ as at $m_S^2=0$, and on these levels $|\varsigma|<3.5\,\Om_0^2$. The remap term supplies between $4$ and $27$ per cent of each shift.
\begin{table}[htbp]
\caption{First-order shifts $\varsigma$ of the levels $\Om_0=n+s_-$ of the light massive channel in the limit $m_S^2\to\infty$, with the remap term included. The frequencies refer to the clock of Sec.~\ref{sec:sector}.}
\label{tab:lightshifts}
\centering
\begin{tabular}{cccccc}
\hline\hline
$l$ & $s_-$ & $n$ & $\Om_0$ & $\varsigma$ & $\varsigma/\Om_0^2$\\
\hline
2 & 3.372 & 0 & 3.372 & $-25.32$ & $-2.227$\\
2 & 3.372 & 1 & 4.372 & $-55.68$ & $-2.913$\\
2 & 3.372 & 2 & 5.372 & $-98.58$ & $-3.415$\\
3 & 4.275 & 0 & 4.275 & $-35.75$ & $-1.956$\\
3 & 4.275 & 1 & 5.275 & $-70.26$ & $-2.525$\\
3 & 4.275 & 2 & 6.275 & $-116.6$ & $-2.963$\\
4 & 5.217 & 0 & 5.217 & $-48.96$ & $-1.799$\\
4 & 5.217 & 1 & 6.217 & $-88.18$ & $-2.282$\\
4 & 5.217 & 2 & 7.217 & $-138.9$ & $-2.667$\\
\hline\hline
\end{tabular}
\end{table}

The drift lift leaves its imprint on this channel. The response-free closure (e) of Sec.~\ref{sec:controls} differs from the drift lift by \eqref{eq:driftdiff} and by the changes of the $D\Xi$ and $\Xi D$ entries quoted after it. Its projection has $\lambda^{-({\rm e})}_{1b}=0$ and, in the limit $m_S^2\to\infty$,
\begin{equation}
\lambda^-_{1a}-\lambda^{-({\rm e})}_{1a}=\frac{2\big[(2\lhat+4)\rho^4+(\lhat+5)\rho^2-\lhat-5\big]}{3(\lhat+2)f} .
\label{eq:lam1minusdrift}
\end{equation}
On the levels of Table~\ref{tab:lightshifts} the drift lift raises $\varsigma$ by between $3.9$ and $9.7$ relative to closure (e), by $4.41$ for the lowest level of $l=2$, about a sixth of its shift. This is the observable of the drift at the physical $m_S^2$.

The coupling of Sec.~\ref{sec:kk} moves this channel further. At large $m_S^2$ the small nonzero root of the characteristic polynomial of $\mathcal N+\delta\mathcal N$ is
\begin{equation}
\lambda_-\;\longrightarrow\;\lhat+2+\frac{2\lhat\,(2q-1)}{q^2+\lhat}=\lhat+2+\frac{4\lhat}{q}+O(q^{-2}),
\end{equation}
which moves $s_-$ by $4\lhat/(q\sqrt{4\lhat+9})$. The term $-4\lhat/m_S^2$ of \eqref{eq:lamseries} moves it by $-4\lhat/(m_S^2\sqrt{4\lhat+9})$ relative to Table~\ref{tab:lightshifts}. With $1/q=4\pi^2\alpha/g_c^2$ and $1/m_S^2=16\pi^4\alpha/g_c^4$ these are $165\,\alpha/g_c^2$ and $-6.51\times10^3\,\alpha/g_c^4$ for $l=2$, against the $O(\alpha)$ shift $\alpha\varsigma/(2\Om_0)=-3.75\,\alpha$ of the lowest level. On the levels of Table~\ref{tab:lightshifts} at $g_c\sim1$ the finite value of $m_S^2$ is therefore the largest of the corrections of order $\alpha$ and $1/q$ computed here. The finite value of $m_S^2$ and the coupling move every level by the same amount, and the $O(\alpha)$ shift grows with $\Om_0$. It exceeds the effect of $h_{a\phi}$ by the factor $4\pi^2/g_c^2$ and has the opposite sign, and together these corrections lower the levels below the values of the limit $m_S^2\to\infty$. The one-loop action of the higher Landau levels moves $s_-$ by $\eta_{\rm H}\lhat/\sqrt{4\lhat+9}$, larger than the effect of $h_{a\phi}$ by the factor $\eta_{\rm H}q/4=g_c^2q/48\pi^2$ (Sec.~\ref{sec:hll-throat}). Virtual transitions to the higher Landau levels, which we have not estimated, could enter at the same order $1/q$ (Sec.~\ref{sec:kk}). All four computed corrections are real, and at first order the light massive channel has no growing mode on any level below the screened channel.

\section{The mouth to order \texorpdfstring{$\alpha$}{alpha}}
\label{sec:mouth}

The mouth carries three layers. At $\alpha^0$ and $m_S^2=0$ it is the Einstein--Maxwell mouth on the magnetic Reissner-Nordstrom geometry of Sec.~\ref{sec:sector}. The Schwinger response of Sec.~\ref{sec:mouthS} adds the third channel at every $m_S^2$. At $O(\alpha)$ the anomaly part of the LLL stress tensor corrects the background and adds the drift lift and the polarisation currents of Sec.~\ref{sec:lll}. This section treats all three together and obtains the first two as the $\alpha^0$ part.

\subsection{Source, background and response at order \texorpdfstring{$\alpha$}{alpha}}
\label{sec:mouthalphasource}

The LLL stress tensor does not vanish in the mouth at this order \cite{Sadhukhan:2026czl}. Its Casimir part is smaller than its anomaly part by $(r_e/\ell)^2=\pi^2\alpha^2/16$. The anomaly part is local. It is the anomaly stress tensor of $\hat g_{ab}=g_{ab}/r^2$ with the static frame $(t,r_*)$ as reference, and $\hat g_{ab}$ is not flat in the mouth. On the extremal mouth it gives
\begin{equation}
\kappa\bar T^{Q\,t}{}_t=-\frac{\alpha r_e^2\,(r^2-8rr_e+8r_e^2)}{3r^6},\qquad
\kappa\bar T^{Q\,r}{}_r=-\frac{\alpha r_e^2\,(r-2r_e)^2}{3r^6},\qquad
\kappa\mathcal P=\kappa\bar T^{Q\,\theta}{}_\theta=\frac{\alpha r_e^2\,(r^2-6rr_e+6r_e^2)}{3r^6},
\label{eq:TQmouthax}
\end{equation}
of the same order as in the throat. This source corrects the mouth background to
\begin{equation}
\dd s^2=-e^{2\alpha\Phi_1}N\,\dd t^2+\frac{\dd r^2}{N}+r^2\mathring g,\qquad
N=1-\frac{2GM}{r}+\frac{r_e^2}{r^2}+\alpha N_1 ,
\label{eq:mouthbgax}
\end{equation}
with
\begin{equation}
N_1=\frac{r_e^2}{3r^2}-\frac{4r_e^3}{3r^3}+\frac{8r_e^4}{9r^4},\qquad
\Phi_1=-\frac23\Big[\log\Big(1-\frac{r_e}{r}\Big)+\frac{r_e}{r}+\frac{r_e^2}{2r^2}\Big],\qquad
GM=r_e\Big(1-\frac{\alpha}{18}\Big)+\delta_M .
\label{eq:mouthbgfun}
\end{equation}
These are the mouth counterparts of $\gamma$ and $\varphi$ of \eqref{eq:gamphi}, and the companion paper derives them. At this order the factor $e^{2\alpha\Phi_1}$ enters only through $1+2\alpha\Phi_1$. The drift lift \eqref{eq:driftlift} and the currents \eqref{eq:Jtheta}--\eqref{eq:Jpol} need only $\bar T^Q$ and $\mathcal P$. On the mouth the kernel of the drift lift is
\begin{equation}
\bar T_a{}^c-\mathcal P\,\delta_a^c=-\frac{q}{48\pi^2r^6}\,\mathrm{diag}\big(\mathfrak w,\;r^2-5rr_e+5r_e^2\big),\qquad \mathfrak w\equiv r^2-7rr_e+7r_e^2 .
\label{eq:Kmouth}
\end{equation}
At $r=r_e$ both entries are $-q/48\pi^2r_e^4$, the large-$\rho$ value of the throat kernel of Sec.~\ref{sec:drift}. The lift is covariant on the mouth. On the pure-gauge data of an axial diffeomorphism it gives $u_c=0$, $\delta T^Q_{a\phi}=\mathcal P\,h_{a\phi}$ and no current. The force balance and charge conservation fix $J^\theta$ and $J^a_{\rm pol}$ through \eqref{eq:Jtheta} and \eqref{eq:Jpol}, as in the throat. The Schwinger response keeps the form \eqref{eq:JS} and \eqref{eq:chieq} with $S=e^{\alpha\Phi_1}N$. We evaluate the stress-tensor sources on the $\alpha^0$ mouth, since $\kappa\bar T^Q$ is already $O(\alpha)$, and we set $GM=r_e$ in the $O(\alpha)$ terms. The shift of $GM$ enters through the $\alpha^0$ system, which is exact in $GM$.

With these sources Step~4 returns $\text{Res}\,\mathcal E_{t\phi}=\text{Res}\,\mathcal M^\theta=0$ at $O(\alpha^0)$ and $O(\alpha)$. The controls of Sec.~\ref{sec:controls} behave as in the throat with one exception. Closure (a) fails and closure (e) passes. Closure (c) passes in the throat at this order but fails in the mouth. With $\delta T^Q_{\mu\nu}=0$ the force balance is violated by
\begin{equation}
\delta\big(\nabla_\mu T_Q^{\mu\phi}\big)=-\mathcal P\,\bar g^{\phi\phi}\,\nabla^ah_{a\phi}+\frac{q\,(r-r_e)^2\,(r^2-9rr_e+12r_e^2)}{48\pi^2r^{11}\sin\theta}\,h_1\,Y' ,
\label{eq:ctrlcmouth}
\end{equation}
with $\nabla^a$ the covariant divergence on the $(t,r)$ metric. The first term is the throat result \eqref{eq:ctrlc}. It is proportional to the $\theta\phi$ Einstein equation and vanishes on shell up to $O(\alpha)$. The second term is new. It does not vanish identically and leaves the residual
\begin{equation}
\text{Res}\,\mathcal E_{t\phi}=\frac{2\alpha\,r_e^2\,(r-r_e)^4\,(r^2-9rr_e+12r_e^2)}{3p\,r^{11}}\,h_1 .
\end{equation}
In the mouth the Bianchi identity rejects a missing response at $O(\alpha)$.

\subsection{Reduction and master equations}
\label{sec:mouthmaster}

At $\alpha^0$ and $m_S^2=0$ the mouth is pure Einstein--Maxwell. On the mouth background $A=1/B=\Delta/r^2$, $R=r$, with all parameters kept general, the $\epsilon^0$ Einstein and Maxwell equations vanish identically. The parity-forbidden components of \eqref{eq:angular} vanish identically as well. The six radial equations are displayed in Appendix~\ref{app:mouth}. $\mathcal E_{\theta\phi}$ is algebraic in $h_0$. After $w_0'\to p\,w_1-E$ the two Maxwell equations $\mathcal M^r$ and $\mathcal M^t$ are algebraic in $w_1$ and $w_0$.

Steps 1--3 give the eliminated amplitudes
\begin{align}
h_0&=\frac{2\,(GMr-r_e^2)\,\Delta}{p\,r^5}\,h_1+\frac{\Delta^2}{p\,r^4}\,h_1' ,
\label{eq:mouthh0}\\
w_1&=-\frac{2}{r^2}\,h_1-\frac{p\,r^4}{\lhat\,\Delta}\,E ,\\
w_0&=-\frac{4\,(GMr-r_e^2)\,\Delta}{p\,r^7}\,h_1-\frac{2\Delta^2}{p\,r^6}\,h_1'
-\frac{2\Delta}{\lhat\,r}\,E-\frac{\Delta}{\lhat}\,E' ,
\label{eq:mouthw0}
\end{align}
and the two second-order equations
\begin{align}
h_1''&=\Big\{\frac{1}{r^2\Delta^2}\Big[\,p^2r^6+(\lhat-2)r^4-2GM(\lhat-6)r^3-20G^2M^2r^2
+(\lhat-12)r^2r_e^2+36GM r r_e^2-14r_e^4\Big]\Big\}\,h_1\nonumber\\
&\quad+\frac{2\,(r^2-5GMr+4r_e^2)}{r\,\Delta}\,h_1'
-\frac{2p\,r^4r_e^2}{\lhat\,\Delta^2}\,E ,
\label{eq:mouthh1pp}\\
E''&=-\frac{2\,l(l-1)(l+1)(l+2)}{p\,r^6}\,h_1
-\frac{2\,(2r^2-3GMr+r_e^2)}{r\,\Delta}\,E'\nonumber\\
&\quad+\Big\{\frac{1}{r^2\Delta^2}\Big[\,p^2r^6+(\lhat-2)r^4-2GM(\lhat-2)r^3
+(\lhat+4)r^2r_e^2-12GM r r_e^2+6r_e^4\Big]\Big\}\,E .
\label{eq:mouthEpp}
\end{align}
Equation \eqref{eq:mouthh0} is the statement that the $\theta\phi$ Einstein equation is $\nabla^ah_{a\phi}=0$ on the two-dimensional $(t,r)$ metric, $p\,h_0=A\,(A h_1)'$. Step 4 returns vanishing residuals for both $\mathcal E_{t\phi}$ and $\mathcal M^\theta$, and the vacuum truncation is consistent for the Einstein--Maxwell part. The closed first-order matrix $\mathsf M$ of Step~5 is given in Appendix~\ref{app:mouth}.

With $\Psi_g=A\,h_1/r$, $\Psi_e=\frac{p\,r_e}{\lhat}\,r^2E$ and $\dd r_*/\dd r=1/A$, the first-derivative matrix vanishes identically, $\tilde P\equiv0$, and
\begin{equation}
\partial_{r_*}^2\begin{pmatrix}\Psi_g\\[1pt]\Psi_e\end{pmatrix}
+\big(\omega^2-V\big)\begin{pmatrix}\Psi_g\\[1pt]\Psi_e\end{pmatrix}=0,
\qquad
V=\frac{\Delta}{r^5}\Big[\Big(\lhat\,r+\frac{4r_e^2}{r}\Big)I_2-\mathsf U\Big],
\qquad
\mathsf U=\begin{pmatrix}6GM&2r_e\\ 2r_e\mu^2&0\end{pmatrix},
\label{eq:Vmouth}
\end{equation}
with $\mu^2\equiv(l-1)(l+2)$ and $I_2$ the $2\times2$ identity matrix. Entry by entry,
\begin{equation}
V=\frac{\Delta}{r^6}
\begin{pmatrix}
\lhat r^2-6GMr+4r_e^2 & -2r_er\\[2pt]
-2r_e\mu^2 r & \lhat r^2+4r_e^2
\end{pmatrix}.
\end{equation}
Unlike the $(K,w)$ system of the companion sector, with $K$ and $w$ the polar metric and gauge amplitudes of \cite{Sadhukhan:2026czl}, the potential is frequency independent and no first-derivative matrix survives. $(\Psi_g,\Psi_e)$ are already good master variables with no Zerilli--Moncrief-type recombination required.

With the Schwinger response the amplitudes are \eqref{eq:wD}. In the mouth $S=\Delta/r^2$, and solving \eqref{eq:Xeq} for the highest derivative gives the third second-order equation of the reduction,
\begin{equation}
X''=\frac{r^2}{\Delta}\,E+\frac{p^2r^4}{\Delta^2}\,X-\frac{2\,(GMr-r_e^2)}{r\,\Delta}\,X' .
\label{eq:Xmouth}
\end{equation}
Multiplied by $r\Delta^2$ this is the $\chi$ equation of Appendix~\ref{app:mouth}. Steps~1--3 of Sec.~\ref{sec:consistency} now run with these amplitudes. The $r\phi$ Einstein equation and the definition $E=p\,w_1-w_0'$ give the two remaining second-order equations. With the Schwinger terms they read
\begin{align}
h_1''&=\big[\text{right-hand side of }\eqref{eq:mouthh1pp}\big]+\frac{2\,m_S^2\,p\,r^2r_e^2}{\lhat\,\Delta^2}\,X ,
\nonumber\\
E''&=\big[\text{right-hand side of }\eqref{eq:mouthEpp}\big]+\frac{m_S^2}{\Delta}\,E-\frac{4\,m_S^2\,r_e^2}{r^4\,\Delta}\,X .
\label{eq:extended}
\end{align}
Together with \eqref{eq:Xmouth} they form a closed second-order system for $\bm Y=(h_1,E,X)^{\!\top}$. In the $h_1$ equation the new term combines with the old one into $-2p\,r^2r_e^2\,\mathcal F_D/(\lhat\Delta^2)$, in accordance with \eqref{eq:wD}. The $E$ equation acquires the Schwinger mass term $m_S^2/\Delta$, which in the tortoise coordinate becomes $m_S^2\Delta/r^4$. Step~4 returns $\mathcal E_{t\phi}=\mathcal M^\theta=0$ identically with these equations. The $\mathcal M^t$ and $\mathcal M^r$ equations are also satisfied identically once $w_0$, $w_1$ and the three second-order equations are inserted.

The $\Xi$ row of the mouth system is \eqref{eq:Xirow} with $\sigma=\Delta/r^4$. The other two rows follow from \eqref{eq:extended} in the same way. In the basis $(\Psi_g,\Psi_e,\Xi)$ the mouth system reads
\begin{align}
\partial_{r_*}^2\Psi_g+\omega^2\Psi_g&=\frac{\Delta}{r^5}\Big[\Big(\lhat\,r-6GM+\frac{4r_e^2}{r}\Big)\Psi_g-2r_e\,\Psi_e+2r_e\,m_S^2\,\Xi\Big],
\nonumber\\
\partial_{r_*}^2\Psi_e+\omega^2\Psi_e&=\frac{\Delta}{r^5}\Big[-2r_e\mu^2\,\Psi_g+\Big((\lhat+m_S^2)\,r+\frac{4r_e^2}{r}\Big)\Psi_e-\frac{4r_e^2\,m_S^2}{r}\,\Xi\Big],
\nonumber\\
\partial_{r_*}^2\Xi+\omega^2\Xi&=\frac{\Delta}{r^5}\,r\,\Psi_e .
\label{eq:gexi}
\end{align}
The bosonised charge $\Xi$ is a massless two-dimensional field sourced by the electric flux. The electric field carries the Schwinger mass$^2$ $m_S^2\Delta/r^4$ and is screened. Passing to $\Psi_D=\Psi_e-m_S^2\Xi$ removes the $\Xi$ entry of the first row. In the second row the terms $-\frac{4r_e^2m_S^2}{r}\Xi$ and $\frac{4r_e^2}{r}m_S^2\Xi$ cancel, and subtracting $m_S^2$ times the third row leaves the mass $\lhat\,r+4r_e^2/r$ and the coupling $\lhat\,m_S^2\,r$. The result is the matrix $\hat V$ of \eqref{eq:VmouthS}. In this form the Schwinger mass sits on the diagonal only in the $\Xi$ row.

In the master variables \eqref{eq:mastervars} the first-derivative matrix again vanishes identically, and the system takes the Schr\"odinger form
\begin{equation}
\partial_{r_*}^2 \bm Z+\big(\omega^2-V\big)\bm Z=0,\qquad \bm Z=(\Psi_g,\Psi_D,\Xi)^{\!\top},\qquad
V=\frac{\Delta}{r^5}\,\hat V,
\end{equation}
\begin{equation}
\hat V=
\begin{pmatrix}
\lhat\,r-6GM+\dfrac{4r_e^2}{r} & -2r_e & 0\\[6pt]
-2r_e\,\mu^2 & \lhat\,r+\dfrac{4r_e^2}{r} & \lhat\,m_S^2\,r\\[6pt]
0 & r & m_S^2\,r
\end{pmatrix}.
\label{eq:VmouthS}
\end{equation}
Gravity couples to the displacement $\Psi_D=\Psi_e-m_S^2\Xi$, the $(\Psi_g,\Psi_D)$ block is exactly the Einstein--Maxwell matrix \eqref{eq:Vmouth}, and the fermions enter only through the $(\Psi_D,\Xi)$ block. The coupling of Sec.~\ref{sec:kk} adds $(\Delta/r^5)\,\delta\hat V$ to $V$, with
\begin{equation}
\delta\hat V=\frac1q\begin{pmatrix}
0 & 0 & \lhat\,m_S^2\,r_e\\[2pt]
0 & 0 & -\dfrac{2\lhat\,m_S^2r_e^2}{r}\\[6pt]
r_e\,\mu^2 & -\dfrac{2r_e^2}{r} & \dfrac{\lhat\,m_S^2r_e^2}{q\,r}
\end{pmatrix}.
\label{eq:dVmouth}
\end{equation}
It couples the gravitational channel to $\Xi$ and adds entries of order $1/q$ to the $\Xi$ row. At $r=r_e$ it equals $r_e\,\delta\mathcal N$ of \eqref{eq:dN}. Without the coupling the statements above hold as written. With it, gravity also couples to $\Xi$ through $\delta\hat V$.

At $O(\alpha)$ the sources of Sec.~\ref{sec:mouthalphasource} enter. With the master variables \eqref{eq:mastervars}, $S=e^{\alpha\Phi_1}N$, $\sqrt{AB}=e^{\alpha\Phi_1}$, $R=r$ and $\dd r_*/\dd r=e^{-\alpha\Phi_1}/N$, the mouth system to first order in $\alpha$ is
\begin{equation}
\bm Z_{r_*r_*}+\tilde P\,\bm Z_{r_*}+\tilde Q\,\bm Z=0,\qquad
\tilde P=\alpha\,\mathrm{diag}\Big(0,\;-\frac{7r_e\,(r-2r_e)(r-r_e)^2}{3r^5},\;0\Big),\qquad
\tilde Q=-p^2I_3-\frac{A}{r^3}\,\hat V-\alpha\,\mathcal T ,
\label{eq:mouthalphamaster}
\end{equation}
with $A=e^{2\alpha\Phi_1}N$, $\hat V$ the matrix \eqref{eq:VmouthS} and
\begin{equation}
\mathcal T=\frac{r-r_e}{3r^8}
\begin{pmatrix}
2r_e^2(r-r_e)^2(2r-7r_e) & 0 & 0\\[4pt]
-2r_e\mu^2\,r\,(r-r_e)\,\mathfrak w & (\lhat r^2+4r_e^2)(r-r_e)\,\mathfrak w-2p^2r^6r_e & \lhat\,m_S^2\,r^2(r-r_e)\,\mathfrak w\\[4pt]
0 & 0 & 0
\end{pmatrix}.
\label{eq:Tmouth}
\end{equation}
With the sign in \eqref{eq:mouthalphamaster}, $\alpha\mathcal T$ adds to the potential. The terms in $A$ are the $\alpha^0$ mouth on the corrected background. The fermion terms sit in the gravitational and displacement rows. Apart from the coupling of Sec.~\ref{sec:kk}, the $\Xi$ row has none, and its whole $O(\alpha)$ content is the correction of $\sigma=A/r^2$, as in the throat. Three entries of $\mathcal T$ carry the factor $\mathfrak w$ of the drift kernel \eqref{eq:Kmouth}. The first-derivative term and the $p^2$ term of $\mathcal T_{DD}$ are the mouth form of the friction term $\tilde P_{1,DD}$ of the throat. All fermion terms vanish at $r=r_e$. As in the companion sector \cite{Sadhukhan:2026czl}, the largest $O(\alpha)$ term near $r_e$ comes from the move of the minimum of $N$ from $r_e$ to $r_h=r_e(1+\alpha/6)$, where $N$ and $N'$ vanish to $O(\alpha)$. It is a fraction $\alpha r_e/[3(r-r_e)]$ of the $\alpha^0$ terms, $8\times10^{-30}$ at the junction of Sec.~\ref{sec:matching}. Further out the $O(\alpha)$ terms are smaller. Near $r\simeq1.4r_e$, where the negative potential $V_{\min}$ of Sec.~\ref{sec:mouthneutral} is deepest, the largest relative entry is $8\times10^{-36}$, both at $m_S^2=0$ and at $m_S^2=\alpha q^2$.

\subsection{Chandrasekhar's axial potentials}
\label{sec:chandra}
At $\alpha^0$ and $m_S^2=0$ the system \eqref{eq:Vmouth} has a decoupled form. The characteristic polynomial of $\mathsf U$ is $\lambda^2-6GM\lambda-4r_e^2\mu^2$, whose roots are Chandrasekhar's
\begin{equation}
\beta_{1,2}=3GM\pm\sqrt{9G^2M^2+4r_e^2\mu^2}\,,
\qquad \beta_1\beta_2=-4r_e^2\mu^2,\qquad \beta_1+\beta_2=6GM .
\end{equation}
The combinations $Z_i=\beta_i\,\Psi_g+2r_e\,\Psi_e$ decouple exactly. The off-diagonal entries vanish identically, and
\begin{equation}
V[Z_i]=\frac{\Delta}{r^5}\Big[(\mu^2+2)\,r-\beta_i+\frac{4r_e^2}{r}\Big]
\;=\;V_i^{(-)},\qquad i=1,2,
\label{eq:Vminus}
\end{equation}
which are Chandrasekhar's decoupled axial Reissner-Nordstrom potentials \cite{Chandrasekhar:1979iz,Chandra} with his charge $Q_*\to r_e$. Chandrasekhar builds his $i$-th potential from the other root, our $\beta_j$ with $j\neq i$. Our $V^{(-)}_1$ and $Z_1$ are therefore his $V^{(-)}_2$ and $Z_2$, with the labelling of \cite{Sadhukhan:2026czl}. In \cite{Sadhukhan:2026czl} duality \cite{Pereniguez:2023wxf} identified the polar potentials $V^{(+)}_i$ for the mouth of the opposite sector, and the decoupled variables were then constructed directly. Here they are derived directly on the magnetic background with a polar gauge field. The duality between the two magnetic sectors and the two electric parities is an output of the calculation rather than an input. In the Schwarzschild limit $r_e\to0$ the mixing vanishes and
\begin{equation}
V_{gg}\to\Big(1-\frac{2GM}{r}\Big)\Big(\frac{\lhat}{r^2}-\frac{6GM}{r^3}\Big),
\qquad
V_{ee}\to\Big(1-\frac{2GM}{r}\Big)\frac{\lhat}{r^2},
\end{equation}
the Regge--Wheeler potential and the flat electromagnetic potential.

\subsection{The neutral channel}
\label{sec:mouthneutral}
The matrix $\hat V$ of \eqref{eq:VmouthS} is not symmetric, but a constant similarity transformation makes it so. With
\begin{equation}
\Sigma=\mathrm{diag}\big(\mu,\;1,\;\sqrt{\lhat\,m_S^2}\,\big),
\qquad
\Sigma\hat V\Sigma^{-1}=
\begin{pmatrix}
\lhat\,r-6GM+\dfrac{4r_e^2}{r} & -2r_e\mu & 0\\[6pt]
-2r_e\mu & \lhat\,r+\dfrac{4r_e^2}{r} & \sqrt{\lhat m_S^2}\,r\\[6pt]
0 & \sqrt{\lhat m_S^2}\,r & m_S^2\,r
\end{pmatrix},
\label{eq:Ssym}
\end{equation}
and the same $\Sigma$ symmetrises the throat mass matrix of Sec.~\ref{sec:alpha0} and the coupling terms $\delta\hat V$ and $\delta\mathcal N$. The rescaled variables $(\mu\Psi_g,\Psi_D,\sqrt{\lhat m_S^2}\,\Xi)$ obey a self-adjoint Schr\"odinger system. This is the form in which the energy criterion of Sec.~\ref{sec:stability} applies. The determinant is
\begin{equation}
\det\hat V=\frac{8\,m_S^2\,r_e^2}{r}\,\big(r^2-3GMr+2r_e^2\big),
\label{eq:detV}
\end{equation}
which is negative between the two roots $r_{1,2}=\frac12\big(3GM\mp\sqrt{9G^2M^2-8r_e^2}\big)$ of the quadratic factor. With the coupling $\det(\hat V+\delta\hat V)=(1+\lhat/2q)^2\det\hat V$, with the same sign. These roots are not horizons, since the mouth has none (Sec.~\ref{sec:sector}). In this subsection we work at $\alpha^0$ and set $\delta_M=0$, where $\Delta=(r-r_e)^2$ has a double zero and $r_{1,2}=r_e,2r_e$. The matrix $\hat V$ is similar to the symmetric matrix \eqref{eq:Ssym}, its eigenvalues are real, and its trace is positive. A negative determinant therefore means exactly one negative eigenvalue, between $r_e$ and $2r_e$ for every $m_S^2>0$. It vanishes at $r=r_e$, where $\hat V\to r_e\mathcal N$ with $\det\mathcal N=0$. In the wormhole $r=r_e$ is not a horizon. It is the inner edge of the mouth region, where the mouth hands over to the throat through the overlap region of Sec.~\ref{sec:matching}. This eigenvalue belongs to the neutral channel, the continuation of the exactly massless $E=0$ mode of the throat, and it does not freeze at large $m_S^2$. For $l=2$ and $m_S^2\to\infty$ the corresponding potential $V_{\min}=\lambda_{\min}\Delta/r^5$, with $\lambda_{\min}$ the smallest eigenvalue of $\hat V$, reaches its minimum $-3.6\times10^{-3}/r_e^2$ near $r=1.4r_e$ and turns positive beyond $2r_e$. This is a negative eigenvalue of a matrix whose eigenvector rotates with $r$, and it does not by itself make the channel attractive. Section~\ref{sec:neutralstab} shows that the coupled channel has an exact zero-energy solution.

The light content of the sector at large Schwinger mass can be isolated in closed form. We write the mouth system in the basis $(\Psi_g,\Psi_e,\tilde\Xi)$, with $\Psi_e=\Psi_D+m_S^2\Xi$ the electric flux along the field lines and $\tilde\Xi=m_S^2\Xi$. In this basis
\begin{equation}
\hat V\big|_{(g,e,\tilde\Xi)}=
\begin{pmatrix}
\lhat\,r-6GM+\dfrac{4r_e^2}{r} & -2r_e & 2r_e\\[6pt]
-2r_e\mu^2 & (\lhat+m_S^2)\,r+\dfrac{4r_e^2}{r} &\,\, -\dfrac{4r_e^2}{r}\\[6pt]
0 & m_S^2\,r & 0
\end{pmatrix}.
\end{equation}
We call the $\Psi_e$ channel the electric channel. Its diagonal potential contains the Schwinger term $m_S^2r$, which screens the electric field along the lines. We call the $\tilde\Xi$ channel the neutral channel. $\tilde\Xi$ is $m_S^2$ times the bosonised scalar of the field lines, and the light mode of this channel is the massless mode of Sec.~\ref{sec:alpha0}, which has no net charge and in which the longitudinal field $E$ vanishes. At large $m_S^2$ the electric channel is much heavier than the other two. For frequencies and inverse lengths small compared with $m_S\sqrt\Delta/r^2$ its derivative terms are negligible, and its equation becomes algebraic. $\Psi_e$ is then determined at every point by the light fields,
\begin{equation}
\Psi_e=\frac{2r_e\mu^2\,\Psi_g+(4r_e^2/r)\,\tilde\Xi}{(\lhat+m_S^2)\,r+4r_e^2/r}\simeq\frac{1}{m_S^2r}\Big(2r_e\mu^2\,\Psi_g+\frac{4r_e^2}{r}\,\tilde\Xi\Big)\to0 .
\label{eq:Psiealg}
\end{equation}
Substituting \eqref{eq:Psiealg} into the other two equations removes $\Psi_e$. For a potential matrix split into light blocks $L$ and heavy blocks $H$, this leaves the Schur complement $V_{LL}-V_{LH}V_{HH}^{-1}V_{HL}$ for the light fields, here with $L=\{\Psi_g,\tilde\Xi\}$ and $H=\{\Psi_e\}$. In the $\Psi_g$ row the term $-2r_e\Psi_e$ is $O(1/m_S^2)$ and drops out. In the $\tilde\Xi$ row the source $m_S^2r\,\Psi_e$ stays finite, since the factor $m_S^2r$ cancels the suppression in \eqref{eq:Psiealg}. At $m_S^2\to\infty$ this leaves the two-channel system
\begin{equation}
\partial_{r_*}^2\begin{pmatrix}\Psi_g\\ \tilde\Xi\end{pmatrix}+\big(\omega^2-V_{\rm light}\big)\begin{pmatrix}\Psi_g\\ \tilde\Xi\end{pmatrix}=0,
\qquad
V_{\rm light}=\frac{\Delta}{r^5}
\begin{pmatrix}
\lhat\,r-6GM+\dfrac{4r_e^2}{r} & 2r_e\\[6pt]
2r_e\mu^2 & \dfrac{4r_e^2}{r}
\end{pmatrix},
\label{eq:Vlight}
\end{equation}
with $\det(r^5V_{\rm light}/\Delta)=8r_e^2(r^2-3GMr+2r_e^2)/r^2$, consistent with \eqref{eq:detV}. Compared with the Einstein--Maxwell mouth, the electric channel with mass $\lhat r+4r_e^2/r$ is replaced by the neutral channel with mass $4r_e^2/r$, and the mixing changes sign. The neutral mode costs almost no energy along the lines. It couples to gravity with the full strength $2r_e\mu^2$ of the magnetic background, which is why the gravitational coupling can drive its potential negative. The constant matrix $\mathrm{diag}(\mu,1)$ symmetrises $V_{\rm light}$. The coupling of Sec.~\ref{sec:kk} leaves \eqref{eq:Vlight} unchanged in the limit $m_S^2\to\infty$ at fixed $m_S^2/q$. At extremality, $GM=r_e$, we write $r=r_e+y$, and $\Delta=y^2$. Near $r=r_e$,
\begin{equation}
\frac{r^5}{\Delta}V_{\rm light}=r_e\begin{pmatrix}\mu^2&2\\ 2\mu^2&4\end{pmatrix}
+y\begin{pmatrix}\lhat-4&0\\0&-4\end{pmatrix}+O(y^2),
\label{eq:VlightNH}
\end{equation}
with eigenvalues $0$ and $r_e(l^2+l+2)$ at $y=0$, which are the $m_S^2\to\infty$ masses of the throat. The zero eigenvalue moves at first order in $y$. Its shift is the projection of the $O(y)$ matrix with the left and right null vectors $(2,-1)$ and $(2,-\mu^2)^{\!\top}$ of the constant matrix, which gives $-8y/(\lhat+2)$. Multiplying by $\Delta/r^5\simeq y^2/r_e^5$ gives the neutral-channel potential $-8y^3/\big((\lhat+2)\,r_e^5\big)$ at $O(y^3)$.

\subsection{Matching to the throat}
\label{sec:matching}

The near-horizon limit of the mouth is the first test. At extremality, $GM=r_e$, $r=r_e+y$, the Einstein--Maxwell potential matrix \eqref{eq:Vmouth} collapses to
\begin{equation}
V\;\longrightarrow\;\frac{y^2}{r_e^4}\,\mathcal N_0,
\qquad
\mathcal N_0=\begin{pmatrix}\mu^2&-2\\ -2\mu^2&\lhat+4\end{pmatrix},
\qquad
\operatorname{eig}\mathcal N_0=\{\,l(l-1),\;(l+1)(l+2)\,\},
\label{eq:NHmouth}
\end{equation}
which under the MMP matching $\rho=\ell y/r_e^2$, $\Om=\omega\ell$ (so that $\ell^2y^2/r_e^4=\rho^2\simeq f$ at $\rho\gg1$) is exactly the throat system of Sec.~\ref{sec:alpha0} at $\alpha=0$ and $m_S^2=0$.

Both regions are written in the form \eqref{eq:tortoiseform}, with the tortoise coordinate $r_*$ in the mouth and $x$ in the throat. In the overlap the two coordinates are related as in \cite{Sadhukhan:2026czl},
\begin{equation}
r_*=\ell\Big(x-\frac\pi2\Big),\qquad \dd r_*=\ell\,\dd x,\qquad p_{\rm mouth}=\frac{p}{\ell},\qquad \ell=\frac{4r_e}{\pi\alpha},
\end{equation}
and the two systems are the same equation in the overlap $1\ll\rho\ll1/\alpha$ if and only if
\begin{equation}
\ell\,\tilde P^{\rm mouth}=\tilde P^{\rm throat},\qquad \ell^2\,\tilde Q^{\rm mouth}=\tilde Q^{\rm throat}.
\label{eq:junctioncond}
\end{equation}
We count orders as in \cite{Sadhukhan:2026czl}. A throat term $\alpha^{n_\alpha}\rho^{n_\rho}$ of $\tilde Q$ has the mouth order $n_{\rm m}=n_\alpha-n_\rho+2$, and the terms that both calculations determine are $\rho^2$ and $\alpha\rho^3$ at $n_{\rm m}=0$ and $\rho$ and $\alpha\rho^2$ at $n_{\rm m}=1$.

The terms with $n_{\rm m}=0$ come from the $\alpha^0$ mouth. Insert $GM=r_e$, $r=r_e+\rho\,r_e^2/\ell$ and $p_{\rm mouth}=p/\ell$ into the Einstein--Maxwell potential \eqref{eq:Vmouth} and expand $\ell^2V^{\rm mouth}$ to first order in $\alpha$. Each power of $y/r_e=\pi\alpha\rho/4$ costs one power of $\alpha$ and one of $\rho$, and
\begin{equation}
\ell^2V^{\rm mouth}
=\rho^2\,\mathcal N_0+\alpha\,\rho^3\,\mathcal C+O(\alpha^2),
\qquad
\mathcal C=
\begin{pmatrix}
-\dfrac{\pi}{2}\big(2l^2+2l-3\big) & \dfrac{5\pi}{2}\\[6pt]
\dfrac{5\pi}{2}\,(l-1)(l+2) & -\pi\big(l^2+l+6\big)
\end{pmatrix},
\label{eq:Cmatch}
\end{equation}
in the $(\Psi_g,\Psi_e)$ basis, with $\mathcal N_0$ the matrix \eqref{eq:NHmouth}. The matrix $\mathcal C$ coincides entry by entry with the coefficient of $\rho^3$ in the throat matrix $V_{1a}$ at $m_S^2=0$, and the $\alpha^0$ terms agree with $f\to\rho^2$. The finite-$m_S^2$ channels match through the near-horizon limit of \eqref{eq:VmouthS}.

The terms with $n_{\rm m}=1$ need the mouth of Secs.~\ref{sec:mouthalphasource} and \ref{sec:mouthmaster} and the dictionary to $O(\alpha)$. As in \cite{Sadhukhan:2026czl} we write both metrics in the throat gauge $AB=r_e^4$ with $t=\ell\tau$. This fixes $\dd r/\dd\rho=(r_e^2/\ell)\,e^{-\alpha\Phi_1}$, and we fix the constant by $R=r$. Then
\begin{equation}
\frac{\ell y}{r_e^2}=\rho+\frac{2}{3\pi}+\frac{\alpha\rho}{3}\Big[1+2\log\frac{\pi\alpha\rho}{4}\Big]+\dots .
\label{eq:sigmaax}
\end{equation}
The logarithm cancels against the redshift $e^{2\alpha\Phi_1}$ in every compared coefficient, and the matching of the background requires the mass $GM=r_e(1-\alpha/18)$ of \eqref{eq:mouthbgfun}. The $\rho$ terms of $\tilde Q$ vanish on both sides. The $\alpha\rho^2$ terms are
\begin{equation}
\tilde Q\big|_{\alpha\rho^2}=
\begin{pmatrix}
\tfrac23\mu^2 & -\tfrac53 & 0\\[2pt]
-\mu^2 & \tfrac13(\lhat+8) & \tfrac13\lhat\,m_S^2\\[2pt]
0 & \tfrac23 & \tfrac23m_S^2
\end{pmatrix}
\label{eq:arho2ax}
\end{equation}
in the throat and in the $O(\alpha)$ mouth alike, for all $l$ and $m_S^2$. The first-derivative matrices also agree at these orders, where both vanish. The uncorrected mouth fails this test. On the MMP dictionary every $\alpha\rho^2$ term of \eqref{eq:VmouthS} vanishes, and a constant shift of the dictionary moves the $\rho$ terms away from zero.

The $\alpha\rho^2$ terms carry the drift. The response-free closure (e) of Sec.~\ref{sec:controls} gives a different throat, and in the basis $(\Psi_g,\Psi_D,\Xi)$
\begin{equation}
\tilde Q^{\rm drift}-\tilde Q^{\rm (e)}\Big|_{\alpha\rho^2}=
\begin{pmatrix}
0&0&0\\[2pt]
\tfrac23\mu^2 & -\tfrac13(\lhat+4) & -\tfrac13\lhat\,m_S^2\\[2pt]
0&0&0
\end{pmatrix}.
\label{eq:arho2diff}
\end{equation}
The mouth computed with closure (e) reproduces the throat computed with closure (e), and the mouth computed with the drift lift reproduces the throat computed with the drift lift. The mixed pairs fail. The drift is present in both regions. The matching tests the corrected background, the response and the reduction against an independent calculation, but it cannot select the closure, since both regions carry the same one.

We also compare the full coefficient matrices, with no expansion in $\rho$, at a definite junction and for the fiducial parameters of \cite{Mondal:2025tht}, as was done for the polar sector in \cite{Sadhukhan:2026czl}. On the mouth side we use the exact $\alpha^0$ matrices, $\tilde P^{\rm mouth}=0$ and $\tilde Q^{\rm mouth}=-p_{\rm mouth}^2I_3-V$, with $V=(\Delta/r^5)(\hat V+\delta\hat V)$ the three-channel potential \eqref{eq:VmouthS} with the coupling \eqref{eq:dVmouth} and $GM=r_e+\delta_M$. On the throat side $\tilde P=\alpha\tilde P_1$ and $\tilde Q=-p^2I_3-f(\mathcal N+\delta\mathcal N)+\alpha\tilde Q_1$, with the $O(\alpha/q)$ terms of Appendix~\ref{app:matrices} in $\tilde Q_1$. Both sides are evaluated in 120-digit arithmetic (Appendix~\ref{app:numerics}). We use the fiducial background of \cite{Mondal:2025tht}, $q=10^{33}$, $\ell_p=1.6\times10^{-33}\,{\rm cm}$ for the Planck length and $g_c=1$, for which $r_e=2.8359\,{\rm cm}$, $\ell=1.4255\times10^{35}\,{\rm cm}$, $\alpha=2.5330\times10^{-35}$, $\delta_M=-5.61\times10^{-70}\,{\rm cm}$ and $m_S^2=\alpha q^2=2.5330\times10^{31}$. The junction sits at $r_{\rm J}=r_e(1+\delta_{\rm J})$ with $\delta_{\rm J}=10^{-6}$, that is $\rho_{\rm J}=\ell(r_{\rm J}-r_e)/r_e^2=5.0265\times10^{28}$ and $\alpha\rho_{\rm J}=1.2732\times10^{-6}$, inside the overlap window. The first throat term not computed is $\alpha^2\rho^4$, against the leading $\rho^2$. We therefore expect a relative mismatch of order $(\alpha\rho_{\rm J})^2\sim10^{-12}$ in every entry, and of order $\alpha\rho_{\rm J}\sim10^{-6}$ without the $O(\alpha)$ throat terms. For $l=2$ and $\Om=1$ the mouth and throat values at $\rho_{\rm J}$ agree to every digit shown in the second column of Table~\ref{tab:junctionax}.
\begin{table}[htbp]
\caption{Coefficients $\tilde Q$ of the master system at the fiducial junction $\rho_{\rm J}$ for $l=2$, $\Om=1$ and $m_S^2=\alpha q^2$, in the basis $(\Psi_g,\Psi_D,\Xi)$. The mouth side is multiplied by $\ell^2$. The last two columns give the relative difference between the two sides with and without the $O(\alpha)$ throat terms.}
\label{tab:junctionax}
\centering
\small
\begin{tabular}{llcc}
\hline\hline
 & & \multicolumn{2}{c}{relative difference}\\
\cline{3-4}
entry & \multicolumn{1}{c}{value at $\rho_{\rm J}$} & with $O(\alpha)$ & $\alpha^0$ only\\
\hline
$gg$ & $-1.010642943\times10^{58}$ & $1.35\times10^{-11}$ & $4.5\times10^{-6}$\\
$gD$ & $\phantom{-}5.053212187\times10^{57}$ & $1.50\times10^{-11}$ & $5.0\times10^{-6}$\\
$g\Xi$ & $-3.839980800\times10^{56}$ & $1.50\times10^{-11}$ & $5.0\times10^{-6}$\\
$Dg$ & $\phantom{-}2.021284875\times10^{58}$ & $1.50\times10^{-11}$ & $5.0\times10^{-6}$\\
$DD$ & $-2.526606599\times10^{58}$ & $1.44\times10^{-11}$ & $4.8\times10^{-6}$\\
$D\Xi$ & $-3.839984640\times10^{89}$ & $1.00\times10^{-11}$ & $4.0\times10^{-6}$\\
$\Xi g$ & $-1.010642437\times10^{25}$ & $1.50\times10^{-11}$ & $5.0\times10^{-6}$\\
$\Xi D$ & $-2.526608620\times10^{57}$ & $1.00\times10^{-11}$ & $4.0\times10^{-6}$\\
$\Xi\Xi$ & $-6.399974400\times10^{88}$ & $1.00\times10^{-11}$ & $4.0\times10^{-6}$\\
\hline\hline
\end{tabular}
\end{table}
These are the values at the physical $m_S^2=\alpha q^2$. The entries $g\Xi$ and $\Xi g$ come from the coupling of Sec.~\ref{sec:kk}. Their own $O(\alpha)$ corrections are the $O(\alpha/q)$ terms of Appendix~\ref{app:matrices}, without which they agree only to $O(\alpha\rho_{\rm J})$. At $m_S^2=0$ the $(\Psi_g,\Psi_D)$ block is unchanged and the $\Xi$ channel reduces to $\ell^2\tilde Q_{\Xi\Xi}=-p^2$ on both sides. Over the scan $l=2,3,4$, $\Om\in\{0,1,5\}$, $m_S^2\in\{0,1,\alpha q^2\}$ the largest relative mismatch is $1.5\times10^{-11}$, taken over the entries that do not vanish. This includes the Schwinger entries $D\Xi$ and $\Xi\Xi$, which are thirty-one orders of magnitude larger than the gravitational ones at the physical $m_S^2$. The entries $g\Xi$ and $D\Xi$ at $m_S^2=0$, and $\Xi\Xi$ at $m_S^2=0$ and $\Om=0$, vanish on both sides. The full $O(\alpha)$ mouth \eqref{eq:mouthalphamaster} on the dictionary \eqref{eq:sigmaax} gives the same relative differences to the digits shown, since its $O(\alpha)$ terms are negligible at this junction (Sec.~\ref{sec:mouthmaster}). The only nonzero entry of the throat first-derivative matrix is $\alpha\tilde P_{1,DD}=-\alpha\rho_{\rm J}/f_{\rm J}\simeq-\alpha/\rho_{\rm J}$ with $f_{\rm J}=1+\rho_{\rm J}^2$, of order $10^{-64}$. Against the $\alpha^0$ mouth, which has $\tilde P=0$, the first condition in \eqref{eq:junctioncond} holds to that accuracy. The $O(\alpha)$ mouth \eqref{eq:mouthalphamaster} has $\ell\tilde P^{\rm mouth}_{DD}=\tfrac{7\pi}{12}(\alpha\rho_{\rm J})^2=3.0\times10^{-12}$ at the junction. This is an $\alpha^2\rho^2$ term, beyond the accuracy of the throat, and the agreement in all three channels refers to $\tilde Q$. In the companion sector $\tilde P$ grows as $\alpha\rho^2$ and its relative mismatch is $O(\alpha\rho)$. Setting $GM=r_e$ instead of $GM=r_e+\delta_M$ leaves the mismatch unchanged. The defect enters the mouth through $\Delta=y^2-2\delta_M r$, at relative order $2|\delta_M|/(r_e\delta_{\rm J}^2)=4\times10^{-58}$.

Moving the junction through $\delta_{\rm J}=10^{-9},\dots,10^{-2}$, the relative mismatch divided by $(\alpha\rho_{\rm J})^2$ stays at $8.33$, $8.88$ and $6.17$ in the $gg$, $DD$ and $\Xi\Xi$ entries until $\alpha\rho_{\rm J}\sim10^{-3}$, where the next order sets in. The $\Xi\Xi$ constant follows in closed form. The only source of mismatch in that row is the expansion of $\ell^2m_S^2\Delta/r^4$ in $y/r_e=\pi\alpha\rho/4$. The throat reproduces the term $-4y/r_e$ of $(1+y/r_e)^{-4}$ and misses $10y^2/r_e^2=\tfrac{5\pi^2}{8}(\alpha\rho)^2$, with $\tfrac{5\pi^2}{8}=6.17$. The numerical comparison is dominated by the $\rho^3$ terms. The drift enters one power of $\rho$ lower, and its relative contribution at $\rho_{\rm J}$ is $O(\alpha)$, far below the residual. The expansion \eqref{eq:arho2ax}--\eqref{eq:arho2diff} is what tests it.

\section{Isospectrality of the two parity sectors}
\label{sec:iso}

For the Reissner-Nordstrom black hole the polar and axial sectors are isospectral. Chandrasekhar's potentials $V^{(\pm)}_i$ of the two parities share their reflection and transmission amplitudes and their quasinormal frequencies \cite{Chandrasekhar:1979iz,Chandra}, and the same holds for the magnetic black hole by duality \cite{Pereniguez:2023wxf}. With both sectors of the MMP wormhole in hand, we ask whether the property survives the coupling to the LLL fermions. It does not. The polar-sector inputs are taken from \cite{Sadhukhan:2026czl}, whose $O(\alpha)$ potential matrices we use in the form given there.

\subsection{Reissner-Nordstrom and the mouth}
\label{sec:isoRN}
Write $\Theta_i=\Delta/\big(r^3(\mu^2r+\beta_i)\big)$ with $\beta_{1,2}$ the roots of $\beta^2-6GM\,\beta-4r_e^2\mu^2=0$ from Sec.~\ref{sec:chandra}. Chandrasekhar's potentials are
\begin{equation}
V^{(\pm)}_i=\pm \beta_i\,\partial_{r_*}\Theta_i+\beta_i^2\Theta_i^2+\mu^2(\mu^2+2)\,\Theta_i .
\end{equation}
With the superpotential $\mathsf W_i=\beta_i\Theta_i+\mu^2(\mu^2+2)/2\beta_i$ they take the form
\begin{equation}
V^{(\pm)}_i=\mathsf W_i^2\pm\partial_{r_*}\mathsf W_i-\frac{\mu^4(\mu^2+2)^2}{4\beta_i^2},
\label{eq:susyform}
\end{equation}
which holds identically in $r$, $GM$ and $l$, for both roots when $r_e\neq0$. At $r_e=0$ the root $\beta_2$ vanishes and $\mathsf W_2$ is singular. Two potentials of this form are Darboux partners. With $\mathcal A_i=\partial_{r_*}+\mathsf W_i$ and up to the common constant, the operator $-\partial_{r_*}^2+V^{(-)}_i$ factorises as $\mathcal A_i^\dagger\mathcal A_i$ and $-\partial_{r_*}^2+V^{(+)}_i$ as $\mathcal A_i\mathcal A_i^\dagger$. The operator $\mathcal A_i$ maps solutions of one problem to solutions of the other at the same frequency. The spectra coincide apart from a possible zero mode and the scattering data coincide up to a phase. This is Chandrasekhar's transformation theory \cite{Chandrasekhar:1975nkd,Chandra}, and it does not rest on the Maxwell field. At $r_e=0$ the roots are $\beta_1=6GM$ and $\beta_2=0$. Then $V^{(-)}_1$ and $V^{(+)}_1$ are the Regge--Wheeler and Zerilli potentials, and $V^{(\pm)}_2$ both reduce to the electromagnetic potential $(1-2GM/r)\lhat/r^2$. The duality rotation $F\to\star F$ leaves the metric alone. It relates the electric and magnetic backgrounds, but it cannot map an axial metric perturbation to a polar one. In Hamiltonian language the Darboux transformation is a canonical transformation between master functions \cite{Lenzi:2021wpc,Lenzi:2026aay}. Quantum corrections to the gauge sector break it \cite{delRio:2024kms}. The fermionic backreaction of the MMP wormhole breaks it in a different way, which we now describe region by region.

At the Einstein--Maxwell level the mouth is isospectral. The decoupled axial potentials \eqref{eq:Vminus} are exactly $V^{(-)}_i$, and the decoupled polar potentials of \cite{Sadhukhan:2026czl} are exactly $V^{(+)}_i$. The two sectors of the magnetic mouth are Darboux partners with the superpotential of \eqref{eq:susyform}, and their quasinormal frequencies coincide. Duality pairs Chandrasekhar's polar sector with our polar-metric/axial-gauge sector and his axial sector with our axial-metric/polar-gauge sector.

With the fermions the mouths of the two sectors are different problems. The axial gauge field $a_\phi\propto\sin\theta\,Y'$ of the polar sector has no component along the field lines, and the LLL does not respond through its charge there. Its stress tensor and the Hall current enter the polar mouth only at $O(\alpha)$ \cite{Sadhukhan:2026czl}. The axial mouth has the three-channel potential \eqref{eq:VmouthS}. Its $(\Psi_g,\Psi_D)$ block is the Einstein--Maxwell matrix, but $\Psi_D$ is not $\Psi_e$, and the block is coupled to $\Xi$ with strength $\lhat m_S^2r$. By \eqref{eq:detV} one eigenvalue is negative between $r_e$ and $2r_e$ for every $m_S^2>0$, while both polar potentials $V^{(+)}_i$ are positive there for $l=2,\dots,10$ \cite{Sadhukhan:2026czl}. The count of channels is different as well. At the physical value of $m_S^2$ the light content of the axial mouth is the two-channel system \eqref{eq:Vlight}, whose near-horizon masses are $0$ and $l^2+l+2$ by \eqref{eq:VlightNH}, against $l(l-1)$ and $(l+1)(l+2)$ for the polar mouth. The mouth spectra of the two sectors differ by amounts that do not vanish as $m_S^2$ grows. The combination $\Psi_e=\Psi_D+m_S^2\Xi$ is screened, and the longitudinal electric field, which carries the Darboux partner of the polar mouth, has been removed from the low-frequency spectrum.

\subsection{The throat}
\label{sec:isothroat0}
At $\alpha=0$ and $m_S^2=0$ the two sectors have identical potentials. Both reduce to \eqref{eq:PTpm}, Sec.~\ref{sec:alpha0} here and the $\alpha=0$ throat equations of \cite{Sadhukhan:2026czl}. At extremality the Darboux term $\pm\partial_{r_*}\mathsf W_i$ of \eqref{eq:susyform} is $O(y^3)$ while $\mathsf W_i^2-\mu^4(\mu^2+2)^2/4\beta_i^2$ is $O(y^2)$, which is why the two potentials coincide in the near-horizon limit rather than merely sharing a spectrum. This is the near-horizon image of the mouth isospectrality.

At finite $m_S^2$ the polar throat is unchanged, since there is no Schwinger response in that sector, and the axial throat has the mass matrix \eqref{eq:Nfull}. The masses are
\begin{equation}
\text{polar:}\ \{\,l(l-1),\ (l+1)(l+2)\,\},
\qquad
\text{axial:}\ \{\,0,\ \lambda_-,\ \lambda_+\,\},
\end{equation}
with $\lambda_\pm$ of \eqref{eq:lampm}. At the fiducial parameters of \cite{Mondal:2025tht}, where $m_S^2=2.5\times10^{31}$, and for $l=2$ the normal-mode offsets $\Om-n=\tfrac12+\tfrac12\sqrt{1+4\lambda}$ of the axial throat are $1$, $3.37$ and $5\times10^{15}$ against $2$ and $4$ for the polar throat. The isospectrality of the Einstein--Maxwell limit is broken already at $\alpha^0$, by the finite Schwinger mass, by an amount that does not vanish at large $m_S^2$. The fermions carry electric charge, and a charged fermion is not invariant under the duality rotation that relates the electric and magnetic backgrounds. The two sectors probe different fermionic operators, the current in one and the stress tensor in the other, and no transformation of the gauge field alone can map one response into the other.

At $m_S^2=0$ the Schwinger sector is switched off and both sectors see the same corrected background and the same background Casimir tensor. The $\alpha=0$ problems are identical. At $O(\alpha)$ the two sectors differ in two ways. Their $\rho^3$ terms differ, and this part of the difference is fixed by the Einstein--Maxwell mouths. Their fermion responses also differ. The polar sector has the conformal-Ward-identity response on the moving field lines with its Hall current, and the axial sector has the drift lift with its polarisation current. The first-order shifts \eqref{eq:shift} can be compared level by level, since the unperturbed eigenfunctions are the same. We take the polar $V_{1a}$ and $V_{1b}$ from \cite{Sadhukhan:2026czl}, in the basis $Z_\pm$ of that paper with masses $(l+1)(l+2)$ and $l(l-1)$. At $\alpha=0$ it reads $Z_+=K-lw$, $Z_-=K+(l+1)w$, with $K$ and $w$ the polar metric and gauge amplitudes. There the LLL stress tensor is lifted on the field lines of the full field strength, and the metric of each line is scaled by the local field strength, which is an assumption of that paper. A lift on the coordinate planes passes the consistency tests there as well and gives different polar shifts \cite{Sadhukhan:2026czl}. The diagonal entries of $V_{1a}$ carry the term $m_\pm^2\log f$ in both sectors with the same coefficient, since it descends from the corrected background. The remap term of \eqref{eq:shift} is also the same in both sectors, since it is built from $V_0$ and $\gamma$ alone. Together they give every diagonal shift the logarithmic part $2\Om_0^2\log2$ in both sectors. The rational parts differ. Table~\ref{tab:shifts} lists the diagonal shifts with the remap term included.

\begin{table}[!t]
\centering
\caption{Diagonal entries $\varsigma$ of the first-order shift matrix of the Poschl-Teller levels of the two parity sectors at $m_S^2=0$, with the remap term included. On a level of a single channel $\varsigma=\dd\Om^2/\dd\alpha\,\big|_{\alpha=0}$. On the levels shared by $Z_+$ and $Z_-$ the shifts are the eigenvalues of \eqref{eq:Smatrix} at $\eta_{\rm H}=0$, and its diagonal entries once the higher Landau levels split these levels (Sec.~\ref{sec:hll-throat}). The polar entries use the lift with the scaled line metric of \cite{Sadhukhan:2026czl}, and those for $l=3,4$ are computed here from its potential matrices by the method of Appendix~\ref{app:numerics}. The frequencies refer to the clock of Sec.~\ref{sec:sector}.}
\label{tab:shifts}
\footnotesize\renewcommand{\arraystretch}{1.2}
\begin{tabular}{cccccc}
\hline\hline
$l$ & channel & $\Om_0$ & $\varsigma_{\rm polar}$ & $\varsigma_{\rm axial}$ & $\varsigma_{\rm polar}-\varsigma_{\rm axial}$\\[5pt]
\hline
2 & $Z_-$ & 2 & $-\tfrac{194}{15}+8\log2=-7.388$ & $-\tfrac{226}{15}+8\log2=-9.521$ & $\tfrac{32}{15}$\\[5pt]
2 & $Z_-$ & 3 & $-\tfrac{381}{10}+18\log2=-25.62$ & $-\tfrac{433}{10}+18\log2=-30.82$ & $\tfrac{26}{5}$\\[5pt]
2 & $Z_-$ & 4 & $-\tfrac{1172}{15}+32\log2=-55.95$ & $-\tfrac{1316}{15}+32\log2=-65.55$ & $\tfrac{48}{5}$\\[5pt]
2 & $Z_+$ & 4 & $-\tfrac{748}{21}+32\log2=-13.44$ & $-\tfrac{5188}{105}+32\log2=-27.23$ & $\tfrac{1448}{105}$\\[5pt]
2 & $Z_+$ & 5 & $-\tfrac{6415}{84}+50\log2=-41.71$ & $-\tfrac{7793}{84}+50\log2=-58.12$ & $\tfrac{689}{42}$\\[5pt]
2 & $Z_+$ & 6 & $-\tfrac{923}{7}+72\log2=-81.95$ & $-\tfrac{5269}{35}+72\log2=-100.6$ & $\tfrac{654}{35}$\\[5pt]
3 & $Z_-$ & 3 & $-\tfrac{909}{35}+18\log2=-13.49$ & $-\tfrac{1851}{70}+18\log2=-13.97$ & $\tfrac{33}{70}$\\[5pt]
3 & $Z_-$ & 4 & $-\tfrac{6124}{105}+32\log2=-36.14$ & $-\tfrac{1252}{21}+32\log2=-37.44$ & $\tfrac{136}{105}$\\[5pt]
3 & $Z_-$ & 5 & $-\tfrac{314}{3}+50\log2=-70.01$ & $-\tfrac{4499}{42}+50\log2=-72.46$ & $\tfrac{103}{42}$\\[5pt]
3 & $Z_+$ & 5 & $-\tfrac{4715}{126}+50\log2=-2.763$ & $-\tfrac{3035}{42}+50\log2=-37.60$ & $\tfrac{2195}{63}$\\[5pt]
3 & $Z_+$ & 6 & $-\tfrac{8548}{105}+72\log2=-31.50$ & $-\tfrac{12896}{105}+72\log2=-72.91$ & $\tfrac{4348}{105}$\\[5pt]
3 & $Z_+$ & 7 & $-\tfrac{46027}{330}+98\log2=-71.55$ & $-\tfrac{185447}{990}+98\log2=-119.4$ & $\tfrac{2153}{45}$\\[5pt]
4 & $Z_-$ & 4 & $-\tfrac{296}{7}+32\log2=-20.11$ & $-\tfrac{13576}{315}+32\log2=-20.92$ & $\tfrac{256}{315}$\\[5pt]
4 & $Z_-$ & 5 & $-\tfrac{1145}{14}+50\log2=-47.13$ & $-\tfrac{10475}{126}+50\log2=-48.48$ & $\tfrac{85}{63}$\\[5pt]
4 & $Z_-$ & 6 & $-\tfrac{944}{7}+72\log2=-84.95$ & $-\tfrac{14372}{105}+72\log2=-86.97$ & $\tfrac{212}{105}$\\[5pt]
4 & $Z_+$ & 6 & $-\tfrac{2269}{77}+72\log2=20.44$ & $-\tfrac{115249}{1155}+72\log2=-49.88$ & $\tfrac{11602}{165}$\\[5pt]
4 & $Z_+$ & 7 & $-\tfrac{30079}{396}+98\log2=-8.029$ & $-\tfrac{312361}{1980}+98\log2=-89.83$ & $\tfrac{80983}{990}$\\[5pt]
4 & $Z_+$ & 8 & $-\tfrac{1226776}{9009}+128\log2=-47.45$ & $-\tfrac{3444328}{15015}+128\log2=-140.7$ & $\tfrac{46144}{495}$\\[5pt]
\hline\hline
\end{tabular}
\end{table}

The differences are rational, of the same order as the shifts and positive on every level of the table. With the scaled polar lift every polar entry of the table lies above the corresponding axial entry. The differences do not depend on the clock, since a change of clock adds the same $\gamma_0\Om_0^2$ in both sectors. On these levels the differences for $Z_+$ grow with $l$. For $Z_-$ they are smaller and largest at $l=2$. The two parity sectors are not isospectral at $O(\alpha)$ even with the Schwinger sector removed. The background pieces of $V_{1a}$ are common, as the shared $\log2$ terms show. The $\rho^3$ coefficients are not common. In the polar sector the diagonal coefficients are larger by $\pi(l-1)/(2l+1)$ for $Z_+$ and by $\pi(l+2)/(2l+1)$ for $Z_-$, and the off-diagonal ones do not vanish. This is allowed, since the Darboux term that separates the two sectors near $r_e$ is $O(y^3)$. These $\rho^3$ terms come from the $\alpha^0$ mouths of the two sectors, and this part of the difference is fixed by the Einstein--Maxwell mouths and not by the fermion response. The rest of the difference comes from the fermion response. Both sectors lift the Casimir stress tensor onto the timelike invariant plane of $F$. The polar sector also scales the metric of each line with the local field strength, an assumption of that paper. In this sector $|F|$ and the density of lines do not change at $O(\epsilon)$, the scaling does not act, and the scaled lift of \cite{Sadhukhan:2026czl} reduces to the drift lift on this plane. The perturbations of the two sectors act on the lift differently. The two responses agree only in the part that mixes the modes of a shared level, as we show next.

The product of the two off-diagonal entries of \eqref{eq:Smatrix} is the same in both sectors on every shared level, and it is the product of the two entries of \eqref{eq:Sentries}. The entries themselves differ. The polar $\mathcal S_{+-}$ is $l/(l+1)$ times the axial one and the polar $\mathcal S_{-+}$ is $(l+1)/l$ times the axial one \cite{Sadhukhan:2026czl}. Both facts follow from the reductions of the off-diagonal entries and a map between the two sectors. In both sectors the off-diagonal entries of $V_{1a}-\Om_0^2V_{1b}$ are $-\tfrac12\mathsf R_{+-}V_{\rm mix}$ and $-\tfrac12\mathsf R_{-+}V_{\rm mix}$, up to terms whose matrix elements between the two modes vanish (Sec.~\ref{sec:shifts} and \cite{Sadhukhan:2026czl}). Here $\mathsf R$ is the projector onto the electric variable along the gravitational one. In this sector it is ${\rm diag}(0,1)$ in $(\Psi_g,\Psi_D)$, and $V_{1b}=-\mathsf R/f$. In the polar sector it maps the amplitudes $(K,w)$ of \cite{Sadhukhan:2026czl} to $(0,\,w-2K/\lhat)$, and the mixing comes from the Hall current, which enters only the equation for $w$. At $\alpha=0$ the polar amplitudes obey $\partial_x^2\bm Y_{\rm pol}+(\Om^2-f\mathcal N_{\rm pol})\bm Y_{\rm pol}=0$, with $\bm Y_{\rm pol}=(K,w)^{\!\top}$ and $\mathcal N_{\rm pol}$ the matrix with rows $(\lhat+2,-2\lhat)$ and $(-2,\lhat)$. The variables
\begin{equation}
\Psi_g\;\leftrightarrow\;\frac{K}{\lhat},
\qquad
\Psi_D\;\leftrightarrow\;w-\frac{2K}{\lhat}=-\frac{2K_F}{\lhat}
\label{eq:dualmap}
\end{equation}
turn this system into the $(\Psi_g,\Psi_D)$ block of \eqref{eq:tortoiseform} at $\alpha=0$ and $m_S^2=0$. Here $K_F$ is the relative decrease of the field strength, $|F|=|\bar F|(1-\epsilon K_FY)$ \cite{Sadhukhan:2026czl}. The $\alpha=0$ matrices fix such a map only up to separate rescalings of $Z_+$ and $Z_-$, which leave the products unchanged. The map \eqref{eq:dualmap} is the one that takes the polar projector onto $\mathsf R$. It also takes the Hall terms of the polar $\tilde P_1$ and $V_{1b}$, $(2/\lhat)(\rho/f)\,\partial_xK_F$ and $(2/\lhat)K_F/f$, onto the friction $-(\rho/f)\,\partial_x\Psi_D$ and the entry $-1/f$ of $V_{1b}$ that the drift gives here. Under it the combinations $Z_\pm$ of the two papers obey $Z_+^{\rm pol}=l\,Z_+$ and $Z_-^{\rm pol}=(l+1)\,Z_-$. With the reductions above, the polar off-diagonal entries are those of this sector with $Z_+$ rescaled by $l$ and $Z_-$ by $l+1$. This gives the ratios $l/(l+1)$ and $(l+1)/l$ and the equal products. The Hall current responds to the change of $|F|$ and the drift to the electric flux along the lines. Duality exchanges the radial magnetic and electric fields, and we read the coincidence as a remnant of it in the response. We have not derived \eqref{eq:dualmap} from the duality rotation. With the lift on the coordinate planes of \cite{Sadhukhan:2026czl} the Hall current responds to $K$ alone, and the polar products are $-4/\mu^2$ times these. With the scaled lift the two sectors mix the modes of a shared level with the same strength, and they differ there only through the diagonal entries of Table~\ref{tab:shifts}. With positive products the eigenvalues are real in both sectors. On the shared levels of $l=2$ the shifts are $(-27.16,-65.62)$, $(-57.93,-115.03)$ and $(-100.27,-179.95)$ in the axial sector against $(-13.38,-56.01)$, $(-41.53,-99.69)$ and $(-81.57,-157.56)$ in the polar sector. Neither sector has a complex first-order shift. The higher Landau levels split the shared levels at $O(\eta_{\rm H})$ in both sectors, by different amounts, and add a breaking of isospectrality larger than these $O(\alpha)$ differences by a factor of order $q$ (Sec.~\ref{sec:hll-iso}).

\section{Energy and linear stability}
\label{sec:stability}

\subsection{Null energy and the stability criterion}
\label{sec:criterion}
The MMP solution exists by a finely balanced violation of the averaged null energy condition along the field-line null rays. These rays are chronal, since the outside path between the mouths is shorter, and the achronal averaged null energy condition \cite{Graham:2007va} is not violated. This sector cannot disturb that balance at linear order. The perturbation \eqref{eq:hansatz} has $\delta g_{ab}=0$ on the $(\tau,\rho)$ block, and $\delta T^{\rm mag}_{ab}=0$ and $\delta T^{Q}_{ab}=0$ there as well. The magnetic cross terms need a $\theta$ or $\phi$ leg, and the fermionic response is purely mixed, $\delta T^Q_{a\phi}$. A radial null vector $k^a$ with $k^\theta=k^\phi=0$ remains null without correction, and
\begin{equation}
\delta\big(T_{\mu\nu}k^\mu k^\nu\big)=\delta T_{ab}\,k^ak^b+2\,\delta T_{a\phi}\,k^a\,k^\phi=0
\end{equation}
term by term. The Casimir budget of the wormhole is first disturbed at $O(\epsilon^2)$ in this sector.

A growing mode of $\partial_x^2\bm Z+(\omega^2-V)\bm Z=0$ has $\omega^2<0$, and when $V$ is symmetric its positivity on the domain excludes such a mode \cite{Wald:1979lth,Sadhukhan:2026czl}. The fermion response is part of this potential, as semiclassical linear stability requires \cite{Anderson:2002fk}. The potentials of this paper are not symmetric as they stand. Where a constant similarity transformation makes them symmetric, the criterion is applied in that basis. For the three-channel systems the transformation is $\Sigma$ of \eqref{eq:Ssym}, and for the light sector and the $\alpha=0$ Einstein--Maxwell block it is $\mathrm{diag}(\mu,1)$. The symmetrised potentials have the eigenvalues of the original ones. The symmetric part $\frac12(V+V^{\top})$ taken in the original basis is not the right object. For the mouth it is indefinite for every $r$ and every $m_S^2>0$, since the off-diagonal entries of $\hat V$ are far from symmetric. For the $O(\alpha)$ throat corrections at $m_S^2=0$ the constant matrix $\mathsf H=\mathrm{diag}(l+2,\,l-1)-\alpha\mu^2\sigma_x/(6(2l+1))$, with $\sigma_x$ the first Pauli matrix, makes $\mathsf H(V_0+\alpha V_{1a})$ and $\mathsf H(1+\alpha V_{1b})$ symmetric in the $Z_\pm$ basis up to $O(\alpha^2)$. The argument below needs only the symmetric parts in the $Z_\pm$ basis, where $V_0$ is diagonal and the non-symmetric terms are $O(\alpha)$, as in \cite{Sadhukhan:2026czl}. Eigenvalues move continuously in $\alpha$, and a real eigenvalue $\Om^2$ bounded away from zero cannot change sign at small $\alpha$. A complex pair can still form at first order on a level shared by two channels, and Sec.~\ref{sec:shifts} examines those levels. The one channel not bounded away from zero is the neutral mode, treated separately below.

\subsection{The mouths}
\label{sec:neutralstab}
The Einstein--Maxwell block of the mouth potential \eqref{eq:Vmouth} decouples into Chandrasekhar's axial potentials \eqref{eq:Vminus}. On the horizonless mouth region $r\ge r_e$ the bracket in \eqref{eq:Vminus} is smallest for the larger root $\beta_1$, equals $r_e\,l(l-1)+O(\delta_M)$ at $r=r_e$ and increases outward, and both potentials are positive for $l\ge2$ \cite{Sadhukhan:2026czl}. With the Schwinger response two eigenvalues of the symmetrised three-channel potential \eqref{eq:Ssym} stay positive. Its $(\Psi_D,\Xi)$ block has trace $(\lhat+m_S^2)r+4r_e^2/r>0$, up to a positive term of order $1/q^2$, and determinant $m_S^2r_e^2(2+\lhat/q)^2>0$, which is $4m_S^2r_e^2$ without the coupling. By Cauchy interlacing the two larger eigenvalues of the full matrix are positive for every $r$, $GM$ and $m_S^2>0$. The potential is still indefinite, and two positive eigenvalues do not by themselves exclude a growing mode of the coupled problem. The finite-element computation below tests the coupled problem directly. The third eigenvalue belongs to the neutral channel. By \eqref{eq:detV} it is negative between the two roots of the quadratic factor, which are $r_e$ and $2r_e$ at $\delta_M=0$.

At $\alpha^0$ the neutral channel has an exact zero-energy solution. For every $GM$, $r_e$, $l$, $m_S^2$ and $q$
\begin{equation}
\bm Z_0=\Big(-\Big(2+\frac{\lhat}{q}\Big)\frac{r_em_S^2}{\mu^2r},\;-m_S^2,\;1\Big)^{\!\top},\qquad
\partial_{r_*}^2\bm Z_0=\frac{\Delta}{r^5}\,\big(\hat V+\delta\hat V\big)\bm Z_0 ,
\label{eq:Z0mouth}
\end{equation}
in the basis $(\Psi_g,\Psi_D,\Xi)$. It has $\Psi_e=0$ and therefore $E=0$. At $r=r_e$ it is the throat null vector $\bm v_0$ of \eqref{eq:nullvectorskk}, in the throat at $\alpha=0$ it continues as the constant zero-energy solution of the free mode $\psi_{\rm zm}$ of Sec.~\ref{sec:zmalpha}, and as $r\to\infty$ it tends to $(0,-m_S^2,1)^{\top}$. It is bounded at both ends. The adjoint operator $\partial_{r_*}^2-V^{\top}$ has the zero mode $\bm L_0(r)=\Sigma^2\bm Z_0/(\lhat m_S^2)$, since $\Sigma^2V=V^{\top}\Sigma^2$, and $\bm L_0(r_e)$ is the throat left null vector $\bm L_0$ of \eqref{eq:nullvectorskk}. The finite-element computation of Appendix~\ref{app:numerics} finds no negative eigenvalue for the light system \eqref{eq:Vlight} at $l=2,3,4$ and for the three-channel system at $l=2$ and $m_S^2=1$, $10^2$ and $10^4$. The box of these runs resolves bound states down to $|\omega^2|\sim10^{-9}/r_e^2$, and the statements below hold at that resolution. The light system is the limit $m_S^2\to\infty$ of the three-channel system. The two eigenvalues of $r^5V/\Delta$ that stay finite differ from those of $r^5V_{\rm light}/\Delta$ by at most a relative $\lhat/m_S^2$ at every $r$ we sampled with $10^{-8}r_e<r-r_e<10^4r_e$ (Appendix~\ref{app:numerics}). The derivative terms of the electric channel are smaller than its Schwinger term by the factor $m_S^2$, near $r_e$ and far away alike, since near $r_e$ both scale as $(r-r_e)^2$. The vector $\bm Z_0$ is an exact zero-energy solution at every $m_S^2$, and $\bm L_0(r)\,\partial_{m_S^2}(\hat V+\delta\hat V)\,\bm Z_0=0$ at every $r$. A change of $m_S^2$ therefore does not move the channel off the threshold at first order. We take the runs at $m_S^2\le10^4$ and the light system to cover the physical value $m_S^2\sim10^{31}$. We have not run that value directly. Scaling the potential by a constant $\mathsf s$ does not produce a bound state for $0\le\mathsf s\le1$. The lowest point $\omega^2_{\rm min}(\mathsf s)$ of the spectrum of $-\partial_{r_*}^2+\mathsf s\,\Sigma V\Sigma^{-1}$ is the infimum of functions linear in $\mathsf s$ and is therefore concave. It vanishes at $\mathsf s=0$ and is not negative at $\mathsf s=1$ within the resolution of the box, and concavity gives $\omega^2_{\rm min}(\mathsf s)\ge0$ in between. For $\mathsf s>1$ a cut-off $\bm Z_0$ is a trial function of negative energy. The mouth therefore sits exactly at the threshold for a bound state. The finite-element runs omit $\delta\hat V$. With it \eqref{eq:Z0mouth} is still an exact zero-energy solution, and for $q=10^{33}$ the eigenvalues of the symmetrised potential move by at most a relative $\lhat/q$, $6\times10^{-33}$ for $l=2$, at these $m_S^2$ (Appendix~\ref{app:numerics}). The lowest finite-element eigenvalue of the light system, $7.0\times10^{-10}/r_e^2$ for $l=2$, lies below the lowest mode $8.2\times10^{-10}/r_e^2$ of the empty box. The single-channel picture fails here. The smallest eigenvalue $V_{\min}$ of the symmetrised light potential, used alone, gives a bound state at $\omega^2r_e^2=-9.4\times10^{-5}$, and the geometric term $|\partial_{r_*}\bm n|^2$ of its rotating eigenvector $\bm n$, the analogue of the Born--Huang diagonal correction \cite{BornHuang:1954,MeadTruhlar:1979}, removes it. Neither number measures how far the coupled problem is from a bound state. That distance is zero at $\alpha^0$.

The $O(\alpha)$ terms of the mouth do not move this threshold, and neither does the one-loop action of the higher Landau levels (Sec.~\ref{sec:hll-mouth}). The vector \eqref{eq:Z0mouth}, with $r_e$ unchanged, solves the corrected mouth system \eqref{eq:mouthalphamaster} at $p=0$ to first order in $\alpha$. This holds with the coupling of Sec.~\ref{sec:kk} included and with its $O(\alpha/q)$ terms of Appendix~\ref{app:matrices} kept. At $1/q\to0$ the mechanism is the following. The friction $\tilde P$ does not act on it, since $\Psi_D$ is constant on $\bm Z_0$. The $\Psi_D$ row of $\mathcal T$ vanishes on $\bm Z_0$. In the $\Psi_g$ row the fermion term and the correction of the background cancel. The first contributes
\begin{equation}
-\alpha\,\mathcal T\bm Z_0=\frac{4\alpha\,m_S^2\,r_e^3\,(r-r_e)^3(2r-7r_e)}{3\mu^2r^9}\,\bm e_g
\label{eq:TZ0}
\end{equation}
to the residual, with $\bm e_g$ the unit vector of the $\Psi_g$ channel, and the second contributes the same term with the opposite sign. In the corrected mouth $N$ has its minimum at $r_h=r_e(1+\alpha/6)$, where $N$ and $N'$ vanish to $O(\alpha)$, and $\bm Z_0$ stays bounded there and at large $r$. At the threshold a small perturbation $\delta V$ of the potential creates a bound state when $\int\bm L_0\,\delta V\,\bm Z_0\,\dd r_*<0$ and does not when it is positive. For a single channel with $\bm Z_0=1$ this is the weak-coupling criterion of \cite{Simon:1976un}, and the threshold case is treated in \cite{Klaus:1980ks}. For the $O(\alpha)$ terms of the mouth the weight vanishes, since $\bm Z_0$ solves the corrected system, and to this order the corrected mouth sits at the threshold for a bound state as well.

The wormhole contains two mouths joined by the throat. At large $\rho$ the throat potential of the neutral channel is $\lambda_{1a}=-(2\pi m_S^2/\mathcal H)\,\rho f+O(1)$. Its integral has the closed form
\begin{equation}
\int_0^{\rho}\lambda_{1a}\,\dd x=-\frac{m_S^2}{2\mathcal H}\Big[4f\arctan\rho+4\rho+\frac{\mu^2\rho}{f}\Big]
=-\frac{\pi m_S^2}{\mathcal H}\,\rho^2+c_l+O(\rho^{-1}),
\qquad c_l=-\frac{\pi m_S^2}{\mathcal H},
\label{eq:lamint}
\end{equation}
with no term in $\log\rho$. The term in $\rho^2$ is the near-horizon form of the $\alpha^0$ mouth potential, which $\bm Z_0$ solves exactly (Sec.~\ref{sec:zmalpha}). The two halves of the throat interior therefore add the finite weight $2\alpha c_l$ in throat units, which is $-\pi^2\alpha^2m_S^2/(2\mathcal H r_e)$ in mouth units. It is attractive and has no logarithmic enhancement. The $O(\alpha^2)$ terms of the mouth act at the same order and are not known. The threshold state also does not decay at large $r$. The ambient region between the two mouths, a distance $d$ apart, enters the rest of this paper only at order $d/\ell$, but it acts on this state, and we have not estimated that effect. Whether the next order turns the neutral channel into a bound state which is a growing mode or turns it into a virtual state is not decided here. A bound state at that order would be slow. On a line a weak attractive weight $\mathcal W$ of order $\alpha^2/r_e$ gives a bound state at $\omega^2\simeq-\mathcal W^2/4$ \cite{Simon:1976un}, and the growth rate is of order $\alpha^2/r_e$. If the field lines close through the ambient region, the threshold state is normalisable on a loop of tortoise length about $\pi\ell+d$, dominated by the throat. Then $\omega^2\simeq-\mathcal W/(\pi\ell+d)$, which gives a rate of order $\alpha^{3/2}/r_e$. At the fiducial parameters $r_e/(c\alpha^2)\simeq1.5\times10^{59}$\,s and $r_e/(c\alpha^{3/2})\simeq7.4\times10^{41}$\,s.

\subsection{The throat}
\label{sec:throatstab}
At $\alpha=0$ the statement is exact. $V=f\,\mathcal N$ with $\mathcal N$ constant, the system diagonalises globally in the eigenbasis of $\mathcal N$, and each eigenchannel is a Poschl-Teller problem with mass$^2$ $\lambda$. By \eqref{eq:disc} all eigenvalues are real for every $m_S^2\ge0$, and for $l\ge2$ the two nonzero ones are positive. They give bounded oscillatory evolution with the normal-mode spectrum $\Om^2=(n+\tfrac12+\tfrac12\sqrt{1+4\lambda})^2$ under the standard AdS$_2$ boundary conditions \cite{Ishibashi:2004wx}. The massless channel has the free spectrum of the closed throat, marginal and propagating, and at $O(\alpha)$ it is controlled by the mouths (Sec.~\ref{sec:zmalpha}). No exponentially growing mode exists in the $\alpha=0$ throat for any value of the Schwinger mass. This holds with the coupling of Sec.~\ref{sec:kk} as well, whose masses stay real and positive (Sec.~\ref{sec:alpha0}), and with the one-loop action of the higher Landau levels (Sec.~\ref{sec:hll-throat}).

At $O(\alpha)$ the argument used for the polar throat in \cite{Sadhukhan:2026czl} applies to the massive channels. With $\Om^2=-p^2$ and $V=V_0+\alpha(V_{1a}+p^2V_{1b})$ in the $Z_\pm$ basis at $m_S^2=0$, multiplying \eqref{eq:schrodinger} by $\bm Z^{\dagger}$ and integrating gives
\begin{equation}
\Om^2=\frac{\mathfrak e}{\mathfrak n},\qquad
\mathfrak e=\int\dd x\,\big[|\partial_x\bm Z|^2+\bm Z^{\dagger}(V_0+\alpha V_{1a})\bm Z\big],\qquad
\mathfrak n=\int\dd x\,\bm Z^{\dagger}(1+\alpha V_{1b})\bm Z .
\label{eq:en}
\end{equation}
To first order in $\alpha$ one has $\mathrm{Re}\,\Om^2=\mathrm{Re}\,\mathfrak e/\mathrm{Re}\,\mathfrak n$, and positivity of the symmetric parts of $V_0+\alpha V_{1a}$ and $1+\alpha V_{1b}$ excludes a real negative $\Om^2$ for every $p$ at once. The minimum eigenvalue of $V_{\rm s}\equiv\tfrac12(V+V^{\!\top})$ was evaluated on $2000$ points of the validity window $\alpha\rho\le0.1$ for $l=2,3,4$, $\alpha\in\{10^{-6},10^{-3},10^{-2}\}$ and $|p|\in\{0,1,3\}$ (Appendix~\ref{app:numerics}). Only $p=0$ enters \eqref{eq:en}, since the frequency-dependent part sits in $\mathfrak n$. The other values of $p$, with $p^2V_{1b}$ added to $V$, check that it does not change the picture. The result is
\begin{equation}
\min\;\mathrm{eig}\,V_{\rm s}\;\ge\;0.978\;\times\;l(l-1)\qquad\text{over the entire scan},
\end{equation}
the worst point sitting at the throat centre for $l=2$, $\alpha=10^{-2}$, $|p|=3$. The symmetric part of $1+\alpha V_{1b}$ has smallest eigenvalue $0.989$ at $\alpha=10^{-2}$. Beyond the window, positivity first fails at $\alpha\rho=0.263$, $0.277$ and $0.285$ for $l=2,3,4$ at $\alpha=10^{-6}$, and about $0.002$ lower at $\alpha=10^{-2}$, with the same values for $p=0$, $3$ and $30$. There $\alpha\rho$ is no longer small and the $O(\alpha)$ form of $V$ itself fails. In the companion sector positivity also fails only beyond the validity window. Within this window the massive channels at $m_S^2=0$ have no mode with real $p$ for $l=2,3,4$. A complex $\Om^2$ is not excluded by positivity. It would allow a mode that grows while it oscillates, at a rate of order $\alpha$, and at first order it can arise only on a level shared by $Z_+$ and $Z_-$. By \eqref{eq:Sentries} the off-diagonal product of the first-order matrix is positive on every such level, the shifts are real, and no oscillatory growth appears at first order. These are statements about the normal modes of the throat, with the normalisable Poschl-Teller conditions at $x=\pm\pi/2$ of Sec.~\ref{sec:shifts}. On the levels computed the shifts are negative but small compared with the gaps, and they do not change the count. First-order perturbation theory needs $\alpha|\varsigma|\ll\Om_0$. On the levels computed $|\varsigma|<7.5\,\Om_0^2$, for the eigenvalues and for the diagonal entries of $\mathcal S$, and there the ratio $|\varsigma|/\Om_0^2$ grows roughly like $2\log\Om_0$ (Sec.~\ref{sec:shifts}). If this growth continues, the condition holds for $\alpha\Om_0\log\Om_0\ll1$. At the physical $m_S^2$ the light massive channel has a positive potential on the validity window and real first-order shifts on every level below the screened channel (Sec.~\ref{sec:lightchannel}).

\section{The higher Landau levels}
\label{sec:hll}

The higher Landau levels are gapped and never excited here. Their vacuum polarisation is the one-loop Euler-Heisenberg action of the gauge field \cite{Heisenberg:1936nmg,Schwinger:1951nm,Dunne:2004nc}, whose logarithm is the running of the coupling. It is not suppressed by $1/q$. As in the companion paper \cite{Sadhukhan:2026czl} we include it to first order in
\begin{equation}
  \eta_{\rm H}\equiv\frac{g_c^2}{12\pi^2},\qquad \frac{\eta_{\rm H}}{\alpha}=\frac q3 .
  \label{eq:etaH}
\end{equation}
The parameter $\eta_{\rm H}$ is $8.4\times10^{-3}$ at $g_c=1$. At first order the two corrections add, since each is sourced on the Einstein--Maxwell solution and enters the equations linearly. We drop the cross terms of order $\alpha\eta_{\rm H}$, and the results of the previous sections stand. The Schwinger response and $m_S^2$ are kept as before, unexpanded. The coupling of Sec.~\ref{sec:kk} is left out in this section. Its product with $\eta_{\rm H}$ is a cross term, and at $\eta_{\rm H}=0$ it adds $\delta\mathcal N$ of \eqref{eq:dN}.

\subsection{The one-loop action in this sector}
\label{sec:hll-action}

With $\mathcal Q=-\tfrac14F_{\mu\nu}F^{\mu\nu}$, $\mathcal R=-\tfrac14F_{\mu\nu}\tilde F^{\mu\nu}$ and $\tilde F=\star F$ the gauge Lagrangian becomes $\mathcal L(\mathcal Q,\mathcal R)$, with subscripts on $\mathcal L$ denoting derivatives. The stress tensor is $\mathcal L_{\mathcal Q}F_{\mu\lambda}F_\nu{}^\lambda+g_{\mu\nu}(\mathcal L-\mathcal R\mathcal L_{\mathcal R})$, and the Maxwell equation becomes $\nabla_\mu(\mathcal L_{\mathcal Q}F^{\mu\nu}+\mathcal L_{\mathcal R}\tilde F^{\mu\nu})=J^\nu$, with $J^\nu$ the LLL currents of the previous sections. For the levels $n\ge1$ of one massless Dirac fermion, on the magnetic background with field strength $\bar B=q/2R^2$ in an orthonormal frame, not the metric function $B$ of \eqref{eq:background},
\begin{equation}
  g_c^2\mathcal L_{\mathcal Q}\equiv\varepsilon_\perp=1-\eta_{\rm H}\log\frac{\bar B}{B_0}-\frac{\eta_{\rm H}}{2},
  \qquad
  g_c^2\bar B^2\mathcal L_{\mathcal Q\mathcal Q}=\eta_{\rm H},
  \qquad
  g_c^2\bar B^2\mathcal L_{\mathcal R\mathcal R}=c_\parallel\eta_{\rm H},
  \label{eq:EHcoef}
\end{equation}
with $c_\parallel=\gamma_{\rm E}-\tfrac12+12\zeta'(-1)\simeq-1.908$ and $\mathcal L=-\tfrac12\bar B^2[1-\eta_{\rm H}\log(\bar B/B_0)]/g_c^2$. The coupling $g_c$ is defined at the field $B_0=q/2r_e^2$ of the throat. With this choice $r_e$ stays the radius of its sphere. The derivation is in the companion paper. In this sector $\delta\mathcal Q=0$ at $O(\epsilon)$, and $\delta\mathcal R$ is $\bar B$ times the electric field along the field lines. The perturbation therefore sees $\mathcal L_{\mathcal Q}$ and $\mathcal L_{\mathcal R\mathcal R}$. An electric field along $\bar B$ enters $\mathcal L$ as $\varepsilon_\parallel E^2/2g_c^2$, with
\begin{equation}
  \varepsilon_\parallel=\varepsilon_\perp+c_\parallel\eta_{\rm H},
  \label{eq:epar}
\end{equation}
and an electric field across $\bar B$ as $\varepsilon_\perp E^2/2g_c^2$. The higher levels respond to the electric field along the lines as a dielectric with permittivity $\varepsilon_\parallel$. The Schwinger term of the lowest level comes on top of it. The backgrounds are those of the companion paper. The throat is AdS$_2\times S^2$ with sphere radius $r_e$ and AdS$_2$ radius $R_{AdS_2}\equiv R_A=r_e/\sqrt{1-\eta_{\rm H}}$. The mouth is
\begin{equation}
  A=\frac1B=1-\frac{2GM}{r}+\frac{r_e^2}{r^2}\Big[1+2\eta_{\rm H}\Big(1+\log\frac{r}{r_e}\Big)\Big],
  \label{eq:hllmouthax}
\end{equation}
extremal at $GM=r_e(1+\eta_{\rm H})$, where $A\simeq(r-r_e)^2/R_A^2$ near $r_e$. Together with the $O(\alpha)$ backgrounds \eqref{eq:throatbg} and \eqref{eq:mouthbgax}, and up to terms of order $\alpha\eta_{\rm H}$, the throat and the mouth are
\begin{equation}
\begin{aligned}
\text{throat:}&\quad \dd s^2=R_A^2\Big[-(f+\alpha\gamma)\,\dd\tau_A^2+\frac{\dd\rho^2}{f+\alpha\gamma}\Big]+r_e^2(1+\alpha\varphi)\,\mathring g,\\
\text{mouth:}&\quad \dd s^2=-(1+2\alpha\Phi_1)N\,\dd t^2+\frac{\dd r^2}{N}+r^2\mathring g,\\
&\quad N=1-\frac{2GM}{r}+\frac{r_e^2}{r^2}\Big[1+2\eta_{\rm H}\Big(1+\log\frac{r}{r_e}\Big)\Big]+\alpha N_1,
\end{aligned}
\label{eq:bgbothax}
\end{equation}
with $\tau_A$ the time of the AdS$_2$ factor and $GM=r_e(1+\eta_{\rm H}-\alpha/18)+\delta_M$. In the mouth $\varepsilon_\perp=1-\eta_{\rm H}/2+2\eta_{\rm H}\log(r/r_e)$, and in the throat $\varepsilon_\perp=1-\eta_{\rm H}/2$.

\subsection{The throat}
\label{sec:hll-throat}

The reduction of Sec.~\ref{sec:consistency} runs with these field equations, and the residuals of Step~4 vanish identically with and without the Schwinger response. In the throat $\varepsilon_\perp$ and $\varepsilon_\parallel$ are constant, the first-derivative matrix vanishes, and the potential is $f\,\mathcal N_{\rm H}$ for the time $\tau_A$ of the AdS$_2$ factor of radius $R_A$. In the basis $(\Psi_g,\Psi_{\rm H},\Xi)$, with
\begin{equation}
  \Psi_{\rm H}=\frac{\varepsilon_\parallel\Psi_e-m_S^2\,\Xi}{\sqrt{\varepsilon_\perp}} ,
  \label{eq:PsiH}
\end{equation}
the mass matrix is
\begin{equation}
  \mathcal N_{\rm H}=\frac{1}{1-\eta_{\rm H}}
  \begin{pmatrix}
    \mu^2 & -2\sqrt{\varepsilon_\perp} & 0\\[2pt]
    -2\mu^2\sqrt{\varepsilon_\perp} & \dfrac{\varepsilon_\perp}{\varepsilon_\parallel}\big(\lhat+4-2\eta_{\rm H}+4c_\parallel\eta_{\rm H}\big) & \dfrac{\sqrt{\varepsilon_\perp}}{\varepsilon_\parallel}\,\lhat\,m_S^2\\[8pt]
    0 & \dfrac{\sqrt{\varepsilon_\perp}}{\varepsilon_\parallel} & \dfrac{m_S^2}{\varepsilon_\parallel}
  \end{pmatrix}.
  \label{eq:NHax}
\end{equation}
At $\eta_{\rm H}=0$ it is \eqref{eq:Nfull}, and $\Psi_{\rm H}$ is $\Psi_D$. The variable $\Psi_{\rm H}$ is the displacement flux along the field lines, the electric flux weighted by the permittivity $\varepsilon_\parallel$ minus the Schwinger charge, divided by $\sqrt{\varepsilon_\perp}$. The matrix $\Sigma$ of \eqref{eq:Ssym} still makes $\mathcal N_{\rm H}$ symmetric. The masses are therefore real. The determinant vanishes for every $l$, $m_S^2$ and $\eta_{\rm H}$. The right null vector is $(-2m_S^2/\mu^2,-m_S^2/\sqrt{\varepsilon_\perp},1)$ in this basis, which is $\bm v_0$ of \eqref{eq:nullvectors} in the basis $(\Psi_g,\Psi_D,\Xi)$, with $E=0$. On it the electric field along the lines vanishes, and $\mathcal L_{\mathcal R\mathcal R}$ does not act. The neutral channel stays exactly massless.

At the formal point $m_S^2=0$ the two nonzero masses are those of the companion sector with $\eta_{\rm H}$ replaced by $(2-c_\parallel)\eta_{\rm H}$,
\begin{equation}
  m_-^2R_A^2=l(l-1)\Big[1+(2-c_\parallel)\eta_{\rm H}\frac{l+1}{2l+1}\Big],
  \qquad
  m_+^2R_A^2=(l+1)(l+2)\Big[1+(2-c_\parallel)\eta_{\rm H}\frac{l}{2l+1}\Big],
  \label{eq:mHax}
\end{equation}
with $2-c_\parallel=3.908$. At large $m_S^2$ the screened channel has $\lambda_+\simeq m_S^2/[(1-\eta_{\rm H})\varepsilon_\parallel]$, and the light channel has
\begin{equation}
  \lambda_-\;\longrightarrow\;\frac{\lhat+2-2\eta_{\rm H}}{1-\eta_{\rm H}}=\lhat+2+\eta_{\rm H}\lhat+O(\eta_{\rm H}^2),
  \label{eq:lamH}
\end{equation}
which does not depend on $c_\parallel$, since the screened electric field drops out of the light channel. The offset of the light channel moves to $s_-+\eta_{\rm H}\lhat/\sqrt{4\lhat+9}$ in the time $\tau_A$. On the clock of Sec.~\ref{sec:sector}, the proper time at $\rho=0$ in units of $r_e$, a level $\Om_0$ moves to $\Om^2=\Om_0^2+\alpha\varsigma+\eta_{\rm H}\varsigma_{\rm H}$ with
\begin{equation}
  \varsigma_{\rm H}=\frac{2\Om_0\,\lhat}{\sqrt{4\lhat+9}}-\Om_0^2
  \label{eq:varsigmaHax}
\end{equation}
for the light channel. The term $-\Om_0^2$ is the change of clock. Since $\Om_0\ge s_->2\lhat/\sqrt{4\lhat+9}$, every $\varsigma_{\rm H}$ of the light channel is negative, and on this clock the higher levels lower the light levels. The shift in $s_-$ is $1.04\,\eta_{\rm H}$ for $l=2$, about $9\times10^{-3}$ at $g_c=1$. It exceeds the shift $4\lhat/(q\sqrt{4\lhat+9})$ from the coupling to $h_{a\phi}$ by the factor $\eta_{\rm H}q/4=g_c^2q/48\pi^2$, and the $O(\alpha)$ shifts of Table~\ref{tab:lightshifts} by more. At $m_S^2=0$ the levels shared by $Z_+$ and $Z_-$ split by $(2-c_\parallel)\eta_{\rm H}\,l(l+1)/[(2l-1)(2l+3)]$ in $\Om$. The offsets grow by $(2-c_\parallel)$ times the amounts of the companion sector. Unlike every $\varsigma_{\rm H}$ of the companion sector, $\varsigma_{\rm H}=2\Om_0\,\partial s/\partial\eta_{\rm H}-\Om_0^2$ is positive on the lowest levels of both channels. For $l=2$ it is $2.25$ and $0.38$ for $Z_-$ at $\Om_0=2$ and $3$, and $5.44$ and $1.80$ for $Z_+$ on the shared levels $\Om_0=4$ and $5$. As in the companion sector, the splitting makes the $O(\alpha)$ shift of each mode on the levels examined the diagonal entry of $\mathcal S$, and the mixing of Sec.~\ref{sec:shifts} is a second-order effect. If the slow growth of $|\varsigma|/\Om_0^2$ found in Sec.~\ref{sec:shifts} continues, the splitting exceeds the $O(\alpha)$ shifts for $\Om_0\log\Om_0\ll(2-c_\parallel)q/12$, and the same holds on every level below the gap of the higher levels. With the $O(\alpha)$ terms of Sec.~\ref{sec:throatstab} the potential of the massive channels at $m_S^2=0$ is $V_0+\eta_{\rm H}V_\eta+\alpha(V_{1a}+p^2V_{1b})$, with $\eta_{\rm H}V_\eta$ the $O(\eta_{\rm H})$ part of $f\,\mathcal N_{\rm H}$ on these channels. In the eigenbasis of \eqref{eq:NHax} at $m_S^2=0$ its $O(\eta_{\rm H})$ part is $(2-c_\parallel)\eta_{\rm H}(1+\rho^2)$ times a positive diagonal matrix, and a constant change of basis leaves the kinetic term alone. The symmetric part therefore stays positive on the window $\alpha\rho\le0.1$, and the point where positivity fails moves out. In the eigenbasis of \eqref{eq:NHax} it moves for $l=2$ from $\alpha\rho=0.263$ to $0.267$ at $\eta_{\rm H}=1/12\pi^2$. The light channel at large $m_S^2$ gains $\eta_{\rm H}\lhat(1+\rho^2)$ in its potential, and its first zero moves from $\alpha\rho=0.248$ to $0.250$. On a shared level at $m_S^2=0$ the first-order matrix is $\alpha\mathcal S+\eta_{\rm H}\,{\rm diag}(\varsigma_{{\rm H}+},\varsigma_{{\rm H}-})$. Its discriminant is a square plus $4\alpha^2\mathcal S_{+-}\mathcal S_{-+}$, which is positive on every shared level by \eqref{eq:Sentries}, and the shifts are real for every ratio of $\eta_{\rm H}$ to $\alpha$. The nonzero throat masses stay positive, and the massive channels of the throat have no growing mode at this order. The neutral channel stays exactly massless, and Sec.~\ref{sec:hll-mouth} treats it with the mouths.

\subsection{The mouth}
\label{sec:hll-mouth}

On \eqref{eq:hllmouthax} the reduction closes, and the residuals of Step~4 vanish identically with and without the Schwinger response, for every $GM$, $l$, $p$ and $\eta_{\rm H}$. In the basis $(\Psi_g,\Psi_e,\Xi)$ the electric channel acquires a first-derivative term, $\partial_{r_*}\log(\varepsilon_\parallel^2/\varepsilon_\perp)$ on $\partial_{r_*}\Psi_e$ and a term proportional to $\eta_{\rm H}m_S^2$ on $\partial_{r_*}\Xi$. Both come from the variation of $\varepsilon_\perp$ along $r$. The variable $\Psi_{\rm H}$ of \eqref{eq:PsiH}, now with $\varepsilon_\perp$ and $\varepsilon_\parallel$ functions of $r$, removes them. In the basis $(\Psi_g,\Psi_{\rm H},\Xi)$ the system is $\partial_{r_*}^2\bm Z+(\omega^2-V)\bm Z=0$ with no first-derivative term, $V=(\Delta/r^5)\hat V$, $\Delta=r^2A$ and
\begin{equation}
  \hat V=
  \begin{pmatrix}
    \lhat\,r-6GM+\dfrac{r_e^2}{r}\Big[4+\eta_{\rm H}\Big(6+8\log\dfrac{r}{r_e}\Big)\Big] & -2r_e\sqrt{\varepsilon_\perp} & 0\\[10pt]
    -2r_e\mu^2\sqrt{\varepsilon_\perp} & \hat V_{DD} & \dfrac{\sqrt{\varepsilon_\perp}}{\varepsilon_\parallel}\,\lhat\,m_S^2\,r\\[10pt]
    0 & \dfrac{\sqrt{\varepsilon_\perp}}{\varepsilon_\parallel}\,r & \dfrac{m_S^2\,r}{\varepsilon_\parallel}
  \end{pmatrix},
  \label{eq:VhatH}
\end{equation}
\begin{equation}
  \hat V_{DD}=\lhat\,r+\frac{4r_e^2}{r}-\frac{\eta_{\rm H}}{r}\Big[(c_\parallel\lhat-1)\,r^2+4GM\,r-r_e^2\Big(1+8\log\frac{r}{r_e}\Big)\Big]+O(\eta_{\rm H}^2).
  \label{eq:VDDH}
\end{equation}
This is the expansion to first order in $\eta_{\rm H}$ of the one-loop form, in which $\varepsilon_\perp$ and $\varepsilon_\parallel$ are kept unexpanded,
\begin{equation}
  \hat V_{DD}=\frac{\varepsilon_\perp}{\varepsilon_\parallel}\,\lhat\,r+\frac{(4\varepsilon_\perp+3\eta_{\rm H})\,r_e^2}{r}+\frac{\eta_{\rm H}}{\varepsilon_\perp}\Big(r-4GM+\frac{17\eta_{\rm H}r_e^2}{2r}\Big)+\frac{3\eta_{\rm H}^2}{\varepsilon_\perp^2}\Big(r-2GM+\frac{5\eta_{\rm H}r_e^2}{2r}\Big).
  \label{eq:VDDexact}
\end{equation}
The two forms differ at $O(\eta_{\rm H}^2)$. At $r=r_e$, $l=2$, $\eta_{\rm H}=0.1$ and $GM=r_e(1+\eta_{\rm H})$ the one-loop form \eqref{eq:VDDexact} gives $11.31\,r_e$ and its expansion \eqref{eq:VDDH} gives $10.90\,r_e$. The potential is frequency independent, and $\Sigma$ of \eqref{eq:Ssym} makes it symmetric for every $r$. The Schwinger mass of the $\Xi$ row is divided by $\varepsilon_\parallel$, the permittivity of the higher levels along the field lines. At $GM=r_e(1+\eta_{\rm H})$ the near-horizon limit gives $V\simeq(y^2/R_A^2)\,\hat V(r_e)/r_e^3$, and $\hat V(r_e)/[r_e(1-\eta_{\rm H})]$ is $\mathcal N_{\rm H}$ of \eqref{eq:NHax}. The mouth and the throat match at this order. We use \eqref{eq:VhatH} with \eqref{eq:VDDexact} in the numerics below, without expanding in $\eta_{\rm H}\log(r/r_e)$.

The neutral channel keeps its exact zero-energy solution. For every $GM$, $l$, $m_S^2$ and $\eta_{\rm H}$, with the one-loop form of $\hat V_{DD}$,
\begin{equation}
  \bm Z_0=\Big(-\frac{2r_e\,m_S^2}{\mu^2r},\;-\frac{m_S^2}{\sqrt{\varepsilon_\perp}},\;1\Big)^{\!\top},
  \qquad
  \partial_{r_*}^2\bm Z_0=V\bm Z_0 ,
  \label{eq:Z0H}
\end{equation}
in the basis $(\Psi_g,\Psi_{\rm H},\Xi)$. It has $\Psi_e=0$, hence $E=0$, and $\Psi_g$ and $\Xi$ are those of \eqref{eq:Z0mouth} without the coupling to $h_{a\phi}$. On it $\mathcal L_{\mathcal R\mathcal R}$ does not act. In the row of $\Psi_g$ the factors $\sqrt{\varepsilon_\perp}$ cancel, and the row reduces to $2A-rA'=\hat V_{gg}/r-\mu^2$, an identity for the background \eqref{eq:hllmouthax}. In the basis $(\Psi_g,\Psi_e,\Xi)$ the row of $\Psi_e$ reduces to a relation between $\Psi_g$ and $\Xi$ whose two coefficients carry the same factor $\varepsilon_\perp/\varepsilon_\parallel$, and $\bm Z_0$ satisfies it as at $\eta_{\rm H}=0$. The row of $\Xi$ holds since its source $\Psi_e$ vanishes. The vector $\bm Z_0$ is bounded at both ends, since $\varepsilon_\perp\ge1-\eta_{\rm H}/2>0$ for $r\ge r_e$. At $r=r_e$ it is the throat null vector. The adjoint zero mode is again $\Sigma^2\bm Z_0/(\lhat m_S^2)$. The higher Landau levels therefore do not move the neutral channel off the threshold, and the weight $\int\bm L_0\,\delta V\,\bm Z_0\,\dd r_*$ of Sec.~\ref{sec:neutralstab} vanishes for them. The finite-element computation of Appendix~\ref{app:numerics}, repeated with \eqref{eq:VhatH}, finds no negative eigenvalue for the light system at $l=2,3,4$ and for the three-channel system at $l=2$ and $m_S^2=1$, $10^2$ and $10^4$, at $\eta_{\rm H}=1/12\pi^2$ and at $\eta_{\rm H}=0.05$. For the light system the elimination of the screened channel leaves the neutral variable $\tilde\Xi/\sqrt{\varepsilon_\perp}$ with the mass term $4r_e^2\varepsilon_\perp/r$ and the extra term $\tfrac14(\partial_{r_*}\log\varepsilon_\perp)^2-\tfrac12\partial_{r_*}^2\log\varepsilon_\perp$, which we keep. The lowest eigenvalue is $6.7\times10^{-10}/r_e^2$ for the light system at $l=2$ and $\eta_{\rm H}=1/12\pi^2$, against $8.0\times10^{-10}/r_e^2$ for the empty box, the same pattern as at $\eta_{\rm H}=0$. At $m_S^2=0$ the $(\Psi_g,\Psi_{\rm H})$ block of $\Sigma\hat V\Sigma^{-1}$ is positive definite for $r\ge r_e$, $l=2,\dots,10$ and $\eta_{\rm H}=0$, $1/12\pi^2$, $0.05$ and $0.1$. Its lowest eigenvalue takes its minimum at $r=r_e$ for every $l$ examined. For these values of $l$ and $\eta_{\rm H}$ the massive channels of the mouth at $m_S^2=0$ therefore have no growing mode at this order.

\subsection{Isospectrality}
\label{sec:hll-iso}

The one-loop action breaks the isospectrality of the two sectors even without the lowest Landau level. At $m_S^2=0$ the throat masses \eqref{eq:mHax} of this sector and the masses of the companion sector, $l(l-1)[1+\eta_{\rm H}(l+1)/(2l+1)]$ and $(l+1)(l+2)[1+\eta_{\rm H}l/(2l+1)]$ \cite{Sadhukhan:2026czl}, agree only if $c_\parallel=1$. The polar-metric sector changes $|F|$ and probes $\mathcal L_{\mathcal Q\mathcal Q}$. This sector changes $\bm E\cdot\bm B$ and probes $\mathcal L_{\mathcal R\mathcal R}$. The masses agree only for $\mathcal L_{\mathcal R\mathcal R}=\mathcal L_{\mathcal Q\mathcal Q}$ on the background, and the one-loop action of a charged fermion does not have this property. The mouths show the same breaking. The master potentials of the two sectors \cite{Nomura:2020tpc}, which extend those of \cite{Moreno:2002gg} to Lagrangians that depend on $\mathcal R$, contain $\mathcal L_{\mathcal Q\mathcal Q}$ in one and $\mathcal L_{\mathcal R\mathcal R}$ in the other. This is the kind of breaking found for one-loop corrections to charged black holes \cite{delRio:2024kms}. At $g_c=1$ the squared masses of the two sectors differ by $0.025\,(l-1)l(l+1)/(2l+1)$ and $0.025\,l(l+1)(l+2)/(2l+1)$, with $0.025=(1-c_\parallel)\eta_{\rm H}$. These differences exceed every $O(\alpha)$ difference of Table~\ref{tab:shifts} by a factor of order $q$.

\section{Conclusion and discussion}
\label{sec:summary}
The axial-metric/polar-gauge sector of the MMP wormhole reduces in both regions to coupled master equations with vanishing first-derivative matrix at $\alpha=0$ and frequency-independent potentials. In the mouth the decoupled potentials are Chandrasekhar's $V^{(-)}_{1,2}$, derived here on the magnetic background, and electric-magnetic duality is an output.

The LLL fermions respond in this sector through their charge. The exact Schwinger current survives in the mouth and gives the longitudinal electric field the MMP screening mass $m_S^2=\alpha q^2$ along the field lines. In the throat, and in the mouth at $O(\alpha)$, the LLL stress tensor tilts into the drift lift, and the polarisation currents $J^\theta$ and $J^a_{\rm pol}$ appear as the counterparts of the Hall current. Given the drift lift, the $\tau,\rho,\theta$ components of the force balance hold identically, the $\phi$ component determines $J^\theta$ and charge conservation fixes $J^a_{\rm pol}$ up to the Schwinger piece and the self-conserved curls that we set to zero. The drift lift rests on assumptions. The tilted worldsheet is the timelike invariant plane of $F$, and the lift is the local Lorentz transformation of the background stress tensor onto it. We assume that the Casimir stress tensor follows this plane, that its response is local and without derivatives, that the twist acts like the drift, and that no further self-conserved current is added. The axial metric perturbation also couples to every LLL state, exactly as the worldsheet gauge field $-\lhat h_a/4R^2$. This is an $O(\lhat/q)$ effect of the order of the $O(\alpha)$ terms, and it couples the gravitational channel to $\Xi$.

With these sources the Bianchi identities close identically at $O(\alpha)$ and the three-channel master system \eqref{eq:schrodinger} is obtained in closed form. At $m_S^2=0$ it is block triangular, and its Einstein--Maxwell--Casimir block is spanned by the combinations \eqref{eq:Zexplicit}. The anomaly part of the LLL stress tensor does not vanish in the mouth. The mouth therefore carries the corrected background, the drift lift and the polarisation currents at $O(\alpha)$, and there the Bianchi identity rejects a missing response. With the full response the coefficients that both regions determine agree, the drift terms \eqref{eq:arho2ax} of order $\alpha\rho^2$ included. The uncorrected mouth fails at that order. The full matrices $\tilde Q$ agree at the fiducial junction to $1.5\times10^{-11}$ in all three channels, which is $(\alpha\rho_{\rm J})^2$ times a number below ten. This includes the entries of the coupling, and the $O(\alpha)$ terms of the mouth are of relative size $8\times10^{-30}$ there. The control experiments delimit the logic. Consistency verifies the closure and pairs $\delta T^Q$ with the currents, but it cannot select the constitutive relation. Covariance fixes the family \eqref{eq:family}, and the kernel $\mathsf K$ is fixed by the assumption of Sec.~\ref{sec:driftderivation}, not by consistency. At $m_S^2=0$ the drift imprint on the throat is the matrix \eqref{eq:driftdiff} together with $\alpha\Om^2V_{1b}$. At finite $m_S^2$ the neutral channel depends on the closure as well. The mouth carries the same imprint \eqref{eq:arho2diff}, and the matching cannot separate the closures.

At $\alpha=0$ the throat is three Poschl-Teller channels. At finite $m_S^2$ there is one screened mode, one light massive mode with $\lambda_-\to l^2+l+2$ and an irrational offset $s_-$, and one exactly massless neutral mode, the gravitationally dressed WZW sector of MMP's Appendix~A. At $m_S^2=0$, a formal point at which the Schwinger sector decouples, the masses are the integers $l(l-1)$ and $(l+1)(l+2)$ of the polar sector. There the $O(\alpha)$ shifts of the massive levels of the closed throat are real on every level. On a level of a single channel they are a rational number plus $2\Om_0^2\log2$. The levels shared by $Z_+$ and $Z_-$ split through a real $2\times2$ problem whose off-diagonal product is positive by \eqref{eq:Sentries}. On the levels computed the higher Landau levels split these levels first and push the mixing to second order. At the physical $m_S^2$ the left-right projection of the $O(\alpha)$ potential on the light massive channel stays finite and matches the mouth, and its first-order shifts are real on every level below the screened channel. On the levels computed all these shifts are negative for the clock of the static observer at the centre of the throat. The frequencies at infinity need $\ell$ to relative order $\alpha$, which the matching does not fix \cite{Sadhukhan:2026czl}. On the lowest light levels at $g_c\sim1$ the finite value of $m_S^2$ lowers the levels by far more than the $O(\alpha)$ terms. The coupling to $h_{a\phi}$ raises them by the fraction $g_c^2/4\pi^2$ of that amount, which still exceeds the $O(\alpha)$ terms. Virtual transitions to the higher Landau levels, which we have not estimated, could enter at the same order $1/q$. The one-loop action of the higher levels raises the offset $s_-$ by $\eta_{\rm H}\lhat/\sqrt{4\lhat+9}$, far more than all of these. It also enlarges the AdS$_2$ radius, and on the clock of the static observer the light levels move down. The coupling keeps the neutral mode exactly massless.

Isospectrality of the two parity sectors is exact in the fermion-free Einstein--Maxwell limit, in the mouth through Chandrasekhar's Darboux pair \eqref{eq:susyform} and in the throat through identical Poschl-Teller potentials. It is broken at $\alpha^0$ by the Schwinger sector of the axial throat, which adds a massless channel and moves the masses to $\{0,\lambda_\pm\}$ while the polar masses stay at $\{l(l-1),(l+1)(l+2)\}$. It is broken in the mouth by the neutral channel, which has no polar counterpart and makes one eigenvalue of the potential matrix negative between $r_e$ and $2r_e$. The one-loop action of the higher Landau levels breaks it in both regions at $O(\eta_{\rm H})$, since its response to an electric field along the field lines, $c_\parallel\eta_{\rm H}$, differs from its response to a change of $|F|$, $\eta_{\rm H}$. At the formal point $m_S^2=0$ it is broken at $O(\alpha)$ in the common Einstein--Maxwell--Casimir block by the rational differences of Sec.~\ref{sec:isothroat0}. With the scaled lift of the polar sector the two sectors mix the modes of a shared level with the same strength, and the breaking sits in the diagonal entries. The mixing comes from the drift here and from the Hall current there, and the two act identically in the variables \eqref{eq:dualmap}. Isospectrality follows from Chandrasekhar's transformation theory and survives only in the fermion-free limit. The fermions couple to $A_\mu$ and not to its dual. The polar gauge field is a worldsheet gauge field for them and the axial one is not, and the parities see different fermionic operators. This is the conclusion reached for one-loop corrections to charged black holes in \cite{delRio:2024kms}, here with a fermionic backreaction that creates the throat itself.

The radial null-energy contraction is untouched at $O(\epsilon)$. In the closed throat the massive channels have no growing normal mode for $l\ge2$ at $\alpha=0$, for every $m_S^2$. At $O(\alpha)$ the first-order matrices of the massive channels at $m_S^2=0$ have real eigenvalues on every level. First-order perturbation theory needs $\alpha|\varsigma|\ll\Om_0$, which on the levels computed amounts to $\alpha\Om_0\log\Om_0\ll1$. The light massive channel at the physical $m_S^2$ has real first-order shifts on every level below the screened channel. These normal modes are those of the closed throat, with the interval and the normalisable conditions of the $\alpha=0$ problem. The positivity that excludes modes with real $p$ is shown for $\alpha\rho\le0.1$ and $l=2,3,4$. Transmission through the mouths is not included, and the stability statements are made region by region and not with a single master variable for the whole wormhole. In the mouth two eigenvalues of the symmetrised potential are positive. The finite-element computation finds no negative eigenvalue for the light system at $l=2,3,4$ and for the three-channel system at $l=2$ and $m_S^2\le10^4$, at the resolution $10^{-9}/r_e^2$ of the box. We have not analysed the massive channels of the $O(\alpha)$ mouth. The neutral channel has the exact zero-energy solution \eqref{eq:Z0mouth}, a persistent current along the field lines, and it sits at the threshold for a bound state through the $O(\alpha)$ terms of the mouth. The throat interior is attractive at the next order. The one-loop action of the higher Landau levels leaves the zero-energy solution exact, and with it the finite-element computation finds no negative eigenvalue (Sec.~\ref{sec:hll-mouth}). In the throat the two corrections add at first order. Together they keep the symmetric part of the potential positive on the window and the shifts of every shared level real (Sec.~\ref{sec:hll-throat}). The $O(\alpha^2)$ terms of the mouth and the effect of the ambient region are not known, and the stability of this channel is open. A bound state at that order would grow on time scales beyond $10^{41}$\,s at the fiducial parameters.

These results rest on a small number of assumptions beyond linearised Einstein--Maxwell theory. We assume $q\gg1$. Corrections to the one-loop stress tensor and current of the lowest level are then suppressed by powers of $1/q$. Only the lowest Landau level responds as a dynamical sector. Higher levels are gapped by $\sim\sqrt q/R$ and the Schwinger mass lies below that gap by the factor $g_c/2\pi$. The truncation also needs $l\ll\sqrt q$, as in \cite{Sadhukhan:2026czl}. Transitions of LLL states to the higher levels induced by the perturbation are neglected. The background $t_{ab}$ is the Casimir tensor on the circle $L_c=\pi$, transported by the anomaly, and corrections of order $d/\ell$ from the length $d$ of the ambient part of each line are dropped. Each mouth is treated as an isolated static mouth. The field of the other mouth and the orbital motion of MMP's Sec.~6 are neglected, which requires $r\ll d$. The orbit that holds the mouths apart in MMP has frequency $\Omega_{\rm orb}=2\sqrt{r_e/d^3}$, and their condition $\Omega_{\rm orb}\ell\ll1$ requires $d\gg(4r_e\ell^2)^{1/3}$, about $6\times10^{23}$\,cm for the fiducial parameters (their eqs.~(6.50) and (6.51)). The finite-element box ends at $r=10^4r_e$, still below $d$ by a factor above $10^{19}$. The orbital motion is slower than every throat level. Its Unruh-like temperature, of order $\Omega_{\rm orb}$, lies far below the level spacing $1/\ell$ and does not excite the throat modes. These arguments do not cover the neutral threshold state of Sec.~\ref{sec:neutralstab}. It sits at zero frequency and does not decay at large $r$. It therefore reaches the region $r\sim d$, and we have not estimated how the other mouth and the orbit act on it. In the mouth the Casimir part is $O(\alpha^3)$ and only the anomaly part is kept. The higher Landau levels enter through the one-loop Euler-Heisenberg action of one massless Dirac fermion, with $g_c$ defined at the field of the throat (Sec.~\ref{sec:hll}). We keep it to first order in $\eta_{\rm H}$. We do not compute the cross terms of order $\alpha\eta_{\rm H}$, the two-loop terms, which are of relative order $\eta_{\rm H}$, or the derivative corrections of the action, which are of relative order $1/q$ and $l(l+1)/q$. The action holds below the gap of the higher levels. The finite-element runs with this action use the box of Sec.~\ref{sec:neutralstab}, with $2000$ nodes for both systems. The LLL stress tensor is taken to follow the tilted worldsheet locally, which with general covariance fixes the drift lift $\mathsf K=\bar T-\mathcal P$ of Sec.~\ref{sec:driftderivation}. Its corrections from the finite width of the LLL states are not computed. The Bianchi identities alone do not fix the kernel, as control (e) shows. The state of each closed field line is the antiperiodic vacuum with vanishing holonomy, which is invariant under charge conjugation and gives $\delta t_{ab}=0$. The Schwinger response is the exact two-dimensional anomaly response on the exact LLL states of Sec.~\ref{sec:kk}, with the width factor \eqref{eq:widthfactor} absorbed into $m_S^2$. It is identically conserved and therefore not forced by consistency, as control (d) shows, and its normalisation is anchored to MMP Appendices A and G. We work to first order in $\epsilon$, in $\alpha$ and in $\eta_{\rm H}$ with $m_S^2$ unexpanded, for $l\ge2$ and $m=0$, and we keep the coupling to $h_{a\phi}$ exactly at $\alpha^0$. The sectors $l=0$ and $l=1$ are not treated.

Several features of this sector would imprint on the echo structure of \cite{Mondal:2025tht}. They are the finite-$m_S^2$ deformation of the throat spectrum, the loss of reflectionlessness, the mass $l^2+l+2$ with its corrections and the massless neutral channel threading the wormhole. At the fiducial parameters these echoes are of structural interest only (Sec.~\ref{sec:intro}). At $\alpha=0$ the neutral channel is free along the whole throat and sees only the mouth structure of Sec.~\ref{sec:mouthneutral}, and its transmission through a mouth is a clean scattering problem. A Landau-level computation of $\langle T_{a\phi}\rangle$ in crossed fields, or an effective-action derivation parallel to that of the Hall current in \cite{Sadhukhan:2026czl}, would test the assumption of Sec.~\ref{sec:driftderivation}. The observable at stake is the shift of the light levels at the physical $m_S^2$ that \eqref{eq:lam1minusdrift} and $\lambda^-_{1b}$ produce, between $3.9$ and $9.7$ in $\varsigma$ on the levels of Table~\ref{tab:lightshifts}.

\appendix

\section{Radial equations of the throat at \texorpdfstring{$\alpha=0$}{alpha=0}}
\label{app:throat0}
These are the six radial equations on exact AdS$_2\times S^2$, with $f\equiv1+\rho^2$. In this appendix and the next a prime denotes $\partial_\rho$. The equations of this appendix and of Appendix~\ref{app:alpha} are written without the coupling of Sec.~\ref{sec:kk}. The coupling adds the source \eqref{eq:Tchi} to $\mathcal E_{\tau\phi}$ and $\mathcal E_{\rho\phi}$ and the $h_a$ terms of \eqref{eq:chieqkk} to the $\chi$ equation. Its effect on the master system is $\delta\mathcal N$ of \eqref{eq:dN} and the $O(\alpha/q)$ terms of Appendix~\ref{app:matrices}.
\begin{align}
2r_e^{2}\,\mathcal{E}_{\tau\phi} &= \big(\lhat + 2\big)\,h_0 - f\,h_0'' + p f\,h_1' + 2 r_e^{2}\,w_0
\end{align}
\begin{align}
2r_e^{2}f\,\mathcal{E}_{\rho\phi} &= -p\,h_0' + \big(p^{2} + f \big(\lhat + 2\big)\big)\,h_1 + 2 r_e^{2} f\,w_1
\end{align}
\begin{align}
2r_e^{2}f\,\mathcal{E}_{\theta\phi} &= p\,h_0 - 2 \rho f\,h_1 - f^{2}\,h_1'
\end{align}
\begin{align}
4r_e^{2}f\,\mathcal{M}^{\tau} &= 2 \lhat\,h_0 + \lhat r_e^{2}\,w_0 - r_e^{2} f\,w_0'' + p r_e^{2} f\,w_1'
\end{align}
\begin{align}
4r_e^{2}\,\mathcal{M}^{\rho} &= -2 \lhat f\,h_1 + p r_e^{2}\,w_0' - r_e^{2} \big(p^{2} + \lhat f\big)\,w_1
\end{align}
\begin{align}
4r_e^{2}f\,\mathcal{M}^{\theta} &= 2 p\,h_0 - 4 \rho f\,h_1 - 2 f^{2}\,h_1' + p r_e^{2}\,w_0 - 2 \rho r_e^{2} f\,w_1 - r_e^{2} f^{2}\,w_1'
\end{align}
Steps 1--3 of the reduction give
\begin{align}
h_0 &= \frac{2 \rho f}{p}\,h_1 + \frac{f^{2}}{p}\,h_1'
\end{align}
\begin{align}
w_1 &= -\frac{2}{r_e^{2}}\,h_1 - \frac{p}{\lhat f}\,E
\end{align}
\begin{align}
w_0 &= -\frac{4 \rho f}{p r_e^{2}}\,h_1 - \frac{2 f^{2}}{p r_e^{2}}\,h_1' - \frac{f}{\lhat}\,E'
\end{align}
\begin{align}
h_1'' &= \frac{p^{2} + \big(\lhat - 8\big) f + 4}{f^{2}}\,h_1 - \frac{6 \rho}{f}\,h_1' - \frac{2 p r_e^{2}}{\lhat f^{2}}\,E
\end{align}
\begin{align}
E'' &= -\frac{2 \lhat \big(\lhat - 2\big)}{p r_e^{2}}\,h_1 + \frac{p^{2} + f \big(\lhat + 4\big)}{f^{2}}\,E - \frac{2 \rho}{f}\,E'
\end{align}
The closed first-order system $\bm y'=\mathsf M\bm y$ of Step~5, $\bm y=(h_1,h_1',E,E')^{\!\top}$, is read off from the last two equations. $\mathsf M_{21}$, $\mathsf M_{22}$ and $\mathsf M_{23}$ are the coefficients of $h_1$, $h_1'$ and $E$ in the $h_1''$ equation, $\mathsf M_{41}$, $\mathsf M_{43}$ and $\mathsf M_{44}$ those of $h_1$, $E$ and $E'$ in the $E''$ equation, $\mathsf M_{12}=\mathsf M_{34}=1$, and the other entries vanish. The second-order system $\bm Y''+P\bm Y'+Q\bm Y=0$ for $\bm Y=(h_1,E)^{\!\top}$ has $P=-\mathrm{diag}(\mathsf M_{22},\mathsf M_{44})$ and $Q_{ij}=-\mathsf M_{2i,2j-1}$.

With the Schwinger response the modified Maxwell equations and the $\chi$ equation are
\begin{align}
4r_e^{2}f\,\mathcal{M}^{\tau} &= 2 \lhat\,h_0 + \lhat r_e^{2}\,w_0 - r_e^{2} f\,w_0'' + p r_e^{2} f\,w_1' - m_S^2 r_e^{2} f\,X'
\end{align}
\begin{align}
4r_e^{2}\,\mathcal{M}^{\rho} &= -2 \lhat f\,h_1 + p r_e^{2}\,w_0' - r_e^{2} \big(p^{2} + \lhat f\big)\,w_1 + m_S^2 p r_e^{2}\,X
\end{align}
\begin{align}
f\,\mathcal{X} &= -f\,E - p^{2}\,X + 2 \rho f\,X' + f^{2}\,X''
\end{align}
The reduction then gives the same $h_0$ and
\begin{align}
w_1 &= -\frac{2}{r_e^{2}}\,h_1 - \frac{p}{\lhat f}\,E + \frac{m_S^2 p}{\lhat f}\,X
\end{align}
\begin{align}
w_0 &= -\frac{4 \rho f}{p r_e^{2}}\,h_1 - \frac{2 f^{2}}{p r_e^{2}}\,h_1' - \frac{f}{\lhat}\,E' + \frac{m_S^2 f}{\lhat}\,X'
\end{align}
\begin{align}
h_1'' &= \frac{p^{2} + \big(\lhat - 8\big) f + 4}{f^{2}}\,h_1 - \frac{6 \rho}{f}\,h_1' - \frac{2 p r_e^{2}}{\lhat f^{2}}\,E + \frac{2 m_S^2 p r_e^{2}}{\lhat f^{2}}\,X
\end{align}
\begin{align}
E'' &= -\frac{2 \lhat \big(\lhat - 2\big)}{p r_e^{2}}\,h_1 + \frac{p^{2} + f \big(\lhat + m_S^2 + 4\big)}{f^{2}}\,E - \frac{2 \rho}{f}\,E' - \frac{4 m_S^2}{f}\,X
\end{align}
\begin{align}
X'' &= \frac{1}{f}\,E + \frac{p^{2}}{f^{2}}\,X - \frac{2 \rho}{f}\,X'
\end{align}

\section{Radial equations and reduction of the throat at \texorpdfstring{$O(\alpha)$}{O(alpha)}}
\label{app:alpha}
Each equation carries on the left the factor that clears the denominators of its $\alpha^0$ part, which is the corresponding equation of Appendix~\ref{app:throat0} with its Schwinger terms. Displayed is that factor times $\partial_\alpha\mathcal E|_{\alpha=0}$.
\begin{align}
2r_e^{2}\,\partial_\alpha\mathcal{E}_{\tau\phi} &= -\frac{4 \lhat + 20 + 3 \rho \big(\lhat + 6\big)\,\arctan\rho}{3}\,h_0 + \frac{4 \rho^{2} + 3 \rho \big(f + 2\big)\,\arctan\rho - 3\,\log f}{3}\,h_0'' \nonumber\\
&\quad + p\big(\rho + f\,\arctan\rho\big)\,h_1 - \frac{p}{3}\Big[4 \rho^{2} + 3 \rho \big(f + 2\big)\,\arctan\rho - 3\,\log f\Big]\,h_1' \nonumber\\
&\quad - \frac{r_e^{2}}{f}\big(2 f + 1 + 2 \rho f\,\arctan\rho\big)\,w_0
\end{align}
\begin{align}
2r_e^{2}f\,\partial_\alpha\mathcal{E}_{\rho\phi} &= \frac{p}{f}\big(\rho + f\,\arctan\rho\big)\,h_0 - \frac{p}{3 f}\Big[4 \rho^{2} + 3 \rho \big(f + 2\big)\,\arctan\rho - 3\,\log f\Big]\,h_0' \nonumber\\
&\quad + \frac{1}{3 f}\Big[4 \rho^{2} p^{2} - 2 f \big(\big(2 \lhat + 10\big) f - 3\big) + 3 \rho \big(\big(\rho^{2} + 3\big) p^{2} - f^{2} \big(\lhat + 6\big)\big)\,\arctan\rho \nonumber\\
&\quad - 3 p^{2}\,\log f\Big]\,h_1 - r_e^{2}\big(2 f - 1 + 2 \rho f\,\arctan\rho\big)\,w_1
\end{align}
\begin{align}
2r_e^{2}f\,\partial_\alpha\mathcal{E}_{\theta\phi} &= \frac{p}{3 f}\Big[4 \rho^{2} + 3 \rho \big(f + 2\big)\,\arctan\rho - 3\,\log f\Big]\,h_0 + \frac{f}{3}\big(11 \rho + 9 f\arctan\rho\big)\,h_1 \nonumber\\
&\quad + \frac{f}{3}\Big[4 \rho^{2} + 3 \rho \big(f + 2\big)\,\arctan\rho - 3\,\log f\Big]\,h_1'
\end{align}
\begin{align}
4r_e^{2}f\,\partial_\alpha\mathcal{M}^{\tau} &= \frac{\lhat}{3 f}\Big[2 f - 11 + 12 \rho\,\arctan\rho - 6\,\log f\Big]\,h_0 \nonumber\\
&\quad + \frac{\lhat r_e^{2}}{6 f}\Big[10 f - 11 + 6 \rho \big(f + 2\big)\,\arctan\rho - 6\,\log f\Big]\,w_0 - r_e^{2}\big(\rho + f\arctan\rho\big)\,w_0' \nonumber\\
&\quad - \frac{r_e^{2} f}{3}\big(4 + 3 \rho\,\arctan\rho\big)\,w_0'' + p r_e^{2}\big(\rho + f\,\arctan\rho\big)\,w_1 + \frac{p r_e^{2} f}{3}\big(4 + 3 \rho\,\arctan\rho\big)\,w_1'
\end{align}
\begin{align}
4r_e^{2}\,\partial_\alpha\mathcal{M}^{\rho} &= \frac{\lhat}{3}\Big[14 f - 11 + 12 \rho \big(f + 1\big)\,\arctan\rho - 6\,\log f\Big]\,h_1 \nonumber\\
&\quad + \frac{p r_e^{2}}{3}\big(4 + 3 \rho\,\arctan\rho\big)\,w_0' \nonumber\\
&\quad - \frac{r_e^{2}}{6}\Big[8 p^{2} - \lhat \big(6 f - 11\big) + 6 \rho \big(p^{2} - \lhat \big(f + 2\big)\big)\,\arctan\rho + 6 \lhat\,\log f\Big]\,w_1
\end{align}
\begin{align}
4r_e^{2}f\,\partial_\alpha\mathcal{M}^{\theta} &= \frac{p}{3 f}\Big[2 f - 11 + 12 \rho\,\arctan\rho - 6\,\log f\Big]\,h_0 + \frac{4 f}{3}\big(10 \rho + \big(9 f - 3\big)\arctan\rho\big)\,h_1 \nonumber\\
&\quad + \frac{f}{3}\Big[14 f - 11 + 12 \rho \big(f + 1\big)\,\arctan\rho - 6\,\log f\Big]\,h_1' \nonumber\\
&\quad + \frac{p r_e^{2}}{6 f}\Big[10 f - 11 + 6 \rho \big(f + 2\big)\,\arctan\rho - 6\,\log f\Big]\,w_0 + 3 r_e^{2} f\big(\rho + f\,\arctan\rho\big)\,w_1 \nonumber\\
&\quad + \frac{r_e^{2} f}{6}\Big[6 f - 11 + 6 \rho \big(f + 2\big)\,\arctan\rho - 6\,\log f\Big]\,w_1'
\end{align}
\begin{align}
f\,\partial_\alpha\mathcal{X} &= -\frac{p^{2}}{3 f}\Big[4 \rho^{2} + 3 \rho \big(f + 2\big)\,\arctan\rho - 3\,\log f\Big]\,X - \frac{f}{3}\big(11 \rho + 9 f\arctan\rho\big)\,X' \nonumber\\
&\quad - \frac{f}{3}\Big[4 \rho^{2} + 3 \rho \big(f + 2\big)\,\arctan\rho - 3\,\log f\Big]\,X''
\end{align}
The eliminated amplitudes and the second-order equations follow order by order. Their $\alpha^0$ parts are those of Appendix~\ref{app:throat0} with the Schwinger terms, and their $\alpha$ derivatives are
\begin{align}
\partial_\alpha h_0 &= -\frac{1}{3 p}\Big[\rho \big(19 f - 8\big) + \big(15 f^{2} + 6 f - 12\big)\arctan\rho - 6 \rho\,\log f\Big]\,h_1 \nonumber\\
&\quad - \frac{2 f}{3 p}\Big[4 \rho^{2} + 3 \rho \big(f + 2\big)\,\arctan\rho - 3\,\log f\Big]\,h_1'
\end{align}
\begin{align}
\partial_\alpha w_1 &= \frac{8 + 6 \rho\,\arctan\rho}{3 r_e^{2}}\,h_1 - \frac{p}{6 \lhat f^{2}}\Big[14 f - 11 + 12 \rho \big(f + 1\big)\,\arctan\rho - 6\,\log f\Big]\,E \nonumber\\
&\quad + \frac{m_S^2 p}{6 \lhat f^{2}}\Big[6 f - 11 + 6 \rho \big(f + 2\big)\,\arctan\rho - 6\,\log f\Big]\,X
\end{align}
\begin{align}
\partial_\alpha w_0 &= \frac{1}{3 p r_e^{2}}\Big[2 \rho \big(27 f - 8\big) + \big(42 f^{2} - 24\big)\arctan\rho - 12 \rho\,\log f\Big]\,h_1 \nonumber\\
&\quad + \frac{2 f}{3 p r_e^{2}}\Big[12 f - 8 + 3 \rho \big(3 f + 4\big)\,\arctan\rho - 6\,\log f\Big]\,h_1' - \frac{\rho + f\arctan\rho}{\lhat}\,E \nonumber\\
&\quad + \frac{2 f - 11 + 12 \rho\,\arctan\rho - 6\,\log f}{6 \lhat}\,E' - \frac{m_S^2}{6 \lhat}\Big[10 f - 11 + 6 \rho \big(f + 2\big)\,\arctan\rho - 6\,\log f\Big]\,X'
\end{align}
\begin{align}
\partial_\alpha h_1'' &= \frac{1}{3 f^{3}}\Big[8 \rho^{2} p^{2} + 30 f^{2} - \big(4 \lhat - 30\big) f - 32 + 6 \rho \big(5 f^{2} + \big(\lhat + p^{2} - 8\big) f + 2 p^{2} + 8\big)\,\arctan\rho \nonumber\\
&\quad - \big(6 p^{2} + \big(3 \lhat - 36\big) f + 24\big)\log f\Big]\,h_1 \nonumber\\
&\quad + \frac{1}{f^{2}}\Big[4 \rho \big(f + 2\big) + \big(4 f^{2} - 6 f + 12\big)\arctan\rho + 6 \rho\,\log f\Big]\,h_1' \nonumber\\
&\quad - \frac{4 p r_e^{2}}{3 \lhat f^{3}}\Big[4 \rho^{2} + 3 \rho \big(f + 2\big)\,\arctan\rho - 3\,\log f\Big]\,E \nonumber\\
&\quad + \frac{2 m_S^2 p r_e^{2}}{3 \lhat f^{3}}\Big[4 f - 8 + 3 \rho \big(f + 4\big)\,\arctan\rho - 6\,\log f\Big]\,X
\end{align}
\begin{align}
\partial_\alpha E'' &= \frac{\lhat \big(\lhat - 2\big)}{3 p r_e^{2} f}\big(22 f + 3 + 18 \rho f\,\arctan\rho\big)\,h_1 \nonumber\\
&\quad + \frac{1}{6 f^{3}}\Big[\big(16 \rho^{2} - 6\big) p^{2} + f \big(\big(2 \lhat - 36\big) \rho^{2} - 9 \lhat - 8 m_S^2 - 80\big) \nonumber\\
&\quad + 12 \rho \big(-3 f^{2} + \big(\lhat + m_S^2 + p^{2} + 4\big) f + 2 p^{2}\big)\,\arctan\rho - \big(12 p^{2} + 6 f \big(\lhat + m_S^2 + 4\big)\big)\log f\Big]\,E \nonumber\\
&\quad - \frac{1}{3 f^{2}}\Big[3 \rho^{3} - 8 \rho + \big(3 f^{2} + 6 f - 12\big)\arctan\rho - 6 \rho\,\log f\Big]\,E' \nonumber\\
&\quad + \frac{m_S^2}{3 f^{3}}\Big[3 p^{2} + 2 f \big(14 f + 11\big) + 24 \rho^{3} f\,\arctan\rho + 12 f\log f\Big]\,X - \frac{m_S^2 \rho}{f^{2}}\,X'
\end{align}
\begin{align}
\partial_\alpha X'' &= \frac{4 \rho^{2} + 3 \rho \big(f + 2\big)\,\arctan\rho - 3\,\log f}{3 f^{2}}\,E + \frac{2 p^{2}}{3 f^{3}}\Big[4 \rho^{2} + 3 \rho \big(f + 2\big)\,\arctan\rho - 3\,\log f\Big]\,X \nonumber\\
&\quad + \frac{1}{3 f^{2}}\Big[\rho \big(3 f + 8\big) + \big(3 \rho^{4} + 9\big)\arctan\rho + 6 \rho\,\log f\Big]\,X'
\end{align}

\section{The \texorpdfstring{$O(\alpha)$}{O(alpha)} coefficient matrices}
\label{app:matrices}
These are the first-derivative matrix $\tilde P_1$ and the potential matrix $\tilde Q_1$ of the tortoise-form system \eqref{eq:tortoiseform} in the basis $(\Psi_g,\Psi_D,\Xi)$, with indices $(g,D,\Xi)$. The only nonzero entry of $\tilde P_1$ is
\begin{align}
\tilde P_{1,DD} &= -\frac{\rho}{f}
\end{align}
and the nonzero entries of $\tilde Q_1$ are
\begin{align}
\tilde Q_{1,gg} &= \frac{1}{3}\Big[\big(8 \lhat - 13\big) \rho^{2} + 4 \lhat - 5 + 3 \rho \big(\big(2 \lhat - 3\big) \rho^{2} + 4 \lhat - 7\big)\,\arctan\rho - \big(3 \lhat - 6\big)\log f\Big]
\end{align}
\begin{align}
\tilde Q_{1,gD} &= -\frac{20 f - 8 + 3 \rho \big(5 f + 4\big)\,\arctan\rho - 6\,\log f}{3}
\end{align}
\begin{align}
\tilde Q_{1,Dg} &= -\frac{\big(\lhat - 2\big)}{3}\Big[18 f - 5 + 3 \rho \big(5 f + 4\big)\,\arctan\rho - 6\,\log f\Big]
\end{align}
\begin{align}
\tilde Q_{1,DD} &= \frac{1}{6 f}\Big[6 p^{2} + f \big(\big(14 \lhat + 88\big) \rho^{2} + 9 \lhat + 68\big) + 12 \rho f \big(\big(\lhat + 6\big) f + \lhat + 4\big)\,\arctan\rho \nonumber\\
&\quad - 6 f \big(\lhat + 4\big)\,\log f\Big]
\end{align}
\begin{align}
\tilde Q_{1,D\Xi } &= \frac{\lhat m_S^2}{6}\Big[14 f - 5 + 12 \rho \big(f + 1\big)\,\arctan\rho - 6\,\log f\Big]
\end{align}
\begin{align}
\tilde Q_{1,\Xi D} &= \frac{8 f - 4 + 6 \rho \big(f + 1\big)\,\arctan\rho - 3\,\log f}{3}
\end{align}
\begin{align}
\tilde Q_{1,\Xi \Xi } &= \frac{m_S^2}{3}\Big[8 f - 4 + 6 \rho \big(f + 1\big)\,\arctan\rho - 3\,\log f\Big]
\end{align}
with $\tilde Q_{1,g\Xi}=\tilde Q_{1,\Xi g}=0$. The coupling of Sec.~\ref{sec:kk} adds to $\tilde Q$ the $\alpha^0$ terms $-f\,\delta\mathcal N$ of \eqref{eq:dN} and, at $O(\alpha/q)$, $\alpha/q$ times
\begin{align}
\delta\tilde Q_{1,g\Xi} &= \frac{\lhat m_S^2}{6}\Big[15\rho^3\arctan\rho+20\rho^2+27\rho\arctan\rho-6\log f+12\Big]
\end{align}
\begin{align}
\delta\tilde Q_{1,D\Xi} &= -\frac{\lhat m_S^2}{3}\Big[18\rho^3\arctan\rho+22\rho^2+30\rho\arctan\rho-6\log f+17\Big]
\end{align}
\begin{align}
\delta\tilde Q_{1,\Xi g} &= \frac{\mu^2}{6}\Big[15\rho^3\arctan\rho+20\rho^2+27\rho\arctan\rho-6\log f+12\Big]
\end{align}
\begin{align}
\delta\tilde Q_{1,\Xi D} &= -\frac{2}{3}\Big[9\rho^3\arctan\rho+12\rho^2+15\rho\arctan\rho-3\log f+8\Big]
\end{align}
with no $O(\alpha/q)$ term in $\delta\tilde Q_{1,\Xi\Xi}$ or in $\tilde P$. In the mouth the coupling replaces $\hat V$ by $\hat V+\delta\hat V$ in \eqref{eq:mouthalphamaster}. The $O(\alpha/q)$ terms are then those of the factor $A/r^3$ multiplying $\delta\hat V$, together with the term $(\alpha/q)\,2\lhat m_S^2r_e^2(r-r_e)^2\mathfrak w/(3r^8)$ in the $D\Xi$ entry of $\tilde Q$, with $\mathfrak w$ of \eqref{eq:Kmouth}. In both regions the coupling enters through the Schwinger action alone, on the corrected background. The reduction mixes it with the drift lift and the polarisation currents, as the factor $\mathfrak w$ of the drift kernel in this term shows. A change of the drift lift or of the polarisation currents caused by the coupling is not included. The Schr\"odinger-form corrections of \eqref{eq:schrodinger} in the same basis are $V_{1b}=\mathrm{diag}(0,-1/f,0)$ and
\begin{align}
(V_{1a})_{gg} &= -\frac{1}{3}\Big[\big(8 \lhat - 13\big) \rho^{2} + 4 \lhat - 5 + 3 \rho \big(\big(2 \lhat - 3\big) \rho^{2} + 4 \lhat - 7\big)\,\arctan\rho - \big(3 \lhat - 6\big)\log f\Big]
\end{align}
\begin{align}
(V_{1a})_{gD} &= \frac{1}{6}\Big[37 f - 13 + 6 \rho \big(5 f + 4\big)\,\arctan\rho - 12\,\log f\Big]
\end{align}
\begin{align}
(V_{1a})_{Dg} &= \frac{\big(\lhat - 2\big)}{6}\Big[39 f - 13 + 6 \rho \big(5 f + 4\big)\,\arctan\rho - 12\,\log f\Big]
\end{align}
\begin{align}
(V_{1a})_{DD} &= -\frac{1}{6 f}\Big[\big(14 \lhat + 88\big) f^{2} - \big(5 \lhat + 23\big) f + 6 + 12 \rho f \big(\big(\lhat + 6\big) f + \lhat + 4\big)\,\arctan\rho \nonumber\\
&\quad - 6 f \big(\lhat + 4\big)\,\log f\Big]
\end{align}
\begin{align}
(V_{1a})_{D\Xi } &= -\frac{\lhat m_S^2}{12}\Big[31 f - 13 + 24 \rho \big(f + 1\big)\,\arctan\rho - 12\,\log f\Big]
\end{align}
\begin{align}
(V_{1a})_{\Xi D} &= -\frac{1}{12}\Big[29 f - 13 + 24 \rho \big(f + 1\big)\,\arctan\rho - 12\,\log f\Big]
\end{align}
\begin{align}
(V_{1a})_{\Xi \Xi } &= -\frac{m_S^2}{3}\Big[8 f - 4 + 6 \rho \big(f + 1\big)\,\arctan\rho - 3\,\log f\Big]
\end{align}
with $(V_{1a})_{g\Xi}=(V_{1a})_{\Xi g}=0$. In the $Z_\pm$ basis \eqref{eq:Zexplicit} at $m_S^2=0$, with indices $(+,-)$,
\begin{align}
(V_{1a})_{++} &= -\frac{1}{6 f (2 l + 1)}\Big[\big(30 l^{3} + 118 l^{2} + 150 l + 50\big) f^{2} - \big(l + 2\big) \big(13 l^{2} + 15 l + 5\big) f + 6 l + 12 \nonumber\\
&\quad + 6 \rho f \big(2 l + 1\big) \big(\big(2 l^{2} + 7 l + 7\big) \rho^{2} + 4 l^{2} + 13 l + 11\big)\,\arctan\rho \nonumber\\
&\quad - 6 f \big(l + 1\big) \big(2 l + 1\big) \big(l + 2\big)\,\log f\Big]
\end{align}
\begin{align}
(V_{1a})_{+-} &= -\frac{\big(l - 1\big) \big(\big(2 l^{2} + 4 l + 1\big) f^{2} - 3 \lhat f - 6\big)}{6 f (2 l + 1)}
\end{align}
\begin{align}
(V_{1a})_{-+} &= -\frac{\big(l + 2\big) \big(\big(2 l^{2} - 1\big) f^{2} - 3 \lhat f - 6\big)}{6 f (2 l + 1)}
\end{align}
\begin{align}
(V_{1a})_{--} &= -\frac{1}{6 f (2 l + 1)}\Big[\big(30 l^{3} - 28 l^{2} + 4 l + 12\big) f^{2} - \big(l - 1\big) \big(13 l^{2} + 11 l + 3\big) f + 6 l - 6 \nonumber\\
&\quad + 6 \rho f \big(2 l + 1\big) \big(\big(2 l^{2} - 3 l + 2\big) \rho^{2} + 4 l^{2} - 5 l + 2\big)\,\arctan\rho - 6 l f \big(2 l + 1\big) \big(l - 1\big)\,\log f\Big]
\end{align}
with $V_{1b}$ in this basis displayed in Sec.~\ref{sec:schrodinger}.

The nonvanishing Step-4 residuals of the truncated closures (a) and (b) of Sec.~\ref{sec:controls} are
\begin{align}
\text{(a)}\quad\mathrm{Res}\,\mathcal{E}_{\tau\phi} &= -\frac{\alpha \rho}{\lhat f}\,E + \frac{\alpha}{\lhat}\,E' + \frac{\alpha m_S^2 \rho}{\lhat f}\,X - \frac{\alpha m_S^2}{\lhat}\,X'
\end{align}
\begin{align}
\text{(b)}\quad\mathrm{Res}\,\mathcal{E}_{\tau\phi} &= \frac{\alpha \rho}{\lhat f}\,E - \frac{\alpha}{\lhat}\,E' - \frac{\alpha m_S^2 \rho}{\lhat f}\,X + \frac{\alpha m_S^2}{\lhat}\,X'
\end{align}
and $\mathrm{Res}\,\mathcal M^\theta=0$ in both.

\section{Radial equations and matrices of the mouth}
\label{app:mouth}
These are the six radial equations of the Einstein--Maxwell mouth of Sec.~\ref{sec:mouthmaster}, each multiplied by the factor that clears its denominators, with $\Delta\equiv r^2-2GMr+r_e^2$ and $\lhat=l(l+1)$. A prime denotes $\partial_r$. The field equations are $\mathcal E_{\mu\nu}=0$ and $\mathcal M^\nu=0$.
\begin{align}
2r^{4}\,\mathcal{E}_{t\phi} &= \big(\lhat r^{2} - 4 GM r + 6 r_e^{2}\big)\,h_0 - r^{2} \Delta\,h_0'' + 2 p r \Delta\,h_1 + p r^{2} \Delta\,h_1' + 2 r^{2} r_e^{2}\,w_0
\end{align}
\begin{align}
2r^{4}\Delta\,\mathcal{E}_{r\phi} &= 2 p r^{5}\,h_0 - p r^{6}\,h_0' + \big(r^{6} p^{2} + \Delta \big(\big(\lhat - 2\big) r^{2} + 4 r_e^{2}\big)\big)\,h_1 + 2 r^{2} r_e^{2} \Delta\,w_1
\end{align}
\begin{align}
2r^{3}\Delta\,\mathcal{E}_{\theta\phi} &= p r^{5}\,h_0 - 2 \Delta \big(GM r - r_e^{2}\big)\,h_1 - r \Delta^{2}\,h_1'
\end{align}
\begin{align}
4\Delta\,\mathcal{M}^{t} &= 2 \lhat\,h_0 + \lhat r^{2}\,w_0 - 2 r \Delta\,w_0' - r^{2} \Delta\,w_0'' + 2 p r \Delta\,w_1 + p r^{2} \Delta\,w_1'
\end{align}
\begin{align}
4r^{4}\,\mathcal{M}^{r} &= -2 \lhat \Delta\,h_1 + p r^{6}\,w_0' - \big(r^{6} p^{2} + \lhat r^{2} \Delta\big)\,w_1
\end{align}
\begin{align}
4r^{5}\Delta\,\mathcal{M}^{\theta} &= 2 p r^{5}\,h_0 + 4 \Delta \big(r^{2} - 3 GM r + 2 r_e^{2}\big)\,h_1 - 2 r \Delta^{2}\,h_1' + p r^{7}\,w_0 - 2 r^{2} \Delta \big(GM r - r_e^{2}\big)\,w_1 \nonumber\\
&\quad - r^{3} \Delta^{2}\,w_1'
\end{align}
The closed first-order system $\bm y'=\mathsf M\bm y$ of Step~5, with $\bm y=(h_1,h_1',E,E')^{\!\top}$, has $\mathsf M_{12}=\mathsf M_{34}=1$, the nonzero entries
\begin{align}
\mathsf M_{21} &= \frac{1}{r^{2} \Delta^{2}}\Big[p^{2} r^{6} + \big(\lhat - 2\big) r^{4} - 2 GM \big(\lhat - 6\big) r^{3} + \big(\lhat r_e^{2} - 20 G^2M^2 - 12 r_e^{2}\big) r^{2} + 36 GM r_e^{2} r \nonumber\\
&\quad - 14 r_e^{4}\Big]
\end{align}
\begin{align}
\mathsf M_{22} &= \frac{2 r^{2} - 10 GM r + 8 r_e^{2}}{r \Delta}
\end{align}
\begin{align}
\mathsf M_{23} &= -\frac{2 p r^{4} r_e^{2}}{\lhat \Delta^{2}}
\end{align}
\begin{align}
\mathsf M_{41} &= -\frac{2 \lhat \big(\lhat - 2\big)}{p r^{6}}
\end{align}
\begin{align}
\mathsf M_{43} &= \frac{1}{r^{2} \Delta^{2}}\Big[r^{6} p^{2} + \Delta \big(\big(\lhat - 2\big) r^{2} + 6 r_e^{2}\big)\Big]
\end{align}
\begin{align}
\mathsf M_{44} &= -\frac{4 r^{2} - 6 GM r + 2 r_e^{2}}{r \Delta}
\end{align}
and all other entries zero.

With the Schwinger response of Sec.~\ref{sec:mouthS} the three Einstein equations and $\mathcal M^\theta$ are unchanged, and $\mathcal M^t$, $\mathcal M^r$ and the frequency-domain $\chi$ equation with residual $\mathcal X$ read
\begin{align}
4\Delta\,\mathcal{M}^{t} &= 2 \lhat\,h_0 + \lhat r^{2}\,w_0 - 2 r \Delta\,w_0' - r^{2} \Delta\,w_0'' + 2 p r \Delta\,w_1 + p r^{2} \Delta\,w_1' - m_S^2 \Delta\,X'
\end{align}
\begin{align}
4r^{4}\,\mathcal{M}^{r} &= -2 \lhat \Delta\,h_1 + p r^{6}\,w_0' - \big(r^{6} p^{2} + \lhat r^{2} \Delta\big)\,w_1 + m_S^2 p r^{4}\,X
\end{align}
\begin{align}
r^{3}\Delta\,\mathcal{X} &= -r^{3} \Delta\,E - p^{2} r^{5}\,X + 2 \Delta \big(GM r - r_e^{2}\big)\,X' + r \Delta^{2}\,X''
\end{align}
As in Appendix~\ref{app:throat0}, these equations are written without the coupling of Sec.~\ref{sec:kk}. Its effect on the master system is $\delta\hat V$ of \eqref{eq:dVmouth}.

\section{Numerical methods}
\label{app:numerics}
The first-order shifts of Sec.~\ref{sec:shifts} and the entries of Table~\ref{tab:shifts} are matrix elements of closed-form potentials between Poschl-Teller modes. We compute them by double-exponential quadrature \cite{Takahasi:1974} on $[-\pi/2,0]$ and $[0,\pi/2]$ in 30-digit arithmetic. The remap term is reduced to one quadrature by parts, as in the numerical appendix of \cite{Sadhukhan:2026czl}. Each diagonal entry is identified as $a+b\log2$ with rational $a$ and $b$ by the PSLQ algorithm \cite{Ferguson:1999}, with tolerance $10^{-22}$ and coefficients below $10^{12}$, and every case gives $b=2\Om_0^2$. The off-diagonal entries are identified as rationals in the same way. The form holds on every level. At $m_S^2=0$ the diagonal entries of $V_{1a}$ are $m^2\log f$ plus an even rational function of $\rho$ plus $\rho\arctan\rho$ times another even rational function, with powers of $f$ as denominators. The off-diagonal entries contain no $\log f$. The proof in the numerical appendix of \cite{Sadhukhan:2026czl} then applies unchanged. Its exact evaluation reproduces both columns of Table~\ref{tab:shifts} and the PSLQ values on the highest levels computed, including the entry with the denominator $22309287$. The growth law of Sec.~\ref{sec:shifts} and the comparison of the products with the companion sector use the levels up to $\Om_0=20$ for $l=2$ and up to $\Om_0=15$ for $l=3$. On these levels $\mathcal S$ is computed in 40-digit arithmetic, with $[-\pi/2,\pi/2]$ cut into sixteen subintervals of equal length, and PSLQ uses tolerance $10^{-30}$ and integer coefficients up to $10^{20}$. It identifies every entry except the diagonal entry of $Z_+$ at $l=2$ and $\Om_0=20$. Recomputed in 60-digit arithmetic on 48 subintervals, that entry is identified with tolerance $10^{-45}$, and its rational part has the denominator $22309287$. The products of the companion sector are computed in the same way from the polar matrices of \cite{Sadhukhan:2026czl}, and on every one of these levels they equal the axial products to 25 digits. The exact evaluation reproduces \eqref{eq:Sentries} on every one of these levels. The polar entries use the polar matrices of \cite{Sadhukhan:2026czl}. The shifts of the light channel in Sec.~\ref{sec:lightchannel} use the same quadrature with $\psi_n=\cos^{s_-}x\,C_n^{(s_-)}(\sin x)$. For non-integer $s_-$ the inner integral of the remap term is not a polynomial, and we evaluate it by quadrature as well. A nested quadrature of the remap term agrees to twelve digits for $n=0$. The ratio of the potential of the light channel to $\lambda_-$ is evaluated on $2001$ points of $0\le\alpha\rho\le0.1$. Its second-order couplings are evaluated at $l=2$ and $\rho=0.7$ and $5$, in 60-digit arithmetic at $m_S^2=10^4$, $10^8$ and $10^{16}$. The zero of the potential of that channel beyond the validity window is bracketed on $400$ points of $0.01\le\alpha\rho\le1$ and refined by Brent's method \cite{Brent:1971}.

The direct solution of Sec.~\ref{sec:shifts} uses the finite-difference scheme of \cite{Sadhukhan:2026czl} with the $Z_\pm$ matrices of this paper, for $l=2$ at $m_S^2=0$, on $3\times10^4$ interior points with Dirichlet conditions at $\alpha\rho=\pm0.1$. As there, the potential is evaluated in two ways. The first uses the remap factor $1+2\alpha\rho\,\iota$ of \eqref{eq:remap}, with $\iota$ from the trapezoidal rule. The second uses $\rho(x)$ integrated from $\dd\rho/\dd x=f+\alpha\gamma$ by the eighth-order method of Prince and Dormand \cite{Prince:1981} to relative accuracy $10^{-12}$. The eigenvalues nearest each $\Om_0^2\le36$ are found by shift-and-invert Arnoldi iteration \cite{Lehoucq:1998}. For both forms $(\Om^2-\Om_0^2)/\alpha$ approaches the eigenvalues of $\mathcal S$ as $\alpha$ decreases. The linear extrapolation from $\alpha=3\times10^{-4}$ and $10^{-4}$ agrees with $\mathcal S$ to better than $0.1$ per cent, and no eigenvalue has an imaginary part.

The positivity scan of Sec.~\ref{sec:throatstab} evaluates the smallest eigenvalue of $V_{\rm s}$ on $2000$ equally spaced points of $0\le\alpha\rho\le0.1$ for each $l$, $\alpha$ and $|p|$. The first zero beyond the window is bracketed on $300$ points of $0.02\le\alpha\rho\le1$ and refined by Brent's method \cite{Brent:1971} to $10^{-7}$. The smallest eigenvalue of the symmetric part of $1+\alpha V_{1b}$ is taken on $2000$ points of $0\le\rho\le10$. For Sec.~\ref{sec:hll-throat} the scan is repeated with $\eta_{\rm H}=1/12\pi^2$, $0.05$ and $0.1$ added, in the eigenbasis of \eqref{eq:NHax} at $m_S^2=0$ and in the basis $Z_\pm$ of $\eta_{\rm H}=0$. With $\eta_{\rm H}>0$ the smallest ratio is never below its value $0.978$ at $\eta_{\rm H}=0$. At $\alpha=10^{-6}$ and $\eta_{\rm H}=1/12\pi^2$ the thresholds in the eigenbasis of \eqref{eq:NHax} are $0.267$, $0.281$ and $0.289$ for $l=2,3,4$. The light channel is scanned on the same grid with \eqref{eq:lam1minus}, and its first zeros at $\alpha=10^{-6}$ and $\eta_{\rm H}=1/12\pi^2$ are $0.250$, $0.274$ and $0.289$.

The junction comparison of Sec.~\ref{sec:matching} evaluates both sides of \eqref{eq:junctioncond} from the closed-form matrices in 120-digit arithmetic. The fiducial numbers follow from $r_e=\sqrt\pi\,q\,\ell_p/g_c$, $\ell=16\pi^{3/2}q^2\ell_p/g_c^3$, $\alpha=g_c^2/4\pi^2q$ and $\delta_M=-g_c^3\ell_p/512\pi^{3/2}q$. The scan over $\delta_{\rm J}$ uses the same evaluation. The entries of the coupling are evaluated in the same way, with $\delta\hat V$ on the mouth side and $\delta\mathcal N$ and the $O(\alpha/q)$ terms on the throat side. The relative size of the $O(\alpha)$ terms of the mouth quoted in Sec.~\ref{sec:mouthmaster} is taken entry by entry. It is the ratio of the $O(\alpha)$ part of each entry of $\tilde Q$ in \eqref{eq:mouthalphamaster} to its $\alpha^0$ part, with the mass shift included. We evaluate it at $p=0$, $l=2$ and $m_S^2=0$ and $\alpha q^2$, in 40-digit arithmetic at $y/r_e=10^{-6}$, $10^{-3}$, $0.5$ and $3$ and in 50-digit arithmetic at $y/r_e=0.4$.

The minimum of $V_{\min}$ quoted in Sec.~\ref{sec:mouthneutral} is bracketed on $10^5$ points of $r_e<r\le3r_e$ and refined by Brent's method. We use the $2\times2$ matrix of the light channels at $m_S^2\to\infty$ and the full $\hat V$ at $m_S^2=10^6$ and $10^8$, and the three results agree to five digits.

The finite-element computation of Sec.~\ref{sec:neutralstab} uses linear elements in the tortoise coordinate $r_*=r+2r_e\log(r/r_e-1)-r_e^2/(r-r_e)$ of the extremal mouth, on nodes spaced logarithmically in $y$. The symmetrised potential is evaluated at the midpoint in $r$ of each element, and the mass matrix is the consistent one. Both ends carry Dirichlet conditions. The light system uses $2000$ nodes on $10^{-5}\le y/r_e\le10^4$ and, as a check, $1200$ nodes on $10^{-4}\le y/r_e\le10^3$. The three-channel system uses $1300$ nodes on the first range. We compute the lowest eigenvalues of the generalised problem formed by the stiffness matrix, which includes the potential, and the mass matrix. We also count the negative eigenvalues of the stiffness matrix. The mass matrix is positive definite, and by Sylvester's law of inertia this count is the number of eigenvalues $\omega^2<0$ on the grid. On the first range the light system at $l=2$ has the lowest eigenvalue $7.0\times10^{-10}/r_e^2$, against $8.2\times10^{-10}/r_e^2$ for the empty box, and these are the values quoted in Sec.~\ref{sec:neutralstab}. The comparison of the light and three-channel eigenvalues in Sec.~\ref{sec:neutralstab} is done in 60-digit arithmetic for $l=2,3,4$ and $m_S^2=10^4$, $10^8$ and $10^{16}$, at $121$ values of $r-r_e$ spaced logarithmically between $10^{-8}r_e$ and $10^4r_e$. The largest relative difference of the two finite eigenvalues is $\lhat/m_S^2$ to within $0.2$ per cent. The move of the eigenvalues of $\Sigma\hat V\Sigma^{-1}$ under $\delta\hat V$ is evaluated in the same way, for $l=2,3,4$ and $m_S^2=1$, $10^2$ and $10^4$, and its largest relative size is $\lhat/q$. The single-channel test solves the same finite-element problem with the smallest eigenvalue of the symmetrised light potential as a single potential, with and without $|\partial_{r_*}\bm n|^2$. The eigenvector $\bm n$ is tabulated on $4\times10^4$ points and $|\partial_{r_*}\bm n|^2$ is obtained there by finite differences.

The finite-element runs of Sec.~\ref{sec:hll-mouth} use the same elements with the potential \eqref{eq:VhatH}, $\hat V_{DD}$ of \eqref{eq:VDDexact} and $GM=r_e(1+\eta_{\rm H})$. The tortoise coordinate $\int\dd r/A$ of \eqref{eq:hllmouthax} is integrated between neighbouring nodes by adaptive quadrature to relative accuracy $10^{-12}$. Both the light and the three-channel systems use $2000$ nodes on $10^{-5}\le y/r_e\le10^4$. At $\eta_{\rm H}=0$ these runs reproduce the lowest eigenvalue $7.0\times10^{-10}/r_e^2$ of the light system at $l=2$. The lowest eigenvalue is found by sparse shift-and-invert iteration at zero shift. At $\eta_{\rm H}=1/12\pi^2$ the lowest eigenvalues, in units of $10^{-10}/r_e^2$, are $6.66$, $7.32$ and $7.55$ for the light system at $l=2,3,4$. For the three-channel system at $m_S^2=1$, $10^2$ and $10^4$ they are $7.81$, $6.73$ and $6.66$, against $8.03$ for the empty box. At $\eta_{\rm H}=0.05$ they are $5.35$, $5.99$, $6.23$ and $7.16$, $5.49$, $5.36$, against $7.42$. The stiffness matrix has no negative eigenvalue in any of these runs. The third digit of these eigenvalues depends on the quadrature of the potential within each element at the level of half a per cent. The light potential with its extra term was checked against the Schur complement of the three-channel potential at $m_S^2=10^{10}$ at four points of the mouth. The positivity of the $m_S^2=0$ block uses $4000$ points with $r-r_e$ spaced logarithmically between $10^{-8}r_e$ and $10^4r_e$.

\bibliography{MMP_perturbations_ax}

@article{Morris:1988cz,
    author = "Morris, M. S. and Thorne, K. S.",
    title = "{Wormholes in space-time and their use for interstellar travel: A tool for teaching general relativity}",
    doi = "10.1119/1.15620",
    journal = "Am. J. Phys.",
    volume = "56",
    pages = "395--412",
    year = "1988"
}

@article{Gao:2016bin,
    author = "Gao, Ping and Jafferis, Daniel Louis and Wall, Aron C.",
    title = "{Traversable Wormholes via a Double Trace Deformation}",
    eprint = "1608.05687",
    archivePrefix = "arXiv",
    primaryClass = "hep-th",
    doi = "10.1007/JHEP12(2017)151",
    journal = "JHEP",
    number="151",
    volume = "12",
    pages = "151",
    year = "2017"
}

@article{Lenzi:2021wpc,
    author = "Lenzi, Michele and Sopuerta, Carlos F.",
    title = "{Master functions and equations for perturbations of vacuum spherically symmetric spacetimes}",
    eprint = "2108.08668",
    archivePrefix = "arXiv",
    primaryClass = "gr-qc",
    doi = "10.1103/PhysRevD.104.084053",
    journal = "Phys. Rev. D",
    volume = "104",
    number = "8",
    pages = "084053",
    year = "2021"
}

@article{Kanai:2025zgq,
    author = "Kanai, Takamasa and Maeda, Kengo and Yoshida, Daisuke",
    title = "{Wormholes as perturbations of near-horizon black hole geometries: No-go theorems within effective field theories}",
    eprint = "2511.21017",
    archivePrefix = "arXiv",
    doi = "10.1103/dsnv-gq47",
    journal = "Phys. Rev. D",
    volume = "113",
    number = "6",
    pages = "064026",
    year = "2026"
}

@article{Koga:2025bqw,
    author = "Koga, Yasutaka and Maeda, Ryota and Saito, Daiki and Uemichi, Keiya and Yoshida, Daisuke",
    title = "{Dynamical formation of charged wormholes}",
    eprint = "2505.20040",
    archivePrefix = "arXiv",
    doi = "10.1103/mtcj-13wm",
    journal = "Phys. Rev. D",
    volume = "113",
    number = "4",
    pages = "044035",
    year = "2026"
}

@article{delRio:2024kms,
    author = "del R{\'i}o, Adri{\'a}n and Ester, Evelyn-Andreea",
    title = "{Electrically charged black hole solutions in semiclassical gravity and dynamics of linear perturbations}",
    eprint = "2401.08783",
    archivePrefix = "arXiv",
    primaryClass = "gr-qc",
    doi = "10.1103/PhysRevD.109.105022",
    journal = "Phys. Rev. D",
    volume = "109",
    number = "10",
    pages = "105022",
    year = "2024"
}

@article{Lenzi:2026aay,
    author = "Lenzi, Michele and Mena Marug{\'a}n, Guillermo A. and M{\'i}nguez-S{\'a}nchez, Andr{\'e}s and Sopuerta, Carlos F.",
    title = "{Hamiltonian formalism, master functions and Darboux transformations for perturbed nonrotating black holes}",
    eprint = "2605.19829",
    archivePrefix = "arXiv",
    doi = "10.1007/s10714-026-03559-6",
    journal = "Gen. Rel. Grav.",
    volume = "58",
    number = "5",
    pages = "56",
    year = "2026"
}

@misc{Stashko:2026kja,
    author = "Stashko, Oleksandr S. and Konoplya, Roman A. and Hertzberg, Mark P.",
    title = "{Perturbations of charged black holes with higher-order interactions}",
    eprint = "2609.09419",
    archivePrefix = "arXiv",
    month = "9",
    year = "2026"
}

@misc{Sharma:2026gvu,
    author = "Sharma, Chanchal and Chakravarti, Kabir and Sarkar, Sudipta",
    title = "{Quasinormal ringdown and echoes in accreting exotic compact objects}",
    eprint = "2607.18919",
    archivePrefix = "arXiv",
    primaryClass = "gr-qc",
    month = "7",
    year = "2026"
}

@article{Liu:2025reu,
    author = "Liu, Hui-Fa and Zeng, Ding-fang",
    title = {{Master functions of Reissner-Nordstr{\"o}m black hole perturbations and their Darboux transformation}},
    eprint = "2505.04407",
    archivePrefix = "arXiv",
    primaryClass = "gr-qc",
    doi = "10.1103/148b-pfqr",
    journal = "Phys. Rev. D",
    volume = "112",
    number = "4",
    pages = "044025",
    year = "2025"
}

@article{deCesare:2024csp,
    author = "de Cesare, Marco and Oliveri, Roberto and Porfyriadis, Achilleas P.",
    title = {{Connecting gravitational perturbations: From Bertotti-Robinson to extreme Reissner-Nordstr{\"o}m black holes}},
    eprint = "2410.23446",
    archivePrefix = "arXiv",
    primaryClass = "gr-qc",
    reportNumber = "CCTP-2024-16, ITCP-IPP 2024/16",
    doi = "10.1103/PhysRevD.111.044028",
    journal = "Phys. Rev. D",
    volume = "111",
    number = "4",
    pages = "044028",
    year = "2025"
}

@misc{Maldacena:2018lmt,
    author = "Maldacena, Juan and Qi, Xiao-Liang",
    title = "{Eternal traversable wormhole}",
    eprint = "1804.00491",
    archivePrefix = "arXiv",
    primaryClass = "hep-th",
    month = "4",
    year = "2018"
}

@article{Maldacena:2018gjk,
    author = "Maldacena, Juan and Milekhin, Alexey and Popov, Fedor",
    title = "{Traversable wormholes in four dimensions}",
    eprint = "1807.04726",
    archivePrefix = "arXiv",
    primaryClass = "hep-th",
    doi = "10.1088/1361-6382/acde30",
    journal = "Class. Quant. Grav.",
    volume = "40",
    number = "15",
    pages = "155016",
    year = "2023"
}

@article{Cardoso:2016rao,
    author = "Cardoso, Vitor and Franzin, Edgardo and Pani, Paolo",
    title = "{Is the gravitational-wave ringdown a probe of the event horizon?}",
    eprint = "1602.07309",
    archivePrefix = "arXiv",
    primaryClass = "gr-qc",
    doi = "10.1103/PhysRevLett.116.171101",
    journal = "Phys. Rev. Lett.",
    volume = "116",
    number = "17",
    pages = "171101",
    year = "2016",
    note = "[Erratum: Phys.Rev.Lett. 117, 089902 (2016)]"
}

@article{Cardoso:2017cqb,
    author = "Cardoso, Vitor and Pani, Paolo",
    title = "{Tests for the existence of black holes through gravitational wave echoes}",
    eprint = "1709.01525",
    archivePrefix = "arXiv",
    primaryClass = "gr-qc",
    doi = "10.1038/s41550-017-0225-y",
    journal = "Nature Astron.",
    volume = "1",
    number = "9",
    pages = "586--591",
    year = "2017"
}

@article{Bueno:2017hyj,
    author = "Bueno, Pablo and Cano, Pablo A. and Goelen, Frederik and Hertog, Thomas and Vercnocke, Bert",
    title = "{Echoes of Kerr-like wormholes}",
    eprint = "1711.00391",
    archivePrefix = "arXiv",
    primaryClass = "gr-qc",
    doi = "10.1103/PhysRevD.97.024040",
    journal = "Phys. Rev. D",
    volume = "97",
    number = "2",
    pages = "024040",
    year = "2018"
}

@misc{Mondal:2025tht,
    author = "Mondal, Rajdeep and Sadhukhan, Abhishake",
    title = "{Echoes of the Maldacena-Milekhin-Popov traversable wormhole}",
    eprint = "2511.22671",
    archivePrefix = "arXiv",
    primaryClass = "gr-qc",
    month = "11",
    year = "2025"
}

@article{Riley:2026avn,
    author = "Riley, Jeff",
    title = "{Electromagnetic, gravitational wave, and static gravitational transmission through throat spacetimes: A constraint-wave asymmetry}",
    eprint = "2604.14238",
    archivePrefix = "arXiv",
    primaryClass = "gr-qc",
    doi = "10.1103/9xyc-f581",
    journal = "Phys. Rev. D",
    volume = "114",
    number = "4",
    pages = "044030",
    year = "2026"
}

@misc{Freivogel:2026ujn,
    author = "Freivogel, Ben and Fumagalli, Alessandro and Toma{\v{s}}evi{\'c}, Marija",
    title = "{How traversable is a traversable wormhole?}",
    eprint = "2606.12528",
    archivePrefix = "arXiv",
    primaryClass = "hep-th",
    reportNumber = "CERN-TH-2026-117",
    month = "6",
    year = "2026"
}

@article{Regge:1957td,
    author = "Regge, Tullio and Wheeler, John A.",
    title = "{Stability of a Schwarzschild singularity}",
    doi = "10.1103/PhysRev.108.1063",
    journal = "Phys. Rev.",
    volume = "108",
    pages = "1063--1069",
    year = "1957"
}

@article{Witten:1983ar,
    author = "Witten, Edward",
    editor = "Stone, M.",
    title = "{Nonabelian Bosonization in Two-Dimensions}",
    reportNumber = "PRINT-83-0934 (PRINCETON)",
    doi = "10.1007/BF01215276",
    journal = "Commun. Math. Phys.",
    volume = "92",
    pages = "455--472",
    year = "1984"
}

@article{Zerilli:1970se,
    author = "Zerilli, Frank J.",
    title = "{Effective potential for even parity Regge-Wheeler gravitational perturbation equations}",
    doi = "10.1103/PhysRevLett.24.737",
    journal = "Phys. Rev. Lett.",
    volume = "24",
    pages = "737--738",
    year = "1970"
}

@article{Vishveshwara:1970cc,
    author = "Vishveshwara, C. V.",
    title = "{Stability of the schwarzschild metric}",
    doi = "10.1103/PhysRevD.1.2870",
    journal = "Phys. Rev. D",
    volume = "1",
    pages = "2870--2879",
    year = "1970"
}

@article{Zerilli:1974ai,
    author = "Zerilli, F. J.",
    title = "{Perturbation analysis for gravitational and electromagnetic radiation in a reissner-nordstroem geometry}",
    doi = "10.1103/PhysRevD.9.860",
    journal = "Phys. Rev. D",
    volume = "9",
    pages = "860--868",
    year = "1974"
}

@article{Moncrief:1974am,
    author = "Moncrief, V.",
    title = "{Gravitational perturbations of spherically symmetric systems. I. The exterior problem.}",
    doi = "10.1016/0003-4916(74)90173-0",
    journal = "Annals Phys.",
    volume = "88",
    pages = "323--342",
    year = "1974"
}

@article{Moncrief:1974gw,
    author = "Moncrief, Vincent",
    title = "{Odd-parity stability of a Reissner-Nordstrom black hole}",
    doi = "10.1103/PhysRevD.9.2707",
    journal = "Phys. Rev. D",
    volume = "9",
    pages = "2707--2709",
    year = "1974"
}

@article{Moncrief:1974ng,
    author = "Moncrief, Vincent",
    title = "{Stability of Reissner-Nordstrom black holes}",
    doi = "10.1103/PhysRevD.10.1057",
    journal = "Phys. Rev. D",
    volume = "10",
    pages = "1057--1059",
    year = "1974"
}

@article{Chandrasekhar:1979iz,
    author = "Chandrasekhar, Subrahmanyan",
    editor = "Wali, K. C.",
    title = "{ON THE EQUATIONS GOVERNING THE PERTURBATIONS OF THE REISSNER-NORDSTROM BLACK HOLE}",
    doi = "10.1098/rspa.1979.0028",
    journal = "Proc. Roy. Soc. Lond. A",
    volume = "365",
    pages = "453--465",
    year = "1979"
}

@book{Chandra,
  author    = {Chandrasekhar, S.},
  title     = {The Mathematical Theory of Black Holes},
  publisher = {Oxford University Press},
  address   = {Oxford},
  year      = {1983},
  note      = {Chap.~5}
}

@article{Martel:2005ir,
    author = "Martel, Karl and Poisson, Eric",
    title = "{Gravitational perturbations of the Schwarzschild spacetime: A Practical covariant and gauge-invariant formalism}",
    eprint = "gr-qc/0502028",
    archivePrefix = "arXiv",
    doi = "10.1103/PhysRevD.71.104003",
    journal = "Phys. Rev. D",
    volume = "71",
    pages = "104003",
    year = "2005"
}

@article{Pereniguez:2023wxf,
    author = "Pere{\~n}iguez, David",
    title = "{Black hole perturbations and electric-magnetic duality}",
    eprint = "2302.10942",
    archivePrefix = "arXiv",
    primaryClass = "gr-qc",
    doi = "10.1103/PhysRevD.108.084046",
    journal = "Phys. Rev. D",
    volume = "108",
    number = "8",
    pages = "084046",
    year = "2023"
}

@article{DeFelice:2023rra,
    author = "De Felice, Antonio and Tsujikawa, Shinji",
    title = "{Can we distinguish black holes with electric and magnetic charges from quasinormal modes?}",
    eprint = "2312.03191",
    archivePrefix = "arXiv",
    primaryClass = "gr-qc",
    reportNumber = "YITP-23-158, WUCG-23-12",
    doi = "10.1103/PhysRevD.109.084022",
    journal = "Phys. Rev. D",
    volume = "109",
    number = "8",
    pages = "084022",
    year = "2024"
}

@article{Atiyah:1963zz,
    author = "Atiyah, M. F. and Singer, I. M.",
    title = "{The index of elliptic operators on compact manifolds}",
    doi = "10.1090/S0002-9904-1963-10957-X",
    journal = "Bull. Am. Math. Soc.",
    volume = "69",
    pages = "422--433",
    year = "1969"
}

@article{Wu:1976ge,
    author = "Wu, Tai Tsun and Yang, Chen Ning",
    title = "{Dirac Monopole Without Strings: Monopole Harmonics}",
    reportNumber = "ITP-SB-76-5",
    doi = "10.1016/0550-3213(76)90143-7",
    journal = "Nucl. Phys. B",
    volume = "107",
    pages = "365",
    year = "1976"
}

@article{Wen:1992ej,
    author = "Wen, X. G. and Zee, A.",
    title = "{Shift and spin vector: New topological quantum numbers for the Hall fluids}",
    reportNumber = "NSF-ITP-92-20",
    doi = "10.1103/PhysRevLett.69.953",
    journal = "Phys. Rev. Lett.",
    volume = "69",
    pages = "953--956",
    year = "1992",
    note = "[Erratum: Phys.Rev.Lett. 69, 3000 (1992)]"
}

@article{Abanov:2014ula,
    author = "Abanov, Alexander G. and Gromov, Andrey",
    title = "{Electromagnetic and gravitational responses of two-dimensional noninteracting electrons in a background magnetic field}",
    eprint = "1401.3703",
    archivePrefix = "arXiv",
    primaryClass = "cond-mat.str-el",
    doi = "10.1103/PhysRevB.90.014435",
    journal = "Phys. Rev. B",
    volume = "90",
    number = "1",
    pages = "014435",
    year = "2014"
}

@article{Gromov:2014gta,
    author = "Gromov, Andrey and Abanov, Alexander G.",
    title = "{Density-curvature response and gravitational anomaly}",
    eprint = "1403.5809",
    archivePrefix = "arXiv",
    primaryClass = "cond-mat.str-el",
    doi = "10.1103/PhysRevLett.113.266802",
    journal = "Phys. Rev. Lett.",
    volume = "113",
    pages = "266802",
    year = "2014"
}

@article{Bradlyn:2014wla,
    author = "Bradlyn, Barry and Read, N.",
    title = "{Low-energy effective theory in the bulk for transport in a topological phase}",
    eprint = "1407.2911",
    archivePrefix = "arXiv",
    primaryClass = "cond-mat.mes-hall",
    doi = "10.1103/PhysRevB.91.125303",
    journal = "Phys. Rev. B",
    volume = "91",
    number = "12",
    pages = "125303",
    year = "2015",
    note = "[Erratum: Phys.Rev.B 93, 239902 (2016)]"
}

@article{Dirac:1931kp,
  author  = "Dirac, P. A. M.",
  title   = "{Quantised singularities in the electromagnetic field}",
  journal = "Proc. R. Soc. A",
  volume  = "133",
  pages   = "60",
  year    = "1931",
  doi     = "10.1098/rspa.1931.0130"
}

@misc{Sadhukhan:2026czl,
    author = "Sadhukhan, Abhishake",
    title = "{Polar-metric/axial-gauge perturbations of the Maldacena-Milekhin-Popov wormhole}",
    eprint = "2609.33511",
    archivePrefix = "arXiv",
    primaryClass = "gr-qc",
    month = "9",
    year = "2026"
}

@article{Schwinger:1962tp,
    author = "Schwinger, Julian S.",
    title = "{Gauge Invariance and Mass. II}",
    doi = "10.1103/PhysRev.128.2425",
    journal = "Phys. Rev.",
    volume = "128",
    pages = "2425--2429",
    year = "1962"
}

@article{Coleman:1975pw,
    author = "Coleman, Sidney R.",
    title = "{The Quantum Sine-Gordon Equation as the Massive Thirring Model}",
    doi = "10.1103/PhysRevD.11.2088",
    journal = "Phys. Rev. D",
    volume = "11",
    pages = "2088",
    year = "1975"
}

@article{Mandelstam:1975hb,
    author = "Mandelstam, Stanley",
    title = "{Soliton Operators for the Quantized Sine-Gordon Equation}",
    doi = "10.1103/PhysRevD.11.3026",
    journal = "Phys. Rev. D",
    volume = "11",
    pages = "3026",
    year = "1975"
}

@article{Michelson:1999zc,
    author = "Michelson, Jeremy and Spradlin, Marcus",
    title = "{Supergravity spectrum on AdS(2) x S**2}",
    eprint = "hep-th/9906056",
    archivePrefix = "arXiv",
    doi = "10.1088/1126-6708/1999/09/029",
    journal = "JHEP",
    number ="09",
    volume = "09",
    pages = "029",
    year = "1999"
}

@article{Alford:1992yx,
    author = "Alford, Mark G. and Strominger, Andrew",
    title = "{S wave scattering of charged fermions by a magnetic black hole}",
    eprint = "hep-th/9202075",
    archivePrefix = "arXiv",
    doi = "10.1103/PhysRevLett.69.563",
    journal = "Phys. Rev. Lett.",
    volume = "69",
    pages = "563--566",
    year = "1992"
}

@article{Lowenstein:1971fc,
    author = "Lowenstein, J. H. and Swieca, J. A.",
    title = "{Quantum electrodynamics in two dimensions}",
    doi = "10.1016/0003-4916(71)90246-7",
    journal = "Annals Phys.",
    volume = "68",
    pages = "172--195",
    year = "1971"
}

@article{ColemanJackiwSusskind:1975,
    author = "Coleman, Sidney R. and Jackiw, R. and Susskind, Leonard",
    title = "{Charge Shielding and Quark Confinement in the Massive Schwinger Model}",
    doi = "10.1016/0003-4916(75)90212-2",
    journal = "Annals Phys.",
    volume = "93",
    pages = "267",
    year = "1975"
}

@article{Callan:1982ah,
    author = "Callan, Jr., Curtis G.",
    title = "{Dyon-Fermion Dynamics}",
    doi = "10.1103/PhysRevD.26.2058",
    journal = "Phys. Rev. D",
    volume = "26",
    pages = "2058--2068",
    year = "1982"
}

@article{Rubakov:1982fp,
    author = "Rubakov, V. A.",
    title = "{Adler-Bell-Jackiw Anomaly and Fermion Number Breaking in the Presence of a Magnetic Monopole}",
    doi = "10.1016/0550-3213(82)90034-7",
    journal = "Nucl. Phys. B",
    volume = "203",
    pages = "311--348",
    year = "1982"
}

@article{Aharonov:1978gb,
    author = "Aharonov, Y. and Casher, A.",
    title = "{Ground state of a spin-1/2 charged particle in a two-dimensional magnetic field}",
    doi = "10.1103/PhysRevA.19.2461",
    journal = "Phys. Rev. A",
    volume = "19",
    pages = "2461--2462",
    year = "1979"
}

@article{Adler:1969gk,
    author = "Adler, Stephen L.",
    title = "{Axial vector vertex in spinor electrodynamics}",
    doi = "10.1103/PhysRev.177.2426",
    journal = "Phys. Rev.",
    volume = "177",
    pages = "2426--2438",
    year = "1969"
}

@article{Bell:1969ts,
    author = "Bell, J. S. and Jackiw, R.",
    title = "{A PCAC puzzle: $\pi^0 \to \gamma \gamma$ in the $\sigma$ model}",
    doi = "10.1007/BF02823296",
    journal = "Nuovo Cim. A",
    volume = "60",
    pages = "47--61",
    year = "1969"
}

@article{Gralla:2014yja,
    author = "Gralla, Samuel E. and Jacobson, Ted",
    title = "{Spacetime approach to force-free magnetospheres}",
    eprint = "1401.6159",
    archivePrefix = "arXiv",
    primaryClass = "astro-ph.HE",
    doi = "10.1093/mnras/stu1690",
    journal = "Mon. Not. Roy. Astron. Soc.",
    volume = "445",
    number = "3",
    pages = "2500--2534",
    year = "2014",
    note = "[Erratum: Mon.Not.Roy.Astron.Soc. 534, 1541 (2024)]"
}

@misc{Girvin:1999,
    title="The Quantum Hall Effect: Novel Excitations and Broken Symmetries",
    author="Steven M. Girvin",
    year="1999",
    eprint="cond-mat/9907002",
    archivePrefix={arXiv},
    primaryClass={cond-mat.mes-hall}
}

@book{Jacksonelec,
  author    = {Jackson, John David},
  title     = {Classical Electrodynamics},
  edition   = {3rd},
  publisher = {Wiley},
  address   = {New York, NY},
  year      = {1999},
  isbn      = {978-0471309321}
}

@book{Synge:1956,
  author    = {John Lighton Synge},
  title     = {Relativity: The Special Theory},
  year      = {1956},
  publisher = {North-Holland Publishing Company},
  address   = {Amsterdam}
}

@book{Landau:1975,
  title     = {The Classical Theory of Fields},
  author    = {Landau, L. D. and Lifshitz, E. M.},
  volume    = {2},
  series    = {Course of Theoretical Physics},
  edition   = {4th},
  year      = {1975},
  publisher = {Pergamon Press},
  address   = {Oxford},
  isbn      = {978-0080181769}
}

@article{Graham:2007va,
    author = "Graham, Noah and Olum, Ken D.",
    title = "{Achronal averaged null energy condition}",
    eprint = "0705.3193",
    archivePrefix = "arXiv",
    primaryClass = "gr-qc",
    doi = "10.1103/PhysRevD.76.064001",
    journal = "Phys. Rev. D",
    volume = "76",
    pages = "064001",
    year = "2007"
}

@article{Simon:1976un,
    author = "Simon, B.",
    title = "{The Bound State of Weakly Coupled Schrodinger Operators in One and Two-Dimensions}",
    doi = "10.1016/0003-4916(76)90038-5",
    journal = "Annals Phys.",
    volume = "97",
    pages = "279--288",
    year = "1976"
}

@article{Ferguson:1999,
  author  = {Ferguson, Helaman R. P. and Bailey, David H. and Arno, Steve},
  title   = {Analysis of {PSLQ}, an integer relation finding algorithm},
  journal = {Math. Comp.},
  volume  = {68},
  pages   = {351--369},
  year    = {1999},
  doi     = {10.1090/S0025-5718-99-00995-3}
}

@article{Chandrasekhar:1975nkd,
    author = "Chandrasekhar, Subrahmanyan",
    title = "{On the equations governing the perturbations of the Schwarzschild black hole}",
    doi = "10.1098/rspa.1975.0066",
    journal = "Proc. Roy. Soc. Lond. A",
    volume = "343",
    number = "1634",
    pages = "289--298",
    year = "1975"
}

@book{Kato:1995,
  author = {Kato, Tosio},
  title = {Perturbation Theory for Linear Operators},
  series = {Classics in Mathematics},
  publisher = {Springer},
  address = {Berlin},
  year = {1995}
}

@article{Takahasi:1974,
  author  = {Takahasi, Hidetosi and Mori, Masatake},
  title   = {Double exponential formulas for numerical integration},
  journal = {Publ. Res. Inst. Math. Sci.},
  volume  = {9},
  number  = {3},
  pages   = {721--741},
  year    = {1974},
  doi     = {10.2977/prims/1195192451}
}

@article{Prince:1981,
  author  = {Prince, P. J. and Dormand, J. R.},
  title   = {High order embedded {Runge-Kutta} formulae},
  journal = {J. Comput. Appl. Math.},
  volume  = {7},
  number  = {1},
  pages   = {67--75},
  year    = {1981},
  doi     = {10.1016/0771-050X(81)90010-3}
}

@book{Lehoucq:1998,
  author    = {Lehoucq, R. B. and Sorensen, D. C. and Yang, C.},
  title     = {{ARPACK} Users' Guide: Solution of Large-Scale Eigenvalue Problems with Implicitly Restarted {Arnoldi} Methods},
  publisher = {SIAM},
  address   = {Philadelphia},
  year      = {1998},
  doi       = {10.1137/1.9780898719628}
}

@article{Brent:1971,
  author  = {Brent, Richard P.},
  title   = {An algorithm with guaranteed convergence for finding a zero of a function},
  journal = {Comput. J.},
  volume  = {14},
  number  = {4},
  pages   = {422--425},
  year    = {1971},
  doi     = {10.1093/comjnl/14.4.422}
}

@article{Ishibashi:2004wx,
    author = "Ishibashi, Akihiro and Wald, Robert M.",
    title = "{Dynamics in nonglobally hyperbolic static space-times. 3. Anti-de Sitter space-time}",
    eprint = "hep-th/0402184",
    archivePrefix = "arXiv",
    doi = "10.1088/0264-9381/21/12/012",
    journal = "Class. Quant. Grav.",
    volume = "21",
    pages = "2981--3014",
    year = "2004"
}

@article{Cooper:1994eh,
    author = "Cooper, Fred and Khare, Avinash and Sukhatme, Uday",
    title = "{Supersymmetry and quantum mechanics}",
    eprint = "hep-th/9405029",
    archivePrefix = "arXiv",
    reportNumber = "LA-UR-94-569",
    doi = "10.1016/0370-1573(94)00080-M",
    journal = "Phys. Rept.",
    volume = "251",
    pages = "267--385",
    year = "1995"
}

@article{Klaus:1980ks,
    author = "Klaus, Martin and Simon, Barry",
    title = "{Coupling constant thresholds in nonrelativistic quantum mechanics. I. Short-range two-body case}",
    doi = "10.1016/0003-4916(80)90338-3",
    journal = "Annals Phys.",
    volume = "130",
    number = "2",
    pages = "251--281",
    year = "1980"
}

@book{BornHuang:1954,
    author = "Born, Max and Huang, Kun",
    title = "{Dynamical Theory of Crystal Lattices}",
    publisher = "Clarendon Press",
    address = "Oxford",
    year = "1954"
}

@article{Fu:2018oaq,
    author = "Fu, Zicao and Grado-White, Brianna and Marolf, Donald",
    title = "{A perturbative perspective on self-supporting wormholes}",
    eprint = "1807.07917",
    archivePrefix = "arXiv",
    primaryClass = "hep-th",
    doi = "10.1088/1361-6382/aafcea",
    journal = "Class. Quant. Grav.",
    volume = "36",
    number = "4",
    pages = "045006",
    year = "2019",
    note = "[Erratum: Class.Quant.Grav. 36, 249501 (2019)]"
}

@article{Konoplya:2016hmd,
    author = "Konoplya, R. A. and Zhidenko, A.",
    title = "{Wormholes versus black holes: quasinormal ringing at early and late times}",
    eprint = "1606.00517",
    archivePrefix = "arXiv",
    primaryClass = "gr-qc",
    doi = "10.1088/1475-7516/2016/12/043",
    journal = "JCAP",
    volume = "12",
    pages = "043",
    year = "2016"
}

@article{Gonzalez:2008wd,
    author = "Gonzalez, J. A. and Guzman, F. S. and Sarbach, O.",
    title = "{Instability of wormholes supported by a ghost scalar field. I. Linear stability analysis}",
    eprint = "0806.0608",
    archivePrefix = "arXiv",
    primaryClass = "gr-qc",
    reportNumber = "UMSNH-IFM-F-2008-18",
    doi = "10.1088/0264-9381/26/1/015010",
    journal = "Class. Quant. Grav.",
    volume = "26",
    pages = "015010",
    year = "2009"
}

@article{Bronnikov:2012ch,
    author = "Bronnikov, K. A. and Konoplya, R. A. and Zhidenko, A.",
    title = "{Instabilities of wormholes and regular black holes supported by a phantom scalar field}",
    eprint = "1205.2224",
    archivePrefix = "arXiv",
    primaryClass = "gr-qc",
    doi = "10.1103/PhysRevD.86.024028",
    journal = "Phys. Rev. D",
    volume = "86",
    pages = "024028",
    year = "2012"
}

@article{Wald:1979lth,
    author = "Wald, Robert M.",
    title = "{Note on the stability of the Schwarzschild metric}",
    doi = "10.1063/1.524181",
    journal = "J. Math. Phys.",
    volume = "20",
    number = "6",
    pages = "1056",
    year = "1979"
}

@article{Gusynin:1999pq,
    author = "Gusynin, V. P. and Miransky, V. A. and Shovkovy, I. A.",
    title = "{Theory of the magnetic catalysis of chiral symmetry breaking in QED}",
    eprint = "hep-ph/9908320",
    archivePrefix = "arXiv",
    reportNumber = "UCTP-103-99",
    doi = "10.1016/S0550-3213(99)00573-8",
    journal = "Nucl. Phys. B",
    volume = "563",
    pages = "361--389",
    year = "1999"
}

@article{Fukushima:2011nu,
    author = "Fukushima, Kenji",
    title = "{Magnetic-field Induced Screening Effect and Collective Excitations}",
    eprint = "1103.4430",
    archivePrefix = "arXiv",
    primaryClass = "hep-ph",
    doi = "10.1103/PhysRevD.83.111501",
    journal = "Phys. Rev. D",
    volume = "83",
    pages = "111501",
    year = "2011"
}

@article{Kharzeev:2010gd,
    author = "Kharzeev, Dmitri E. and Yee, Ho-Ung",
    title = "{Chiral Magnetic Wave}",
    eprint = "1012.6026",
    archivePrefix = "arXiv",
    primaryClass = "hep-th",
    doi = "10.1103/PhysRevD.83.085007",
    journal = "Phys. Rev. D",
    volume = "83",
    pages = "085007",
    year = "2011"
}

@article{Moreno:2002gg,
    author = "Moreno, Claudia and Sarbach, Olivier",
    title = "{Stability properties of black holes in self-gravitating nonlinear electrodynamics}",
    eprint = "gr-qc/0208090",
    archivePrefix = "arXiv",
    doi = "10.1103/PhysRevD.67.024028",
    journal = "Phys. Rev. D",
    volume = "67",
    pages = "024028",
    year = "2003"
}

@article{Chew:1956cgl,
    author = "Chew, G. F. and Goldberger, M. L. and Low, F. E.",
    title = "{The Boltzmann equation and the one-fluid hydromagnetic equations in the absence of particle collisions}",
    doi = "10.1098/rspa.1956.0116",
    journal = "Proc. R. Soc. Lond. A",
    volume = "236",
    pages = "112--118",
    year = "1956"
}

@article{MeadTruhlar:1979,
    author = "Mead, C. Alden and Truhlar, Donald G.",
    title = "{On the determination of Born--Oppenheimer nuclear motion wave functions including complications due to conical intersections and identical nuclei}",
    doi = "10.1063/1.437734",
    journal = "J. Chem. Phys.",
    volume = "70",
    pages = "2284",
    year = "1979"
}

@article{Anderson:2002fk,
    author = "Anderson, Paul R. and Molina-Par{\'\i}s, Carmen and Mottola, Emil",
    title = "{Linear response, validity of semiclassical gravity, and the stability of flat space}",
    eprint = "gr-qc/0209075",
    archivePrefix = "arXiv",
    doi = "10.1103/PhysRevD.67.024026",
    journal = "Phys. Rev. D",
    volume = "67",
    pages = "024026",
    year = "2003"
}

@article{Heisenberg:1936nmg,
    author = "Heisenberg, W. and Euler, H.",
    title = "{Consequences of Dirac's theory of positrons}",
    eprint = "physics/0605038",
    archivePrefix = "arXiv",
    doi = "10.1007/BF01343663",
    journal = "Z. Phys.",
    volume = "98",
    number = "11-12",
    pages = "714--732",
    year = "1936"
}

@article{Schwinger:1951nm,
    author = "Schwinger, Julian S.",
    editor = "Milton, K. A.",
    title = "{On gauge invariance and vacuum polarization}",
    doi = "10.1103/PhysRev.82.664",
    journal = "Phys. Rev.",
    volume = "82",
    pages = "664--679",
    year = "1951"
}

@inbook{Dunne:2004nc,
    author = "Dunne, Gerald V.",
    editor = "Shifman, M. and Vainshtein, A. and Wheater, J.",
    title = "{Heisenberg-Euler effective Lagrangians: Basics and extensions}",
    booktitle = "{From fields to strings: Circumnavigating theoretical physics. Ian Kogan memorial collection (3 volume set)}",
    publisher = "World Scientific",
    eprint = "hep-th/0406216",
    archivePrefix = "arXiv",
    doi = "10.1142/9789812775344_0014",
    pages = "445--522",
    month = "6",
    year = "2004"
}

@article{Nomura:2020tpc,
    author = "Nomura, Kimihiro and Yoshida, Daisuke and Soda, Jiro",
    title = "{Stability of magnetic black holes in general nonlinear electrodynamics}",
    eprint = "2004.07560",
    archivePrefix = "arXiv",
    primaryClass = "gr-qc",
    reportNumber = "KOBE-COSMO-20-08",
    doi = "10.1103/PhysRevD.101.124026",
    journal = "Phys. Rev. D",
    volume = "101",
    number = "12",
    pages = "124026",
    year = "2020"
}

@article{Blazquez-Salcedo:2018ipc,
    author = "Bl{\'a}zquez-Salcedo, Jose Luis and Chew, Xiao Yan and Kunz, Jutta",
    title = "{Scalar and axial quasinormal modes of massive static phantom wormholes}",
    eprint = "1806.03282",
    archivePrefix = "arXiv",
    primaryClass = "gr-qc",
    doi = "10.1103/PhysRevD.98.044035",
    journal = "Phys. Rev. D",
    volume = "98",
    number = "4",
    pages = "044035",
    year = "2018"
}

@article{Azad:2022qqn,
    author = "Azad, Bahareh and Bl{\'a}zquez-Salcedo, Jose Luis and Chew, Xiao Yan and Kunz, Jutta and Yeom, Dong-han",
    title = "{Polar modes and isospectrality of Ellis-Bronnikov wormholes}",
    eprint = "2212.12601",
    archivePrefix = "arXiv",
    primaryClass = "gr-qc",
    doi = "10.1103/PhysRevD.107.084024",
    journal = "Phys. Rev. D",
    volume = "107",
    number = "8",
    pages = "084024",
    year = "2023"
}
\end{document}